\documentclass[%
 reprint,
superscriptaddress,
 amsmath,amssymb,
prx,
floatfix,
]{revtex4-2}

\usepackage{mathtools}
\usepackage{graphicx}% Include figure files
\usepackage{dcolumn}% Align table columns on decimal point
\usepackage{bm}% bold math

\usepackage[caption=false]{subfig}

\usepackage{algorithm}
\usepackage[noend]{algpseudocode}
\algrenewcommand\algorithmicdo{}

\makeatletter
\renewcommand{\ALG@name}{Procedure}
\makeatother

\newcommand{\argmax}{\mathop{\rm arg~max}\limits}

\usepackage[linktocpage=true,
  colorlinks=true, 
  pdfborder={0 0 0},
  linkcolor=blue,
  citecolor=red,
  filecolor=yellow,
  urlcolor=blue,
  bookmarks,
  pdfauthor={},
]{hyperref}

\usepackage{orcidlink}

\usepackage{ifthen}
\newcounter{is_qcircuit_used}
\newcounter{are_figs_merged}
\begin{document}

\preprint{APS/123-QED}

\title{
Exhaustive search for optimal molecular geometries using\\
imaginary-time evolution on a quantum computer
}

\author{Taichi Kosugi\orcidlink{0000-0003-3379-3361}}
\email{kosugi.taichi@gmail.com}
\affiliation{
Laboratory for Materials and Structures,
Institute of Innovative Research,
Tokyo Institute of Technology,
Yokohama 226-8503,
Japan
}

\affiliation{
Quemix Inc.,
Taiyo Life Nihombashi Building,
2-11-2,
Nihombashi Chuo-ku, 
Tokyo 103-0027,
Japan
}

\author{Hirofumi Nishi\orcidlink{0000-0001-5155-6605}}
\affiliation{
Laboratory for Materials and Structures,
Institute of Innovative Research,
Tokyo Institute of Technology,
Yokohama 226-8503,
Japan
}

\affiliation{
Quemix Inc.,
Taiyo Life Nihombashi Building,
2-11-2,
Nihombashi Chuo-ku, 
Tokyo 103-0027,
Japan
}

\author{Yu-ichiro Matsushita\orcidlink{0000-0002-9254-5918}}

\affiliation{
Laboratory for Materials and Structures,
Institute of Innovative Research,
Tokyo Institute of Technology,
Yokohama 226-8503,
Japan
}

\affiliation{
Quemix Inc.,
Taiyo Life Nihombashi Building,
2-11-2,
Nihombashi Chuo-ku, 
Tokyo 103-0027,
Japan
}

\affiliation{
Quantum Material and Applications Research Center,
National Institutes for Quantum Science and Technology,
2-12-1, Ookayama, Meguro-ku, Tokyo 152-8552, Japan
}

\date{\today}

\begin{abstract}
This study proposes a nonvariational scheme for geometry
optimization of molecules for the first-quantized eigensolver,
which is a recently proposed framework for quantum chemistry using
probabilistic imaginary-time evolution (PITE).
In this scheme, the nuclei in a molecule are treated as classical point charges while the electrons are treated as quantum mechanical particles.
The electronic states and candidate geometries are encoded as a superposition of many-qubit states,
for which a histogram created from repeated measurements gives the global minimum of the energy surface.
We demonstrate that the circuit depth per step scales as
$\mathcal{O} (n_e^2 \mathrm{poly}(\log n_e))$
for the electron number $n_e,$ which can be reduced to
$\mathcal{O} (n_e \mathrm{poly}(\log n_e))$
if extra $\mathcal{O} (n_e \log n_e)$ qubits are available.
Moreover, resource estimation implies that the total computational time of our scheme starting from a good initial guess may exhibit overall quantum advantage in molecule size and candidate number.
The proposed scheme is corroborated using numerical simulations.
Additionally, a scheme adapted to variational calculations is examined that prioritizes saving circuit depths for noisy intermediate-scale quantum (NISQ) devices.
A classical system composed only of charged particles is considered as a special case of the scheme.
The new efficient scheme will assist in achieving scalability in practical quantum chemistry on quantum computers.
\end{abstract}

\maketitle 

\section{Introduction}
\label{sec:introduction}

Modern computational designs for materials \cite{bib:5744},
proteins \cite{bib:5745}, and drug discovery \cite{bib:5747} often include atomistic simulations instead of coarse-grained models for distinguishing microscopic subtleties.
Electronic-structure calculations based on the density functional theory  \cite{bib:76, bib:77} or wave function theory \cite{Helgaker}
must be performed to optimize the geometries of solids and molecules in their ground states to ensure that simulations are as quantitatively reliable as possible.
Although target systems with a diverse number of atoms and elements are found in physics, chemistry, and biochemistry,
there are two main approaches for determining the optimal geometry of a molecule using a classical computer: energy- and force-based.

The energy-based approach is based on the calculated total energies of all the candidate geometries.
The procedure in a naive form typically begins by determining the discretization of the positions for each nucleus and
calculating the total energies of all possible geometries.
This approach leads to an exhaustive search for the optimal geometry among all candidates and the search can be easily parallelized for many classical computers. However, the required computational resources grow exponentially with respect to the size of the target molecule.
This extensive scaling makes the naive energy-based approach impractical for systems of practical interest.

The force-based approach is based on the forces acting on the nuclei within the Born--Oppenheimer (BO) approximation.
This optimization procedure for a target molecule is performed by calculating the total energy and forces acting on the constituent nuclei.
More precisely, the procedure typically calculates the Hellmann--Feynman forces \cite{bib:32}.
If necessary, the Pulay forces are calculated to compensate for the incompleteness of the adopted basis set \cite{bib:5743}.
These forces can be calculated using only a small amount of additional computational resources for the total-energy calculation.
The nuclear positions are iteratively updated until convergence according to the forces.
The steepest-descent and conjugate-gradient methods are force-based approaches in the simplest forms.
However, the updating process used in these methods is not parallelizable in principle.
In addition, the search is prone to becoming stuck in a local minimum on the energy surface.
Various elaborate force-based approaches have been proposed to achieve the efficient and robust optimization of molecular geometries.
For details, refer to Ref. \cite{bib:5742}.

While quantum computation has been regarded as a promising alternative for storing many-electron wave functions living in a huge Hilbert space \cite{bib:4828} since long before the advent of quantum computers,
we find that geometry optimization of electronic systems is still going through the phase of establishing basic techniques, on the contrary to classical computation.
Hirai et al.~\cite{bib:5752} proposed recently a method within the first-quantized formalism \cite{bib:5373, bib:5372, bib:5328} by finding the lowest-energy geometry based on the imaginary-time evolution (ITE) with variational parameters \cite{bib:4797, bib:4802, bib:4807} for nonadiabatically coupled electrons and nuclei.
Their approach, which we refer to as the variational ITE (VITE) in what follows, is a kind of the variational quantum eigensolver (VQE) \cite{bib:4470, bib:4517}.
The major difference between our approach described later and their approach exists in how the qubits for nuclear degrees of freedom are used:
we use them to encode the nuclear positions as classical data instead of their femtometer-scale wave functions,
so that we perform exhaustive search for the optimum among candidates via quantum parallelism.
We point out here that a quantum algorithm for force-based geometry optimization has been proposed \cite{bib:6046}.

Since the prevalent paradigm of electronic-structure calculations 
on classical computers has been developed primarily for computing the total energies of systems built up of electrons and nuclei,
we might overlook the important fact, that is,
there is no need for knowing the values of the total energies of candidate geometries to find the optimal one.
We can find it only by knowing which geometry has the unknown lowest energy.
Given this fact and the first-quantized eigensolver (FQE) \cite{bib:5737},
this study presents a quantum algorithm for efficient geometry optimization that outperforms classical algorithms.
FQE is a recently proposed framework based on probabilistic ITE (PITE) for nonvariational energy minimization in quantum chemistry \cite{bib:5737}.
For a brief review of generic PITE,
see Appendix \ref{sec:review_of_PITE}.
The second-quantized formalism is useful for calculating the dynamical properties related to the excitation processes of a molecule, where the electron number can increase and decrease \cite{bib:5005, bib:5163}. However, the first-quantized formalism for finding the ground state offers better scaling of operation numbers \cite{bib:5737}.
This characteristic is inherited even when geometry optimization is involved, as will be demonstrated later.

\section{Results}

\subsection{Exhaustive search for optimal geometries}
\label{sec:exhaustive_search}

\begin{figure}
\begin{center}
\includegraphics[width=8cm]{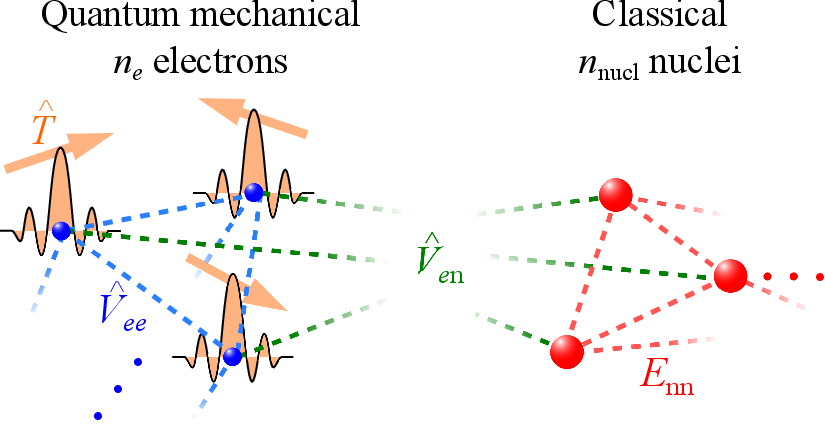}
\end{center}
\caption{
Setup of a geometry optimization problem considered in this paper.
We treat the $n_e$ electrons contained in a target molecule as quantum mechanical particles having the kinetic energies $\hat{T}$,
while the $n_{\mathrm{nucl}}$ nuclei as fixed classical point charges.
The Hamiltonian of the total system involves
the electron-electron interactions
$\hat{V}_{ee},$
the electron-nucleus interactions
$\hat{V}_{e \mathrm{n}},$
and the nucleus-nucleus interactions
$E_{\mathrm{n n}}.$
The electrons can feel an external field $\hat{V}_{\mathrm{ext}}.$
}
\label{fig:particles}
\end{figure}

Let us consider a molecular system consisting of $n_e$ electrons as quantum mechanical particles and
$n_{\mathrm{nucl}}$ nuclei as classical point charges fixed at
$\boldsymbol{R}_\nu \ (\nu = 0, \dots, n_{\mathrm{nucl}} - 1),$
as depicted in Fig.~\ref{fig:particles}.
These two kinds of particles interact with each other via pairwise interactions $v$ dependent only on the distance between two particles.
The Hamiltonian is given by
\begin{gather}
    \mathcal{H}
    \left(
        \{
            \boldsymbol{R}_\nu
        \}_\nu
    \right)
    =
        \underbrace{
            \sum_{\ell = 0}^{n_e - 1}
            \frac{\hat{\boldsymbol{p}}_\ell^2}{2 m_e}
        }_{\equiv \hat{T}}
        +
        \underbrace{
            \frac{1}{2}
            \sum_{\substack{ \ell, \ell' = 0 \\ (\ell \ne \ell')}}^{n_e - 1}
            v
            \left(
                |
                \hat{\boldsymbol{r}}_\ell
                -
                \hat{\boldsymbol{r}}_{\ell'}
                |
            \right)
        }_{\equiv \hat{V}_{ee}}
    \nonumber \\
        +
        \underbrace{
            \sum_{\ell = 0}^{n_e - 1}
            \sum_{\nu = 0}^{n_{\mathrm{nucl}} - 1}
            -Z_\nu
            v
            \left(
                |
                \hat{\boldsymbol{r}}_\ell
                -
                \boldsymbol{R}_\nu
                |
            \right)
        }_{\equiv \hat{V}_{e \mathrm{n}}}
        \nonumber \\
        +
        \underbrace{
            \frac{1}{2}
            \sum_{\substack{ \nu, \nu' = 0 \\ (\nu \ne \nu')}}^{n_{\mathrm{nucl}} - 1}
            Z_{\nu}
            Z_{\nu'}
            v
            \left(
                |
                \boldsymbol{R}_\nu
                -
                \boldsymbol{R}_{\nu'}
                |
            \right)
        }_{\equiv E_{\mathrm{nn}}}
    +
        \underbrace{
            \sum_{\ell = 0}^{n_e - 1}
            v_{\mathrm{ext}}
            \left(
                \hat{\boldsymbol{r}}_\ell
            \right)
        }_{\equiv \hat{V}_{\mathrm{ext}}}
    ,
    \label{struct_opt_using_FQE:def_Hamiltonian}
\end{gather}
where the nuclear positions appear as parameters.
$\hat{T}$ is the kinetic-energy operator of electrons having the mass $m_e = 1.$
All the quantities in this paper are in atomic units unless otherwise stated.
$\hat{\boldsymbol{r}}_{\ell}$
and
$\hat{\boldsymbol{p}}_\ell$
are the position and momentum
operators, respectively, of the $\ell$th electron.
$Z_\nu$ is the charge of the $\nu$th nucleus, while that of an electron is $-1.$
We can introduce a position-dependent external field $v_{\mathrm{ext}}$ felt by each electron.
Although we have adopted the common interaction $v$ for 
$\hat{V}_{\mathrm{e e}}, \hat{V}_{e \mathrm{n}},$
and $\hat{V}_{\mathrm{n n}}$ for simplicity,
distinct interactions for them could be introduced with only small modifications to the following discussion.
Also, the formulations for one- and two-dimensional spaces will be possible similarly to the three-dimensional case.

We encode the $n_e$-electron wave function in real space
by using $n_{q e}$ qubits for each direction per electron,
as usual in the first-quantized formalism \cite{bib:5373, bib:5372, bib:5328, bib:5824, bib:5737, bib:5658},
or equivalently the grid-based formalism.
We refer to the $3 n_e n_{q e}$ qubits collectively as the electronic register.
We generate uniform grid points in a cubic simulation cell of size $L$
to encode the normalized many-electron spatial wave function $\psi$ by using the register as
\begin{gather}
    | \psi \rangle
    =
        \Delta V^{n_e/2}
        \sum_{
            \boldsymbol{k}_0,
            \dots, 
            \boldsymbol{k}_{n_e - 1}
        }
        \psi (
            \boldsymbol{r}^{(\boldsymbol{k}_0)},
            \dots,
            \boldsymbol{r}^{(\boldsymbol{k}_{n_e - 1})}
        )
    \cdot
    \nonumber \\
    \cdot
        | \boldsymbol{k}_0
        \rangle_{3 n_{q e}}
        \otimes \cdots \otimes
        | \boldsymbol{k}_{n_e - 1} 
        \rangle_{3 n_{q e}}
    ,
    \label{many_electron_state} 
\end{gather}
where
$\boldsymbol{k}_\ell$ is the three integers specifying 
the position eigenvalue
$
(
k_{\ell x} \boldsymbol{e}_x +
k_{\ell y} \boldsymbol{e}_y +
k_{\ell z} \boldsymbol{e}_z
) \Delta x
$
for the $\ell$th electron.
$\Delta x \equiv L/N_{q e}$ is the spacing of
$N_{q e} \equiv 2^{n_{q e}}$ grid points for each direction.
We introduced the volume element
$\Delta V \equiv \Delta x^3$
for the normalization of $| \psi \rangle.$

We construct a composite system consisting of 
the electrons and nuclei
and define an appropriate Hamiltonian,
for which we perform energy minimization based on PITE
to find the optimal combination
$\{ \Delta \boldsymbol{R}_\nu^{(\mathrm{opt})} \}_{\nu}$
of displacements
from the original positions $\{ \boldsymbol{R}_{\nu 0} \}_{\nu}.$
To this end,
we first decide upon the largest possible displacement
$\Delta R_{\nu \mu \mathrm{max}} \ (\mu = x, y, z)$
in each direction $\mu$ for each nucleus $\nu.$
We introduce $n_{q \mathrm{n}}$ qubits for encoding the displacement in each direction for each nucleus.
Specifically, we define the $x$ position operator $\hat{\mathcal{R}}_{\nu x}$ of the $\nu$th nucleus such that
each of the computational basis
$| j_{\nu x} \rangle_{n_{q \mathrm{n}}} \ (j_{\nu x} = 0, \dots, 2^{n_{q \mathrm{n}}} - 1)$
is the eigenstate as follows:
\begin{gather}
    \hat{\mathcal{R}}_{\nu x}
    | j_{\nu x} \rangle_{n_{q \mathrm{n}}}
    \equiv
        \left(
            R_{\nu 0 x}
            +
            j_{\nu x}
            \frac{\Delta R_{\nu x \mathrm{max}}}{N_{q \mathrm{n}}}
        \right)
        | j_{\nu x} \rangle_{n_{q \mathrm{n}}}
    ,
    \label{struct_opt_using_FQE:def_displaced_pos_of_nucl}
\end{gather}
where
$N_{q \mathrm{n}} \equiv 2^{n_{q \mathrm{n}}}.$
The operators $\hat{\mathcal{R}}_{\nu y}$ and
$\hat{\mathcal{R}}_{\nu z}$ for the $y$ and $z$ positions,
respectively, are defined similarly. 
We refer to the $3 n_{\mathrm{nucl}} n_{q \mathrm{n}}$
qubits for the nuclear positions as the nuclear register.
There exists one-to-one correspondence between
the $N_{q \mathrm{n}}^{3 n_{\mathrm{nucl}}}$ computational basis vectors
and the possible molecular geometries.
It is noted that
$n_{q \mathrm{n}}$ is a parameter that determines the resolution of the search for the optimal geometry and has no direct relation to the physical properties of the nuclei.
Also, we emphasize here that we have introduced the nuclear register and the operators
$\{ \hat{\boldsymbol{\mathcal{R}}}_\nu \}_\nu$ 
not for encoding quantum states of nuclei,
but for encoding the data for the nuclei as distinguishable classical particles.
Having defined the nuclear position operators,
we rewrite the Hamiltonian in
Eq.~(\ref{struct_opt_using_FQE:def_Hamiltonian})
by replacing the nuclear positions as
$c$-numbers with the corresponding operators:
$
        \mathcal{H}
        (
            \{
                \boldsymbol{R}_{\nu}
            \}_\nu
        )
    \rightarrow
        \mathcal{H}
        (
            \{
                \hat{\boldsymbol{\mathcal{R}}}_\nu
            \}_\nu
        )
    ,
$
leading to the new Hamiltonian for the
$(3 n_e n_{q e} + 3 n_{\mathrm{nucl}} n_{q \mathrm{n}})$-qubit system.
$E_{n n}$ has become an operator $\hat{V}_{\mathrm{n n}}.$

\begin{figure*}
\begin{center}
\includegraphics[width=14cm]{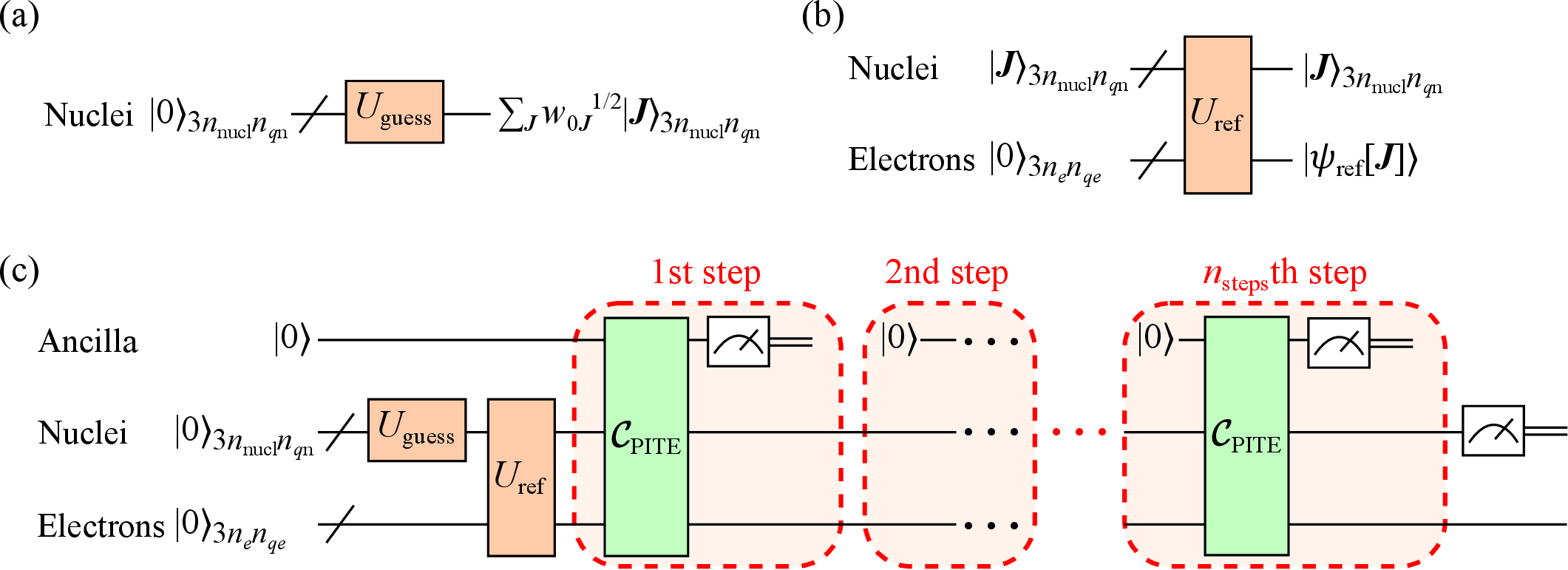}
\end{center}
\caption{
Relevant partial circuits and entire procedure.
(a)
Initial-guess gate $U_{\mathrm{guess}}$ for assigning 
the weight $w_{0 \boldsymbol{J}}$ to each molecular geometry
specified by $\boldsymbol{J}.$
(b)
Reference state gate $U_{\mathrm{ref}}$ for generating
the reference electronic state
$| \psi_{\mathrm{ref}} [\boldsymbol{J}] \rangle$
for the geometry specified by $\boldsymbol{J}.$
(c)
Circuit $\mathcal{C}_{\mathrm{opt}}$ for the entire optimization procedure within FQE.
It contains $n_{\mathrm{steps}}$ PITE steps for energy minimization
governed by the Hamiltonian $\mathcal{H}$ of the system made up of  the electrons and the nuclei.
The ancilla qubit is for observing the success or failure state at each PITE step.
}
\label{fig:circuit_for_opt_using_pite}
\end{figure*}

The preparation of an initial state consists of $U_{\mathrm{guess}}$ and $U_{\mathrm{ref}}$ gates.
$U_{\mathrm{guess}}$ generates the superposition of $N_{\mathrm{cand}}$ possible geometries having nonzero desired weights,
as in Fig.~\ref{fig:circuit_for_opt_using_pite}(a).
$U_{\mathrm{ref}}$ is designed to generate the desired reference electronic state
for the indistinguishable electrons \cite{bib:4825, bib:5389}
in the specified geometry,
as in Fig.~\ref{fig:circuit_for_opt_using_pite}(b).
Possible implementation of the initial-state preparation
that expedites the convergence of subsequent energy minimization
is outlined in
Appendix \ref{sec:methods_circuits_and_measurements_U_ref}.
By using these two gates,
we construct the circuit $\mathcal{C}_{\mathrm{opt}}$ for the entire optimization procedure within FQE,
as shown in Fig.~\ref{fig:circuit_for_opt_using_pite}(c).
For details, see
Appendix \ref{sec:methods_circuits_and_measurements_min_energy}.
The state of the composite system undergoing this circuit 
is written of the form
\begin{gather}
    | \Psi \rangle
    =
        \sum_{\boldsymbol{J}}
        \sqrt{w_{\boldsymbol{J}}}
        | \psi [ \boldsymbol{J} ] \rangle
        \otimes
        | \boldsymbol{J} \rangle_{3 n_{\mathrm{nucl}} n_{q \mathrm{n}}}
    ,
    \label{struct_opt_using_FQE:generic_state_using_psi_J}
\end{gather}
where $\boldsymbol{J}$ is the collective notation of
$3 n_{\mathrm{nucl}} n_{q \mathrm{n}}$ integers specifying one of the candidate geometries.
$| \psi [ \boldsymbol{J} ] \rangle$
is the normalized trial electronic state for the geometry $\boldsymbol{J},$ whose weight is $w_{\boldsymbol{J}}.$
When we perform a measurement on the nuclear register
comprising $| \Psi_s \rangle$ of the form
in Eq.~(\ref{struct_opt_using_FQE:generic_state_using_psi_J})
immediately after the $s$th step,
the probability for observing the molecular geometry corresponding to a specific $\boldsymbol{J}$ is 
clearly $w_{s \boldsymbol{J}},$
which is the weight of geometry contained in $| \Psi_s \rangle.$
The composite state having undergone sufficiently many PITE steps will thus provide the lowest-energy geometry with the highest probability:
\begin{gather}
    \boldsymbol{J}^{(\mathrm{opt})}
    =
        \argmax_{\boldsymbol{J}}
        w_{n_{\mathrm{steps}} \boldsymbol{J}}
        ,
\end{gather}
from which the optimal displacements
$\{ \Delta \boldsymbol{R}_\nu^{(\mathrm{opt})} \}_{\nu}$
are calculated from
Eq.~(\ref{struct_opt_using_FQE:def_displaced_pos_of_nucl}).
In practice, $\boldsymbol{J}^{(\mathrm{opt})}$ can be found by drawing a histogram of observed values of $\boldsymbol{J}$ from repeated measurements.
Our scheme is also applicable to a geometry optimization problem for point charges as a classical system (see Appendix \ref{sec:methods_circuits_and_measurements_classical}).

Let us consider a plausible case of $N_{\mathrm{cand}}$ candidate geometries for which good reference states are available
from sophisticated classical calculations.
As considered in Appendix \ref{sec:estimation_of_num_of_steps},
the energy shift technique by Nishi et al.~\cite{nishi2023analyzing} leads to the required number of steps for obtaining the optimal state with a tolerance $\delta$ estimated to be
\begin{align}
    n_{\mathrm{steps}} (\delta)
    =
        \mathcal{O}
        \left(
            \frac{1}{\Delta E_{\mathrm{cand}} \Delta \tau}
            \log
            \frac{N_{\mathrm{cand}} }{\delta}
        \right)    
    ,
    \label{required_num_of_steps_for_opt}
\end{align}
where
$\Delta E_{\mathrm{cand}}$ is the energy difference between the optimal and second optimal geometries.
$\Delta \tau$ is the amount of each imaginary-time step.
For a practical PITE circuit, an upper bound on $\Delta \tau$ needs to be respected in order for the Taylor expansion of the ITE operator to be justified (see Appendix \ref{sec:review_of_PITE}).

In the actual optimization procedure for a given molecule,
we will be confronted with a dilemma:
while a more accurate prediction of the optimal geometry requires finer discretization of nuclear displacements,
such discretization inevitably leads to smaller energy differences between ``neighboring'' candidate geometries,
which are more difficult to detect via the finite number of PITE steps.
The histogram of observed geometries will thus exhibit a shape formed by multiple maxima,
each of which has a finite width around it
and corresponds to possibly one of the local minima on the energy surface of the molecule.
If we want to predict one of the local-minima geometries more accurately,
we should start newly an optimization procedure by restricting the nuclear displacements within the vicinity of the local minimum,
only for which the nuclear register is spent.

Kassal et al.\cite{bib:5328} demonstrated that
nonadiabatic treatment of nuclei as quantum mechanical particles in a molecule as well as the electrons,
is computationally much more efficient for a chemical-reaction simulation than the BO approximation, except for the smallest molecules.
On the other hand, one finds that the classical treatment of nuclei in our approach for geometry optimization is more efficient than the nonadiabatic treatment for the following reasons.
If we used the $3 n_{\mathrm{nucl}} n_{q \mathrm{n}}$ qubits for the nuclei as quantum mechanical particles
to encode their wave function,
the grid spacing in the simulation cell has to be on the order of femtometer (fm) to detect the finite width of wave function of each nucleus.
The grid spacing $\Delta x$ for electronic wave function has also to be on the same order for a reliable simulation,
while that may be on the order of \AA {} in our original approach.
The required number $n_{q \mathrm{n}}$
of qubits for the nonadiabatic treatment is thus
larger than that for the classical treatment roughly by
$\log_2 (\mathrm{\AA/fm}) \approx 16.6,$
which is also the case for $n_{q e}.$
Furthermore, we will then give up the superposition of candidate geometries since the nuclear register has already been reserved for the many-nucleus wave function.
Therefore we have to perform the energy minimization starting from some single initial geometry.
These considerations indicate that
the classical treatment of nuclei is practically more favorable than the nonadiabatic one unless
the result of optimization is affected qualitatively by the nonadiabatic treatment.

\subsection{Circuit depths}

\begin{figure*}
\begin{center}
\includegraphics[width=13cm]{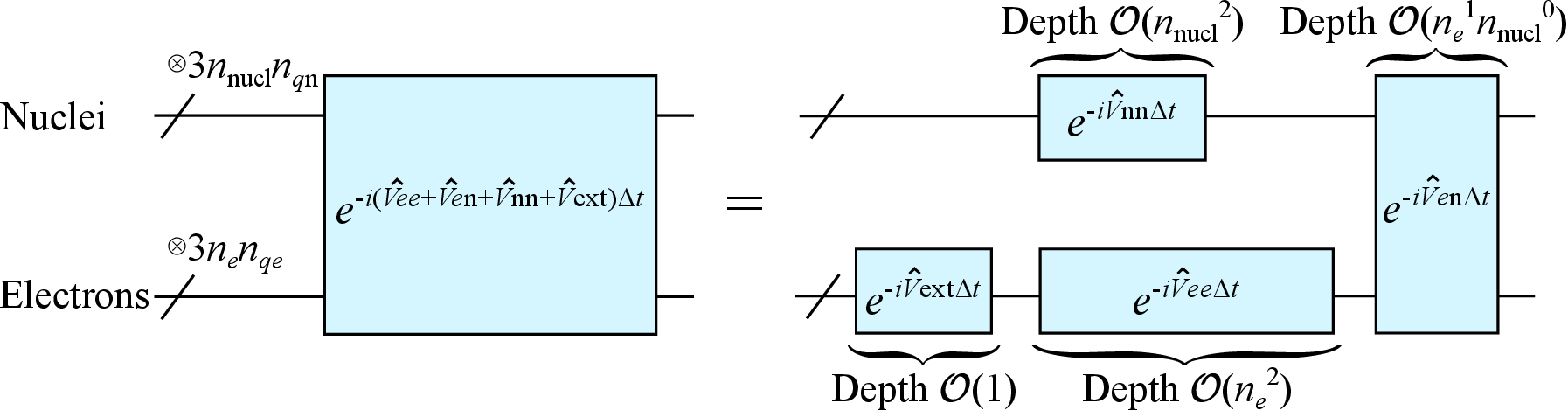}
\end{center}
\caption{
RTE operator $e^{-i \mathcal{H} \Delta t}$ for a time step $\Delta t$ required in the $\mathcal{C}_{\mathrm{PITE}}$ circuit
is decomposed into the kinetic part and the position-dependent part via the first-order Suzuki--Trotter.
The latter is shown in the left part of this figure,
which is further decomposed into the evolution operators generated by 
$
\hat{V}_{e e},
\hat{V}_{e \mathrm{n}},
\hat{V}_{\mathrm{n n}},
$
and
$
\hat{V}_{\mathrm{ext}}
$
separately, as shown in the right part.
The scaling of depths with respect to the numbers of electrons and nuclei are also shown.
}
\label{circuit:real_time_evol_for_opt_with_pite}
\end{figure*}

The PITE circuit $\mathcal{C}_{\mathrm{PITE}}$ consists mainly of the controlled real-time evolution (RTE) operators. \cite{bib:5737} 
We implement the RTE operator $e^{-i \mathcal{H} \Delta t}$ for a time step $\Delta t$ by employing the first-order Suzuki--Trotter as usual to decompose it approximately into the kinetic part
$e^{-i \hat{T} \Delta t}$ and the position-dependent part
$
\exp
[ -i (
\hat{V}_{e e} + 
\hat{V}_{e \mathrm{n}} + 
\hat{V}_{\mathrm{n n}} + 
\hat{V}_{\mathrm{ext}}
)
\Delta t
]
.
$
While the former can be implemented
using the quantum Fourier transform (QFT)-based techniques \cite{bib:5391, bib:5384, bib:5737} as in the electrons-only cases,
the latter is further decomposed exactly into the four parts,
as shown in
Fig.~\ref{circuit:real_time_evol_for_opt_with_pite}.
The evolution $e^{-i \hat{V}_\kappa \Delta t} \ (\kappa = ee, e \mathrm{n}, \mathrm{nn})$
is implemented by applying the pairwise phase gate $U_\kappa (\Delta t)$ that acts diagonally as
\begin{gather}
	U_{\kappa} (\Delta t)
	\left(
		| \boldsymbol{s} \rangle
		\otimes
		| \boldsymbol{s}' \rangle
	\right)
	=
		e^{-i v (\boldsymbol{s}, \boldsymbol{s}') \Delta t}
		\left(
			| \boldsymbol{s} \rangle
			\otimes
			| \boldsymbol{s}' \rangle
		\right)
  \label{def_pairwise_phase_gate}
\end{gather}
to every pair of interacting particles.
$| \boldsymbol{s} \rangle$ and $| \boldsymbol{s}' \rangle$ are the position eigenstates of the particles with the interaction energy $v (\boldsymbol{s}, \boldsymbol{s}').$
On the other hand,
$e^{-i \hat{V}_{\mathrm{ext}} \Delta t}$
is implemented by applying the phase gate $U_{\mathrm{ext}} (\Delta t)$ that acts diagonally as
$
U_{\kappa} (\Delta t)
| \boldsymbol{k} \rangle_{3 n_{q e}}
=
\exp
(
-i
v_{\mathrm{ext}} (\boldsymbol{r}^{(\boldsymbol{k})}) 
\Delta t
)
| \boldsymbol{k} \rangle_{3 n_{q e}}
$
to each electron.
The details of their implementation and the scaling of circuit depths with respect to the particle numbers are explained in
Appendix \ref{sec:impl_of_phase_gates:def_phase_gates}.
It is clear from 
Fig.~\ref{circuit:real_time_evol_for_opt_with_pite}
that
the partial circuits for $e^{-i \hat{V}_{ee} \Delta t}$
and $e^{-i \hat{V}_{\mathrm{nn}} \Delta t}$
are deeper than those for
$e^{-i \hat{V}_{e \mathrm{n}} \Delta t}$
and
$e^{-i \hat{V}_{\mathrm{ext}} \Delta t}$
from the viewpoint of scaling with respect to $n_e$ and $n_{\mathrm{nucl}}$.

While we will be focusing on the first-order Suzuki--Trotter with the fixed $\Delta t$ below,
it is possible instead to employ a generic $p$th-order product formula with controlling the error $\varepsilon$ originating from
the noncommutativity between the kinetic and position-dependent parts of the Hamiltonian.
Specifically, the depth per PITE step takes on a factor of
$
\mathcal{O}
(\widetilde{\alpha}_{\mathrm{comm}}^{1/p} \Delta t^{1+1/p}/\varepsilon^{1/p}),
$
where $\widetilde{\alpha}_{\mathrm{comm}}$ is a function of $L$ and $\Delta x$ \cite{bib:6196}.

\begin{figure}
\begin{center}
\includegraphics[width=8cm]{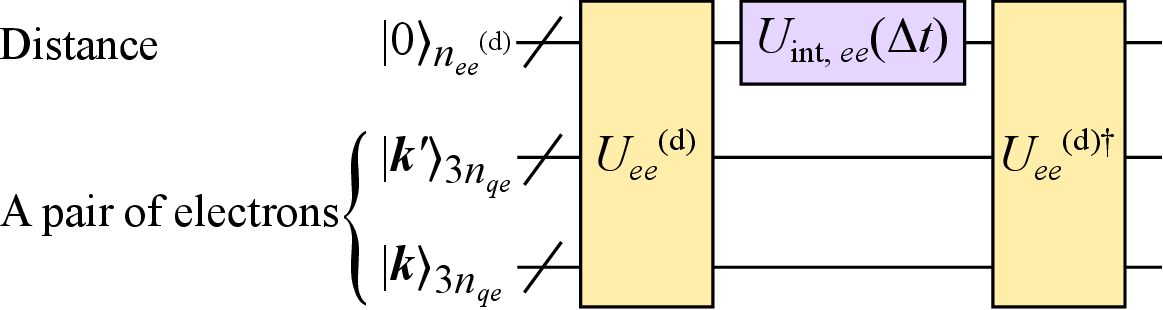}
\end{center}
\caption{
Circuit that implements the pairwise $e$-$e$ phase gate
$U_{e e} (\Delta t)$
as a building block of $e^{-i \hat{V}_{e e} \Delta t}$ operation.
$U_{e e}^{(\mathrm{d})}$ defined in
Eq.~(\ref{struct_opt_using_FQE:def_dist_gate_for_e-e})
computes the distance between two electrons at
$\boldsymbol{r}^{(\boldsymbol{k})}$ and
$\boldsymbol{r}^{(\boldsymbol{k}')},$
which is then stored into the distance register consisting of
$n_{e e}^{(\mathrm{d})}$ qubits.
The interaction phase gate $U_{\mathrm{int}, e e} (\Delta t)$
defined in
Eq.~(\ref{struct_opt_using_FQE:def_interaction_phase_gate})
refers to the distance register to generate the phase required for the evolution coming from the electron pair.
The inverse of $U_{e e}^{(\mathrm{d})}$ performs
uncomputation for disentangling the distance register from the electronic register. 
}
\label{fig:el_el_evol_using_dist_reg}
\end{figure}

Although our PITE circuit does not assume specific implementation of the pairwise phase gates comprising
$
e^{ -i \hat{V}_{e e} \Delta t},
e^{ -i \hat{V}_{e \mathrm{n}} \Delta t},
$
and
$
e^{ -i \hat{V}_{\mathrm{n n}} \Delta t},
$
we propose here a plausible alternative by exploiting the fact that
the pairwise interaction $v$ is common to these three types of evolution
and depends only on the distance between particles.
By dividing the task we have to do into the computation of distances between the particles and that of the phases for evolution,
we find the systematic construction of the circuits,
as explained in
Appendix \ref{sec:impl_of_phase_gates:impl_using_dist_regs}.
Fig.~\ref{fig:el_el_evol_using_dist_reg}
shows the circuit that implements
the pairwise $e$-$e$ phase gate $U_{e e} (\Delta t),$ defined in
Eq.~(\ref{def_phase_gate_for_ee_pair})
as a building block of $e^{-i \hat{V}_{e e} \Delta t}$ operation.
The pairwise phase gates
$U_{e \mathrm{n}}^{(\nu)} (\Delta t)$
and
$U_{\mathrm{n n}}^{(\nu, \nu')} (\Delta t)$
as building blocks of
$e^{-i \hat{V}_{e \mathrm{n}} \Delta t}$ and
$e^{-i \hat{V}_{\mathrm{n n}} \Delta t},$
respectively,
can also be implemented similarly.
For example,
the circuit for computing the distance can be implemented efficiently by combining the addition \cite{bib:5379, bib:5396, bib:5397}, multiplication \cite{bib:5405, bib:5398, bib:5399}, and square root \cite{Hadfield}.

If we approximate the functional shape of the interaction $v$
as a simple or a piecewisely defined polynomial,
the interaction phase gates
$U_{\mathrm{int}, \kappa} (\Delta t) \ (\kappa = ee, e \mathrm{n}, \mathrm{nn}),$
defined in
Eq.~(\ref{struct_opt_using_FQE:def_interaction_phase_gate}),
can be implemented with polynomial depths \cite{bib:5390, bib:5384} in the numbers
$n^{(\mathrm{d})}_\kappa$ of qubits for the distance registers.
(See also Appendix \ref{sec:phase_gate_for_piecewise_poly}.)
Although $n^{(\mathrm{d})}_\kappa$
can be set independently of $n_{q e}$ and $n_{q \mathrm{n}}$,
it is suitable to set them such that the resolutions induced by the former are on the same order as by the latter:
$
n^{(\mathrm{d})}_{e e} = \mathcal{O} (n_{q e}),
n^{(\mathrm{d})}_{e \mathrm{n}} = \mathcal{O} (\mathrm{max} (n_{q e}, n_{q \mathrm{n}}))
,
$
and
$
n^{(\mathrm{d})}_{\mathrm{n n}}
=
\mathcal{O} (n_{q \mathrm{n}}).
$
These considerations tell us that
the pairwise phase gates
$
U_{e e} (\Delta t),
U_{e \mathrm{n}}^{(\nu)} (\Delta t),
$
and
$U_{\mathrm{n n}}^{(\nu, \nu')} (\Delta t)$
can be implemented with polynomial depths in
$n_{q e}$ and $n_{q \mathrm{n}}.$

As discussed in Ref.~\cite{bib:5737},
the number of qubits for the electronic wave function
with a resolution $\Delta x$ typically scales as
$n_{qe} = \mathcal{O} (\log (n_e^{1/3}/\Delta x)).$
On the other hand,
that for the nuclear displacements scales as
$
n_{q \mathrm{n}}
=
\mathcal{O} (\log (\Delta R_{\mathrm{max}}/\Delta R))
$
for typical values of a resolution $\Delta R$ and the maximal displacement $\Delta R_{\mathrm{max}}.$
Recalling the fact that $n_e$ is much larger than $n_{\mathrm{nucl}}$ despite their common scaling 
for a generic molecule,
we find that $e^{-i \hat{V}_{ee} \Delta t}$ dominates the scaling of circuit depth of the entire position-dependent evolution when $\Delta R_{\mathrm{max}}$ and $\Delta R$ are fixed.
In fact, the $e^{-i \hat{V}_{\mathrm{nn}} \Delta t}$ circuit does not contribute to the total depth since it and
$e^{-i \hat{V}_{ee} \Delta t},$ which is much deeper than it,
can be performed in parallel,
as seen in Fig.~\ref{circuit:real_time_evol_for_opt_with_pite}.
The scaling coming from the electron-electron interactions,
given by Eq.~(\ref{circuit:depth_scaling_of_evol_e-e}),
is dominant even in the entire RTE circuit:
\begin{gather}
    \mathrm{depth} (e^{-i \mathcal{H} \Delta t})
    =
        \mathcal{O}
        \left(
            n_e^2
            \mathrm{poly}
            \left(
                \log
                \frac{n_e^{1/3}}{\Delta x}
            \right)
        \right)
    .
    \label{circuit:depth_scaling_of_entire_RTE}
\end{gather}
For details, see Appendix \ref{sec:impl_of_phase_gates:depths}.
Since the single PITE step contains the controlled RTE operations,
its depth exhibits the same scaling:
$
\mathrm{depth} (\mathcal{C}_{\mathrm{PITE}})
=
\mathcal{O}
(\mathrm{depth} (e^{-i \mathcal{H} \Delta t})).
$

If the same number $3 n_e n_{q e}$
of extra qubits as in the electronic register are available,
the scaling of depth for $e^{-i \hat{V}_{e e} \Delta t}$ can be reduced.
Specifically, $n_e^2$ on the RHS in
Eq.~(\ref{circuit:depth_scaling_of_evol_e-e})
becomes $n_e$ via the technique described in
Appendix \ref{sec:impl_of_phase_gates:efficient_el_el_evol}.
(See also Ref.~\cite{bib:5824})
The scaling of depth for $e^{-i \hat{V}_{\mathrm{n n}} \Delta t}$ can be reduced similarly if the same number of qubits as in the nuclear register are available.
With these techniques,
the depth of the entire RTE circuit is
\begin{gather}
    \mathrm{depth} (e^{-i \mathcal{H} \Delta t})
    =
        \mathcal{O}
        \left(
            n_e
            \mathrm{poly}
            \left(
                \log
                \frac{n_e^{1/3}}{\Delta x}
            \right)
        \right)
    ,
    \label{circuit:reduced_depth_scaling_of_entire_RTE}
\end{gather}
instead of
Eq.~(\ref{circuit:depth_scaling_of_entire_RTE}).
It is noted that, if the number of available extra qubits is
$\mathcal{O} (n_e^2 n_{q e}),$
the technique proposed in Ref.~\cite{bib:5824} leads to more drastic reduction of the depth:
$n_e^2$ on the RHS in
Eq.~(\ref{circuit:depth_scaling_of_evol_e-e})
becomes 1.

Let us estimate the computational cost for finding the optimal geometry for the case considered above
[see Eq.~(\ref{required_num_of_steps_for_opt})],
where the good reference states are available for the $N_{\mathrm{cand}}$ candidates.
From the required number of steps for a tolerance $\delta$ and the depth for the single step
[see Eq.~(\ref{circuit:reduced_depth_scaling_of_entire_RTE})],
the total depth scales as
\begin{align}
    \mathrm{depth} (\mathcal{C}_{\mathrm{PITE}})
    n_{\mathrm{steps}} (\delta)
    =
        \mathcal{O}
        \left(
            n_e
            \mathrm{poly} (\log n_e)
            \log
            \frac{N_{\mathrm{cand}} }{\delta}
        \right)    
\end{align}
with respect to $n_e, N_{\mathrm{cand}},$ and $\delta.$
The RHS of this equation imposes a lower bound on the coherence time of hardware being used.
Since the expected number $n_{\mathrm{meas}} (\delta)$
of measurements performed until we reach the optimal state
(see Appendix \ref{sec:estimation_of_num_of_steps})
is larger than $n_{\mathrm{steps}} (\delta)$ due to the probabilistic nature,
the scaling of computational time apart from $U_{\mathrm{ref}}$ is estimated to be
\begin{gather}
    \mathrm{depth} (\mathcal{C}_{\mathrm{PITE}})
    n_{\mathrm{meas}} (\delta)
    \nonumber \\
    =
        \mathcal{O}
        \left(
            n_e
            \mathrm{poly} (\log n_e)
            N_{\mathrm{cand}}
            \log
            \frac{N_{\mathrm{cand}} }{\delta}
        \right)
        .
        \label{computational_time_for_geom_opt}
\end{gather}
As for energy-based geometry optimization on a classical computer,
$N_{\mathrm{cand}}$ total-energy calculations are needed
and each of them involves the construction of Hamiltonian matrix
of dimension $N_{q e}^{3 n_e}.$
The classical-operation number for finding the optimal geometry is thus at least
$\mathcal{O} ((N_{q e}^{3 n_e})^2 N_{\mathrm{cand}} )$
whether using the good reference states or not.
This should be compared with the quantum scaling in
Eq.~(\ref{computational_time_for_geom_opt}).
Specifically, the scaling in $n_e$ for classical computational time is exponential,
while that for quantum computational time is at most polynomial.
The scaling in $N_{\mathrm{cand}}$ for classical computation is linear,
while that for quantum computation is $\mathcal{O} (N_{\mathrm{cand}} \log N_{\mathrm{cand}}).$
These observations imply that our optimization scheme with a fixed number of candidates exhibits quantum advantage in molecule size ($n_e$ and $n_{\mathrm{nucl}}$).
When the candidate number also varies independently of molecule size,
the quantum scaling is still at most polynomial.
Since the quantum scaling in $N_{\mathrm{cand}}$ is worse than the classical one only logarithmically,
it may not cause serious disadvantage that would cancel the advantage in $n_e.$
In this sense, our scheme may offer overall quantum advantage when molecule size and candidate number vary,
as long as we have implementation of 
$U_{\mathrm{guess}}$ and $U_{\mathrm{ref}}$ that do not spoil this quantum scaling.
Although the pursuit of efficient preparation of reference states is a crucial and challenging task not only for our optimization scheme but also for all the first-quantized schemes,
we do not go into further details than
Appendix \ref{sec:methods_circuits_and_measurements_U_ref}.

Quantum amplitude amplification (QAA) \cite{bib:4884, bib:4878},
known as a generalization of Grover's search algorithm,
can raise the success probability at each PITE step \cite{Nishi_QAA}.
This technique is also applicable for multiple steps by delaying the measurements,
as demonstrated by Nishi {\it et al.}\cite{2023arXiv230803605N} recently.
If we introduce the QAA technique to our optimization scheme,
the total success probability undergoes quadratic speedup, that is, it changes from 
$\sim 1/N_{\mathrm{cand}}$ to $\sim 1/\sqrt{N_{\mathrm{cand}}}.$
The scaling of computational time in terms of the candidate number is then
$\mathcal{O} (\sqrt{N_{\mathrm{cand}}} \log N_{\mathrm{cand}})$
instead of Eq.~(\ref{computational_time_for_geom_opt}).
The optimization scheme for this case offers quantum advantage with respect to $N_{\mathrm{cand}}$ itself,
in addition to $n_e.$

It should be noted that, for a case where all the possible displacements of all the nuclei are candidates ($N_{\mathrm{cand}} = N_{q \mathrm{n}}^{3 n_{\mathrm{nucl}} }$) with uniform initial weights,
the quantum scaling of computational time is exponential in $n_{\mathrm{nucl}}$ as well as the classical scaling.
This comes from the exponential decrease in the initial weight of the optimal geometry following the increase in the molecule size,
lowering the success probability at each step.
A situation in which such quantum computation is demanded is, however, actually unlikely.
It is because the uniform distribution of weights for 
the $N_{q \mathrm{n}}^{3 n_{\mathrm{nucl}} }$ geometries
means that we are completely ignorant of the relative stability among them.
The modern sophisticated techniques for electronic-structure calculations and molecular dynamics are, as assumed in our resource estimation,
able to enumerate a very small number (compared to $N_{q \mathrm{n}}^{3 n_{\mathrm{nucl}} }$) of promising candidates by spending moderate classical resources.
Implementation of $U_{\mathrm{guess}}$ that assigns significant weights to those candidates will be a practical strategy.

\subsection{PITE simulation for a model LiH molecule}

We consider here an effective model of a lithium hydride molecule in one-dimensional space used in Ref.~\cite{bib:3919}.
This model regards the $1s$ electrons of the Li atom to be frozen so that the system consists of the two valence electrons,
the H ion with $Z_{\mathrm{H}} = 1,$
and the Li ion with $Z_{\mathrm{Li}} = 1.$
The interactions between the particles are modelled basically by the soft-Coulomb interaction
$
v_{\mathrm{soft}} (r; \lambda)
\equiv
1/\sqrt{\lambda^2 + r^2}
,
$
where
$r$ is the distance between two particles and
the parameter $\lambda$ measures the softness of the interaction.
This family of potentials is often used for avoiding the singular behavior of the bare-Coulomb potential \cite{bib:5679}.
The adopted values for the interactions are as follows:
$
v_{e e} (r)
=
v_{\mathrm{soft}} (r; \lambda_{e e})
$
between the electrons with $\lambda_{e e}^2 = 0.6,$
$
v_{e \mathrm{H}} (r)
=
v_{\mathrm{soft}} (r; \lambda_{e \mathrm{H}})
$
between each electron and the H ion 
with $\lambda_{e \mathrm{H}}^2 = 0.7,$
$
v_{e \mathrm{Li}} (r)
=
v_{\mathrm{soft}} (r; \lambda_{e \mathrm{Li}})
$
between each electron and the Li ion 
with $\lambda_{e \mathrm{Li}}^2 = 2.25,$
and
$
v_{\mathrm{LiH}} (r)
=
v_{\mathrm{soft}} (r; \lambda_{\mathrm{LiH}})
$
between the ions with
$
\lambda_{\mathrm{LiH}}^2
\equiv
\lambda_{e \mathrm{H}}^2
+
\lambda_{e \mathrm{Li}}^2
-
\lambda_{e e}^2
=
2.35.
$
The potential felt by each electron is thus
$
v_{e \mathrm{n}} (x)
=
-Z_{\mathrm{H}}
v_{e \mathrm{H}} (| x - X_{\mathrm{H}} |)
-Z_{\mathrm{Li}}
v_{e \mathrm{Li}} (| x - X_{\mathrm{Li}} |)
,
$
where
$X_{\mathrm{H}}$ and $X_{\mathrm{Li}}$ are
the positions of the H and Li ions, respectively.
The details of the following simulations are described in
Appendix \ref{sec:details_of_LiH}.

Figure \ref{fig:LiH_energy_and_el_dens}(a) shows
the energy eigenvalues of the molecule as functions of the bond length
$d \equiv | X_{\mathrm{Li}} - X_{\mathrm{H}} |$
obtained by numerical diagonalization of the Hamiltonian matrix.
By using $n_{q e} = 6$ qubits per electron for a simulation cell with $L = 15,$
we obtained the equilibrium bond length $d_{\mathrm{eq}} = 1.55,$ in reasonable agreement with that in the earlier paper \cite{bib:3919}.
Figure \ref{fig:LiH_energy_and_el_dens}(b) shows
the electron densities of the energy eigenstates
obtained by numerical diagonalization for
$d = d_{\mathrm{eq}}, 4.$
For $d = d_{\mathrm{eq}},$
the electrons are localized near the H ion to exhibit the single-peak shape.
For $d = 4,$ on the other hand, they are localized at each ion, indicative of dissociation.
We found for both bond lengths that
the ground state $| \phi_{\mathrm{gs}} \rangle$ and
the second excited state $| \phi_{\mathrm{ex2}} \rangle$ 
are symmetric under exchange of the spatial coordinates $x_0$ and $x_1$ of the two electrons,
while 
the first excited state $| \phi_{\mathrm{ex1}} \rangle$ is antisymmetric under the exchange.
Recalling that our encoding of wave functions does not incorporate explicitly the spin parts
[see Eq.~(\ref{many_electron_state})],
the ground state and the second excited state are spin-singlet states,
while the first excited state is a spin-triplet state.

\begin{figure}
\begin{center}
\includegraphics[width=9cm]{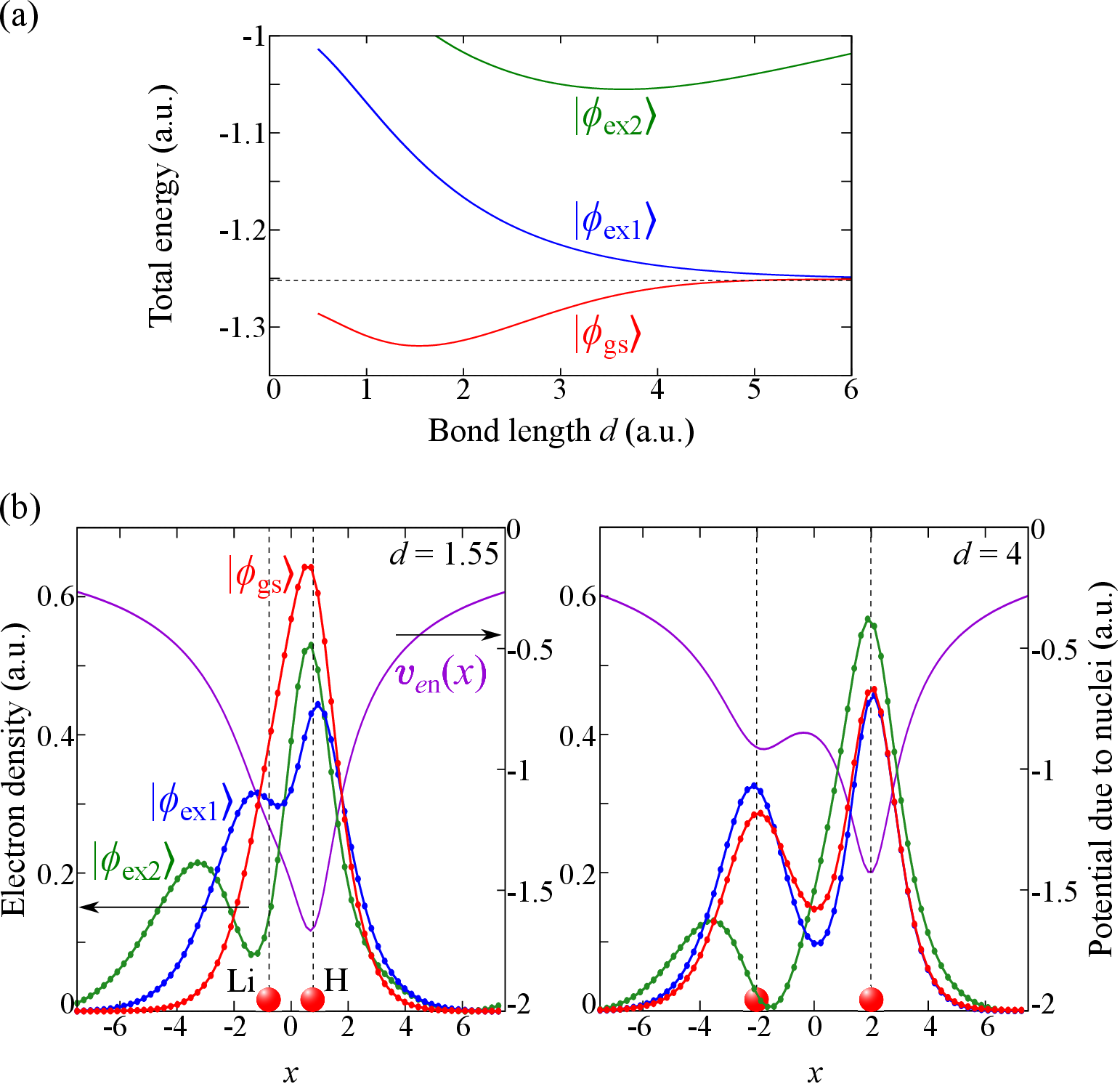}
\end{center}
\caption{
Energy curves and electron density of the model LiH molecule.
(a)
Energy eigenvalues of the LiH model system as functions of the bond length $d$ obtained by numerical diagonalization.
$
| \phi_{\mathrm{gs}} \rangle,
| \phi_{\mathrm{ex1}} \rangle,
$
and $| \phi_{\mathrm{ex2}} \rangle$ are
the ground state, the first excited state,
and the second excited state, respectively.
The horizontal dashed line indicates the dissociation limit,
that is, the sum of total energies for the isolated H and Li atoms.
(b)
Left panel shows the electron densities of the energy eigenstates for the equilibrium bond length $d_{\mathrm{eq}}.$
The vertical lines indicate the positions of the ions.
The potential $v_{e \mathrm{n}}$ felt by each electron due to the nuclei is also shown.
The right panel is a similar plot for $d = 4.$
The $x$ coordinates in the figures have been shifted so that the midpoint of the bond is at the origin.
}
\label{fig:LiH_energy_and_el_dens}
\end{figure}

We performed simulations of geometry optimization among eight candidates represented by $n_{q \mathrm{n}} = 3$ qubits.
To be specific,
we tried the bond lengths 
$d_J = 0.55 + 0.5 J \ (J = 0, \dots, 7).$
The amount of imaginary-time step does not need to be constant.
For example, we can define it for the $k$th PITE step as
$
    \Delta \tau_k
    =
        (1 - e^{-k/\kappa})
        (\Delta \tau_{\mathrm{max}} - \Delta \tau_{\mathrm{min}})
        +
        \Delta \tau_{\mathrm{min}}
        ,
$
so that it changes gradually from $\Delta \tau_{\mathrm{min}}$
to $\Delta \tau_{\mathrm{max}}.$
$\kappa$ determines the rate of change.
We adopted
$
\Delta \tau_{\mathrm{min}} = 0.2,
\Delta \tau_{\mathrm{max}} = 0.3,
$
and $\kappa = 8$ for the following simulations.

\begin{figure*}
\begin{center}
\includegraphics[width=16cm]{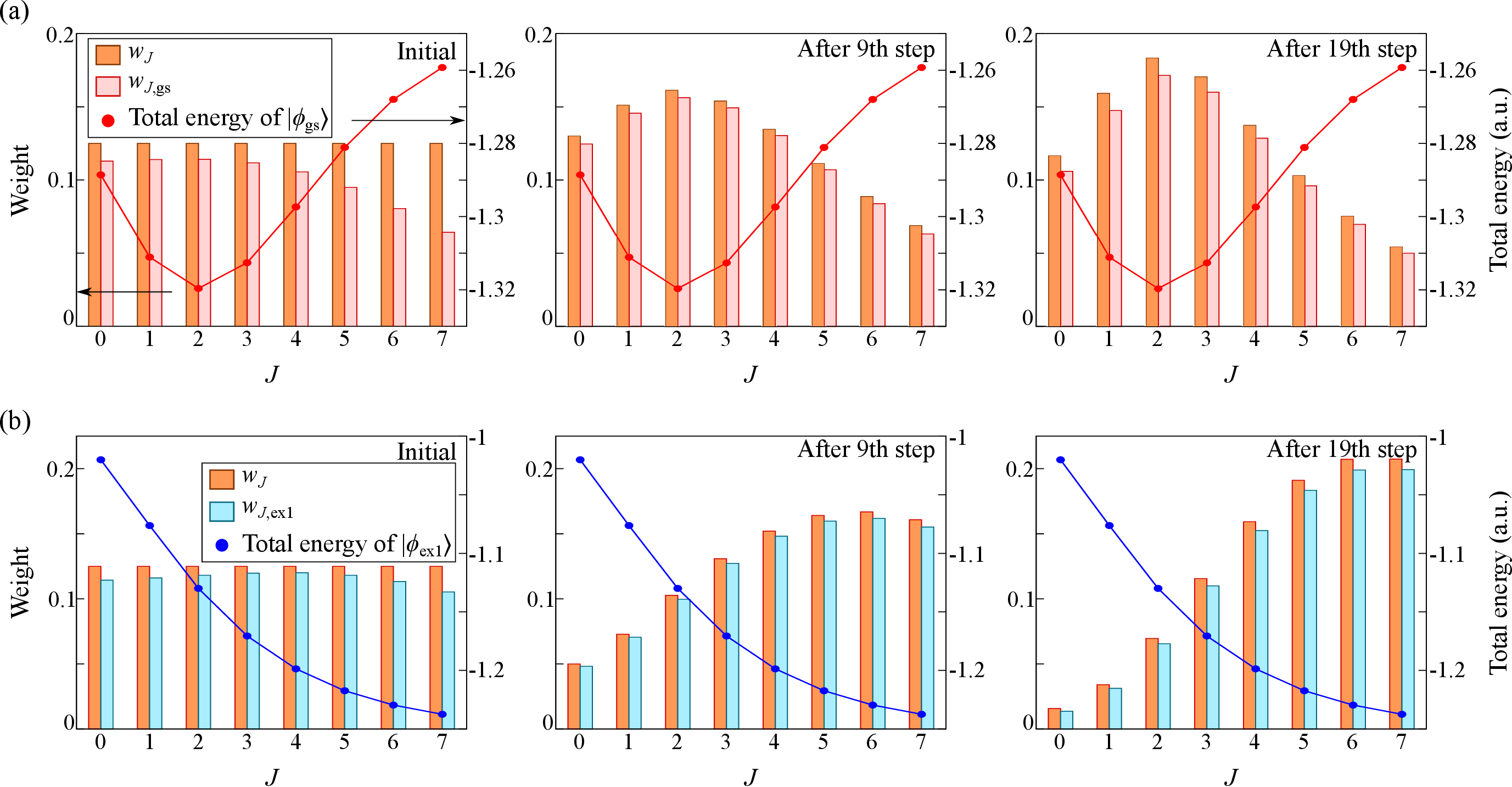}
\end{center}
\caption{
Simulation results of geometry optimization for the LiH molecule.
(a)
Those for eight candidates
starting from the symmetric spatial wave function in
Eq.~(\ref{init_wave_func_symm}).
The boxes show the weight $w_J$ of each geometry $J$ during the PITE steps.
The weight $w_{J, \mathrm{gs}}$ of the ground state for each geometry is also shown.
The red circles represent the total energies of the ground states $| \psi_{\mathrm{gs}} \rangle$ for the geometries.
(b)
The results of simulation
starting from the antisymmetric spatial wave function in
Eq.~(\ref{init_wave_func_antisymm}),
indicating the absence of equilibrium bond length for the first excited state $| \psi_{\mathrm{ex1}} \rangle.$
}
\label{fig:LiH_opt_weights}
\end{figure*}

To find the optimal bond length for the ground state,
we assigned a uniform weight distribution to the candidate geometries,
for which we generated the initial spatial wave functions
\begin{gather}
    \Psi_{\mathrm{s}} (x_0, x_1)
    \propto
        \exp
        \left(
            - 
            \frac{
            (x_0 - X_{\mathrm{m}})^2
            +
            (x_1 - X_{\mathrm{m}})^2}{w^2} 
        \right)
    \label{init_wave_func_symm}
\end{gather}
for the geometries.
$X_{\mathrm{m}} \equiv (X_{\mathrm{H}} + X_{\mathrm{Li}})/2$ is the midpoint of the bond and $w = 3$ is the width of the wave function.
Since $\Psi_{\mathrm{s}}$ is symmetric under exchange of the electrons,
it is for obtaining a spin singlet state.
Figure \ref{fig:LiH_opt_weights}(a)
shows the weight $w_J$ of each geometry $J$ during the steps contained in the state
$| \Psi \rangle$ for the composite system of the electrons and nuclei.
The weight $w_{J, \mathrm{gs}}$
of the ground state $| \psi_{\mathrm{gs}} \rangle$ for each geometry is also shown in the figure.
It is seen that the uniform distribution of weights in the initial state undergoes the deformation via the PITE steps.
It has the peak around the geometry for $J = 2$
already after the 9th step,
corresponding to the equilibrium bond length $d_{\mathrm{eq}}.$
This peak structure becomes more prominent after the 19th step.
These observations corroborate the validity of our generic scheme.

Using the fact that the ground state and the first excited state $| \psi_{\mathrm{ex1}} \rangle$ of this system have the different symmetry,
we can perform geometry optimization for the first excited state.
To this end, we adopted the initial spatial wave functions
\begin{gather}
    \Psi_{\mathrm{a}} (x_0, x_1)
    \propto
        \frac{x_0 - x_1}{w}
        \Psi_{\mathrm{s}} (x_0, x_1)
    \label{init_wave_func_antisymm}
\end{gather}
for the geometries.
Since $\Psi_{\mathrm{a}}$ is antisymmetric under exchange of the electrons,
it is for obtaining a spin triplet state.
The results are shown in Fig.~\ref{fig:LiH_opt_weights}(b).
In contrast to the case of the ground state,
the resultant weight distribution does not have a peak between $J = 0$ and $7,$
which lets the observer recognize that there exists no equilibrium bond length among the candidate geometries.

Although the non-optimal geometries in
Fig.~\ref{fig:LiH_opt_weights}(a)
were found to have the significant weights even after the 19th step,
our scheme worked thanks to the detectable peak in the histogram.
This means that a severe tolerance $\delta$ for quashing the near-optimal geometries that would lead to more steps threatening the coherence time
[see Eq.~(\ref{required_num_of_steps_for_opt})]
is not necessary for this small system.
If it is also the case for a generic large molecule whose energy surface possibly has many local minima,
one practical strategy is to continue to pile up data points on a histogram using a moderate tolerance until the optimal and near-optimal geometries become detectable via statistical data processing.
How practical compromise between the tolerance for PITE steps and the number of data points for a histogram is met and quantum advantage taking it into account should be examined in the future.

\subsection{VITE simulation for a model H$_{2}^+$ molecule}

Since the essence of our approach is the superposition of nuclear-register states where the candidate geometries are encoded,
geometry optimization based on VITE instead of PITE is straightforwardly formulated.
To demonstrate that,
we consider here an effective model of a hydrogen molecular ion in a one-dimensional space used in Ref.~\cite{bib:5799}.
The charge of each ion is $Z_\mathrm{H} = 1$.
Here, we also use the soft-Coulomb interaction to model the interactions between the particles.
We adopt the softness $\lambda_{e \mathrm{H}}^2 = 1$
for the interaction $v_{e \mathrm{H}} (r)$ between the electron and each ion and $\lambda_{\mathrm{H H}}^2 = 1$
for $v_{\mathrm{H H}} (r)$ between the ions.
The potential felt by the electron is thus 
$
v_{e \mathrm{n}}(x)
= 
- Z_{\mathrm{H}} v_{e \mathrm{H}} (|x - X_{\mathrm{H} \alpha}|) 
- Z_{\mathrm{H}} v_{e \mathrm{H}} (|x - X_{\mathrm{H} \beta}|)
,
$
where $X_{\mathrm{H}\alpha}$ and $X_{\mathrm{H}\beta}$ are the positions of the H ions.

The VITE approach is explained briefly in Appendix \ref{sec:review_of_VITE}.
Figure \ref{fig:circuit_for_opt_of_H2_using_vite} shows our ansatz circuit for geometry optimization of the H$_2^+$ model system.
We adopted the hardware-efficient connectivity \cite{bib:4292} for the circuit simulations \cite{Suzuki2021qulacsfast},
which is desirable for noisy intermediate-scale quantum (NISQ) devices due to shallow circuit depths. 
In addition, the accuracy of the quantum computation systematically improves by incrementing the repetition $d$ of the layer. 
Here, we use the full coupling model; C$Z$ gates connect every pair of qubits for entangling all qubits. 
We allocated $n_{q \mathrm{nucl}} = 3$ qubits for encoding the nuclear positions and $n_{q e} = 6$ qubits for encoding the single-electron wave function in a simulation cell with $L = 15.$
As demonstrated below, the VITE-based scheme can, despite the absence of $U_{\mathrm{guess}}$ and $U_{\mathrm{ref}}$,
find the optimal geometry going through more than a thousand of steps,
while the PITE-based scheme finds the optimal one in much fewer steps (see Appendix \ref{sec:details_of_H2p}).
Such many steps are practically possible since the circuit depth is related not to the number of steps,
but to the depth of the ansatz.
This feature renders the VITE-based scheme NISQ-friendly,
in contrast to the PITE-based one.

The VITE calculation was performed for candidates
whose bond lengths were specified by
$d_J= 0.5+(7.5/8)J$ ($J=0, \ldots ,7$). 
We simulated the updating process of variational parameters with $d = 12$ for 6000 VITE steps with $\Delta \tau = 0.01$. 
All the initial values of the variational parameters were set to random values.
The expected energy of the trial state $|\Psi\rangle$ at each VITE step measured from the numerically exact ground state energy is shown in Fig.~\ref{fig:H2_vite_results}(a). 
We recognize the monotonic but slow decrease in the energy difference. 
Figure~\ref{fig:H2_vite_results}(b) shows the weights $w_{J}$ of candidate geometries contained in the trial wave function at each VITE step. 
The weight of the most stable geometry labeled by $J=2$ monotonically increases and reaches close to unity at the final step. 
The second most stable structure, $J=3$, is amplified once in the first 1500 steps and then turns to decrease. 
We draw the electronic wave function component contained in the most stable state, $J=2$, in Fig. \ref{fig:H2_vite_results}(c). 
The ground state $|\phi_{\mathrm{gs}} \rangle$ for the geometry $J=2$ quickly increases, and the excited states decrease to zero within 1000 steps.
These results support that our ideas of encoding candidate geometries for optimization work also for the variational scheme.
The convergence of nuclear states was rather slow compared to that of the electronic states for the individual geometries,
as seen in Figs.~\ref{fig:H2_vite_results}(b) and (c).
This observation reflects the generic fact that
the continuous energy of classical nuclei leads to a small energy difference between neighboring candidate geometries, as discussed in Sect.~\ref{sec:exhaustive_search}.

\begin{figure}
\begin{center}
\includegraphics[width=8cm]{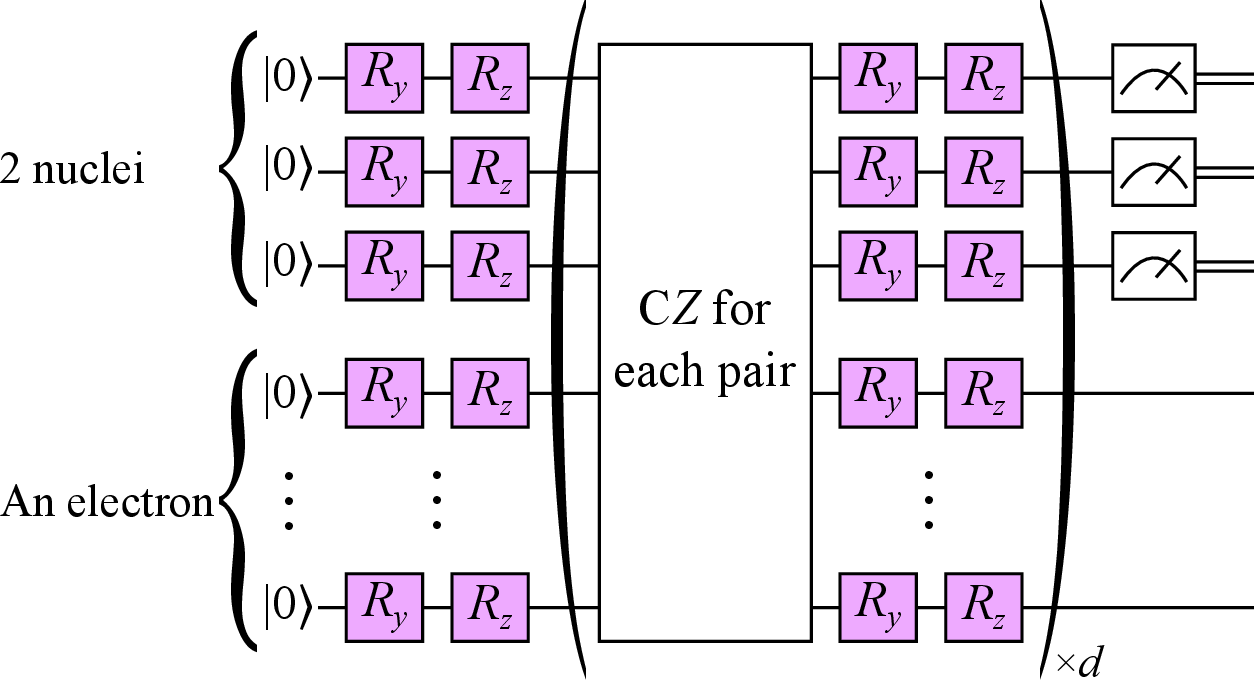}
\end{center}
\caption{
Ansatz circuit for geometry optimization of the H$_2^+$ model system based on the VITE approach.
The part inside the parentheses are applied to the nuclear and electronic registers $d$ times.
The purple boxes stand for single-qubit rotations
 whose angles are specified by distinct variational parameters.
The nuclear register is measured at the end of the circuit to find the optimal bond length.
}
\label{fig:circuit_for_opt_of_H2_using_vite}
\end{figure}

\begin{figure}
\begin{center}
\includegraphics[width=7cm]{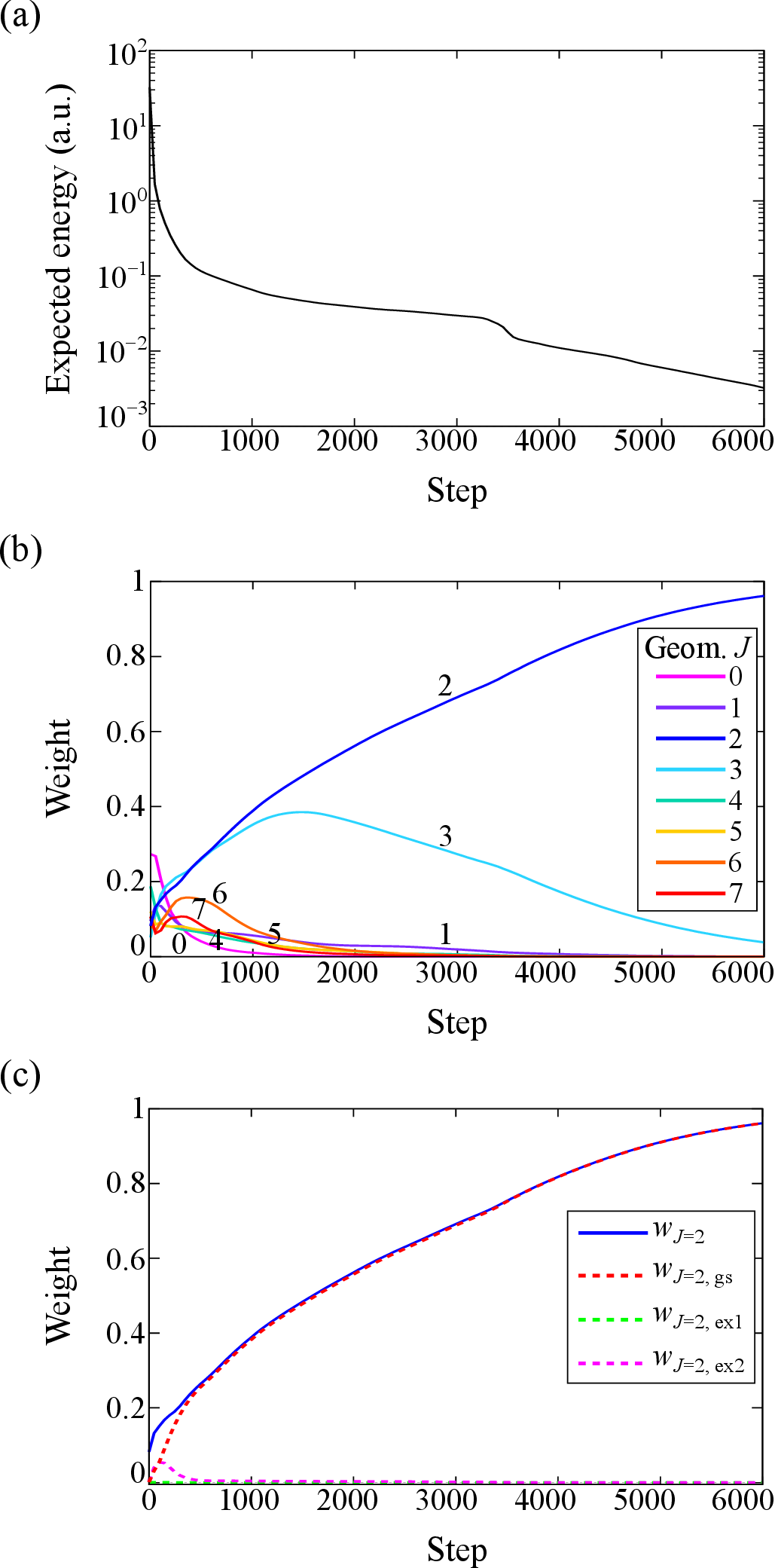}
\end{center}
\caption{
Geometry optimization process for the H$_2^+$ molecule.
(a)
Expected energy of the trial state $| \Psi \rangle$ at each VITE step measured from the lowest energy eigenvalue of the molecule.
(b)
The weights $w_J$ of candidate geometries contained in the trial state at each step.
(c)
The weight $w_{J = 2, \mathrm{gs}}$ of the ground state
$| \phi_{\mathrm{gs}} \rangle$ for the geometry $J = 2$ contained in the trial state at each step.
Those of the first- and second-excited states,
$w_{J = 2, \mathrm{ex1}}$ and $w_{J = 2, \mathrm{ex2}}$,
are also shown.
}
\label{fig:H2_vite_results}
\end{figure}

\subsection{PITE simulation for a classical C$_6$H$_6$-Ar system}

\begin{figure*}
\begin{center}
\includegraphics[width=16cm]{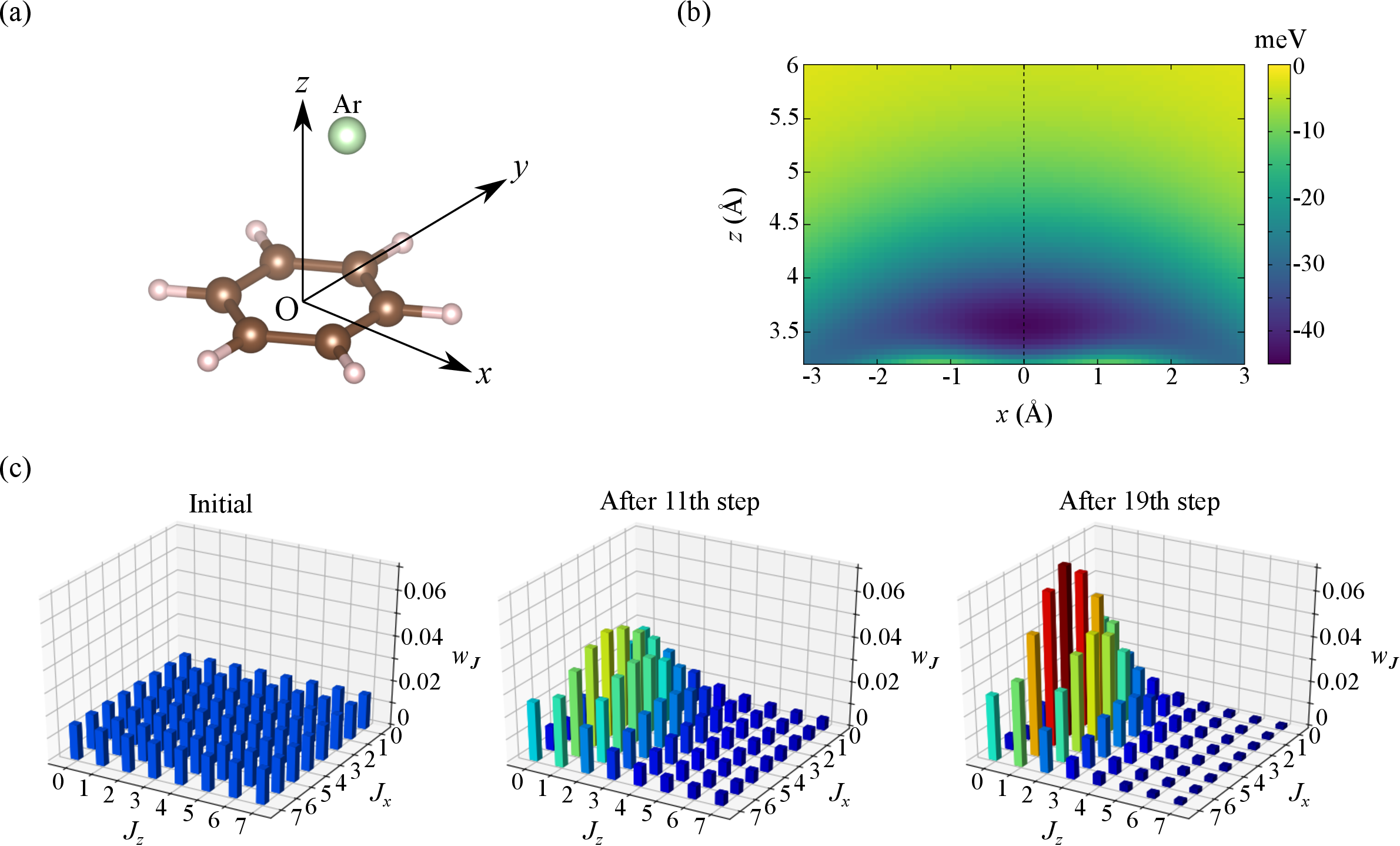}
\end{center}
\caption{
Simulation results for the C$_6$H$_6$-Ar system.
(a) Classical target system for geometry optimization based on PITE approach.
The C$_6$H$_6$ molecule lies on the $xy$ plane with its center of mass located at the origin. Two of the C-H bonds are along the $y$ axis.
We consider an Ar atom on the $xz$ plane.
(b)
The interaction energy as a function of the position of the Ar atom.
(c)
The weight of each candidate during the PITE steps contained in the state of nuclear register.
}
\label{fig:C6H6-Ar_opt}
\end{figure*}

As stated in
Appendix \ref{sec:methods_circuits_and_measurements_classical},
our scheme is also applicable to a geometry optimization problem for point charges as a classical system.
It is known that the improved Lennard--Jones (ILJ) \cite{bib:5834, bib:5835} potentials describe the experimental data well for hydrocarbon molecules interacting with rare-gas atoms.
We adopt here these model potentials to consider a classical system consisting of a benzene molecule interacting weakly with an argon atom \cite{bib:5834},
as depicted in Fig.~\ref{fig:C6H6-Ar_opt}(a).
We perform simulations of geometry optimization for this system by using our PITE scheme.

The C-C and C-H bond lengths are fixed at 1.39 \AA ${}$ and 1.09 \AA \cite{bib:5837}, respectively, throughout the simulations.
The explicit expressions for the ILJ potentials are provided in Appendix \ref{sec:details_of_C6H6_Ar}.
Figure \ref{fig:C6H6-Ar_opt}(b) shows the interaction energy between the C$_6$H$_6$ molecule and the Ar atom on the $xz$ plane as a function of the position of the Ar atom.
The interaction energy takes a minimum value at $z = 3.57$ \AA ${}$ with $x = y = 0$ \AA \cite{bib:5834}.

We performed simulations of geometry optimization among 64 candidates represented by $n_{q \mathrm{n}} = 3$ qubits for each of the $x$ and $z$ coordinates of the Ar atom.
Each of the candidates is specified by two integers
$\boldsymbol{J} = (J_x, J_z)$ with $J_x, J_z = 0, \dots, 7,$
which generate the coordinates
$x_{\boldsymbol{J}} = -2.4 + 0.8 J_x$ \AA ${}$ and
$z_{\boldsymbol{J}} = 3.2 + 0.4 J_z$ \AA.
We used a constant amount $\Delta \tau = 0.004$ meV$^{-1}$ of each PITE step in the following simulations.

In each simulation of the circuit shown in Fig.~\ref{fig:pite_steps_for_classical_opt},
we assigned a uniform weight distribution to the candidate geometries for an initial state.
Figure \ref{fig:C6H6-Ar_opt}(c)
shows the weight $w_{\boldsymbol{J}}$ of each candidate during the steps contained in the state of nuclear register.
It is seen that the uniform distribution of weights in the initial state undergoes the deformation via the steps, as expected.
The largest weight is already seen after the 11th step at
$\boldsymbol{J} = (3, 1),$
which is closer to the true optimal geometry than any other candidate is.
This peak structure becomes more prominent after the 19th step,
as seen in the figure.

\section{Discussion}
\label{sec:conclusions}

In summary, this study proposed a nonvariational scheme for geometry optimization of a molecule within the framework of FQE,
where the electrons and nuclei are treated as quantum mechanical particles and classical point charges, respectively.
The scheme encodes their information as a many-qubit state,
for which repeated measurements give the global minimum among all the candidate geometries.
We demonstrated that the total computational time may exhibit overall quantum advantage in terms of molecule size and candidate number.
The circuit depth of RTE operation,
which is the central component of each PITE step,
was found to scale as
$\mathcal{O} (n_e^2 \mathrm{poly}(\log n_e))$
for the electron number $n_e.$
This can be reduced to
$\mathcal{O} (n_e \mathrm{poly}(\log n_e))$ if the same number of extra qubits as in the original circuit are available.
If $\mathcal{O} (n_e^2 \log n_e)$ extra qubits are available,
the depth can be reduced to
$\mathcal{O} (\mathrm{poly}(\log n_e)).$
The validity of the new scheme was verified through numerical simulations.
The scheme will assist in achieving scalability in practical quantum chemistry on quantum computers.
Additionally, this approach will support the realization of geometry optimization using NISQ devices.

There may be room for elaborating the sampling strategy for candidate geometries for this scheme to be more efficient from a practical perspective.
That is, adaptively changing the range and resolution of nuclear displacements under the constraint of a fixed total number of measurements may more accurately determine the optimal geometry,
which could be examined in the future.

\section*{Data availability}

The datasets generated and analyzed during the current study are available from the corresponding author on reasonable request.

\section*{Code availability}

The code developed for the current study is available from the corresponding author
on reasonable request.

\begin{acknowledgments}
This work was supported by MEXT as ``Program for Promoting Researches on the Supercomputer Fugaku'' (JPMXP1020200205) and JSPS KAKENHI as ``Grant-in-Aid for Scientific Research(A)'' Grant Number 21H04553. The computation in this work has been done using (supercomputer Fugaku provided by the RIKEN Center for Computational Science/Supercomputer Center at the Institute for Solid State Physics in the University of Tokyo).
\end{acknowledgments}

\section*{Competing interests}

The authors declare no competing interests.

\section*{Author contributions}

T.K. developed the methods and wrote the simulation code.
H.N. and Y.M. discussed our approach with T.K. from the viewpoint of quantum chemistry and solid-state physics.
All the authors contributed equally to the manuscript
preparation and presentation of results.

\begin{widetext}
\appendix

\section{Review of PITE approach}
\label{sec:review_of_PITE}

\subsection{Generic circuit}

In our previous work \cite{bib:5737},
we proposed the PITE approach
that implements probabilistically
the nonunitary ITE operator
$m_0 e^{-\mathcal{H} \Delta \tau}$
by using a single ancilla.
$\mathcal{H}$ is the Hamiltonian of a target system.
$\Delta \tau$ is the amount of an imaginary-time step
and $m_0$ is an adjustable parameter satisfying the conditions $0 < m_0 < 1$ and $m_0 \ne 1/\sqrt{2}.$
The generic PITE circuit for a single step is shown in
Fig.~\ref{fig:generic_PITE_circuit},
where a single-qubit gate
\begin{gather}
    W
    \equiv
    \frac{1}{\sqrt{2}}
    \begin{pmatrix}
        1 & -i \\
        1 & i
    \end{pmatrix}
\end{gather}
and
$
\theta_0
\equiv
\mathrm{sgn} (m_0 - 1/\sqrt{2})
\cdot
\arccos [(m_0 + \sqrt{1 - m_0^2})/\sqrt{2}]
$
are used.
As seen in the circuit,
each PITE step is implemented by using the real-time evolution (RTE) gates
for the renormalized real-time step
$\Delta t \equiv s_1 \Delta \tau,$
where $s_1 \equiv m_0/\sqrt{1 - m_0^2}.$
If the measurement outcome of the ancilla qubit is $| 0 \rangle$,
the correctly evolved state (the success state)
$\propto e^{-\mathcal{H} \Delta \tau} | \psi \rangle$
for an arbitrary input state $| \psi \rangle$ within the first order of $\Delta \tau$ has been obtained.
If the observed ancillary state is $| 1 \rangle,$
on the other hand, the input state has become the failure state.
The PITE step has to be repeated until the initial state becomes satisfactorily close to the ground state.
The formalism of PITE can also be understood in the context of block encoding \cite{bib:5728}.
The PITE circuit for obtaining the ground state of a molecular system for a fixed geometry within the first-quantized formalism
has been provided
in Fig.~2(b) of the original paper \cite{bib:5737}.
Finding the ground state of a molecule under an external uniform magnetic field is also possible \cite{bib:6103}.

When we have drawn a failure state unfortunately on the way to the ground state,
there exist two alternatives basically:
one is continuing the steps without worrying about the failure
and the other is restarting from a new first step.
The former might be better as long as
the number of steps required for reaching the ground state is small compared to the coherence time of hardware being used.
We adopt the latter in this study.
The probability for obtaining the success states throughout
$n_{\mathrm{steps}}$ steps is
$\langle \psi | (m_0 e^{-\mathcal{H} \Delta \tau})^{2 n_{\mathrm{steps}}} | \psi \rangle,$
which decreases exponentially as the iterations proceed.
For alleviating this inherent drawback of the PITE approach,
the quantum amplitude amplification \cite{bib:4884, bib:4878},
known as a generalization of Grover's search algorithm,
can be employed \cite{Nishi_QAA}.
It is also important to prepare an initial state having a large overlap with the ground state so that the required number of PITE steps is as small as possible.

\begin{figure}
\begin{center}
\includegraphics[width=8cm]{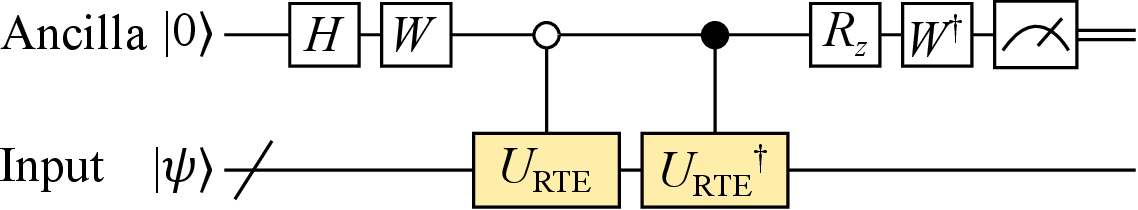}
\end{center}
\caption{
First-order PITE circuit for a generic system governed by its Hamiltonian $\mathcal{H}.$
It contains
the real-time evolution gate
$U_{\mathrm{RTE}} \equiv \exp(-i \mathcal{H} \Delta t)$
for an imaginary-time step $\Delta \tau$ and
$\Delta t \equiv s_1 \Delta \tau$.
$R_z \equiv R_z (-2 \theta_0)$ is the single-qubit $z$-rotation.
$H$ is the Hadamard gate.
This circuit can perform the ITE within the first order of $\Delta \tau.$
}
\label{fig:generic_PITE_circuit}
\end{figure}

\subsection{Estimation of required number of steps with known ground-state energy}

The required number of steps for reaching the ground state of a single-qubit system within a specified tolerance was derived in our previous study \cite{bib:5737}.
Here, we consider a case where we know the ground-state energy $\lambda_0$ of a target many-qubit system and want to obtain the ground state.
We expand the initial state in terms of the energy eigenstates as
$| \psi \rangle = \sum_{j = 0} c_j | \phi_j \rangle,$
where $c_j$ is the coefficient for the $j$th eigenstate $| \phi_j \rangle$ belonging to the eigenvalue $\lambda_j.$
We can assume that the energy eigenvalues are in ascending order without loss of generality.
To get rough estimation,  
we assume here that the PITE circuit implements the exact ITE operator $m_0 e^{- \mathcal{H} \Delta \tau}$ (no errors with respect to $\Delta \tau$).
Mathematically more rigorous resource estimation
for the first-order PITE is possible by taking the errors in terms of $\Delta \tau$ into account and has been reported by 
Nishi {\it et al.} \cite{nishi2023analyzing}.
In the present study, we only point out that a rough upper bound 
$\Delta \tau \ll |\langle \mathcal{H} \rangle|^{-1}$
is imposed on the amount of time step for the first-order PITE to be justified.
$\langle \mathcal{H} \rangle$ is the expected energy of a trial state.

If we have an estimation of the ground state energy $\lambda_0,$
we can introduce an energy shift $\Delta E \equiv - (\Delta \tau)^{-1} \ln m_0 + \lambda_0$ to the Hamiltonian:
$\mathcal{H} \to \mathcal{H}' \equiv \mathcal{H} - \Delta E,$
rendering the success probability at each step higher.
This shift is effectively implemented by changing the rotation angle of the $R_z$ gate for the ancilla.
The PITE circuit is now for the shifted ITE operator
$m_0 e^{-\mathcal{H}' \Delta \tau}.$ 
Although $\Delta \tau$ is assumed to be constant during the steps in the present study,
this technique was originally introduced for the first-order PITE with variable time steps \cite{nishi2023analyzing}.
The trial state immediately after $k$ steps for which all the measurement outcomes are success states is
$
| \psi_k \rangle
=
\mathcal{N}^{-1}
\sum_{j = 0} 
c_j
e^{-k \Delta \lambda_j \Delta \tau}
| \phi_j \rangle,
$
where
$\Delta \lambda_j \equiv \lambda_j - \lambda_0$
and
$\mathcal{N} =
(
\sum_{j = 0} 
| c_j |^2
e^{-2 k \Delta \lambda_j \Delta \tau}
)^{1/2}
$
is the normalization constant.
The weight of the ground state $| \phi_0 \rangle$ contained in $| \psi_k \rangle$ is thus
\begin{align}
    w^{(\mathrm{gs})}_k
    &=
        \frac{1}{
            1
            +
            \sum_{j = 1} |c_j/c_0|^2
            e^{-2 k \Delta \lambda_j \Delta \tau}
        }
    .
\end{align}
Since 
$
\sum_{j = 1} |c_j/c_0|^2
e^{-2 k \Delta \lambda_j \Delta \tau}
\leq
e^{-2 k \Delta \lambda_1 \Delta \tau}
(1 - w_0)/w_0
$
for the initial weight $w_0 \equiv |c_0|^2$ of the ground state,
one can understand that,
if $k$ exceeds
\begin{align}
    n_{\mathrm{steps}} (\delta)
    =
        \frac{1}{2 \Delta \lambda_1 \Delta \tau}
        \ln
        \frac{(1 - w_0)(1 - \delta)}{w_0 \delta}
    \label{required_num_steps_for_generic_PITE}
\end{align}
for a tolerance $\delta > 0,$
we have $w^{(\mathrm{gs})}_k > 1 - \delta.$
The probability for finding the success states throughout the $k$ steps is
$
P_k
=
\langle \psi_0 | (m_0 e^{-\mathcal{H}' \Delta \tau})^{2 k} | \psi_0 \rangle
=
|c_0|^2 + \sum_{j = 1} |c_j|^2 e^{-2 k \Delta \lambda_j \Delta \tau}
.
$
Eq.~(\ref{required_num_steps_for_generic_PITE})
indicates that increasing simply $\Delta \tau$ allows us to reach the ground state in fewer steps.
In a practical situation where the first-order PITE is adopted,
however, we have to take the upper bound on $\Delta \tau$ into account, as mentioned above.
There will thus be a task for finding a moderate $\Delta \tau.$

The expected number of measurements performed in the exact PITE until we reach the ground state within the tolerance is estimated to be
\begin{align}
    n_{\mathrm{meas}} (\delta)
    &\approx
        \frac{n_{\mathrm{steps}} (\delta)}{P_{n_{\mathrm{steps}} (\delta)}}
    \nonumber \\
    &\approx
        \frac{n_{\mathrm{steps}} (\delta)}{ |c_0|^2 }
    \nonumber \\
    &=
        \frac{1}{2 \Delta \lambda_1 \Delta \tau w_0}
        \ln
            \frac{(1 - w_0) (1 - \delta)}{w_0 \delta}    
    .
    \label{required_num_meas_for_generic_PITE}
\end{align}
We will use this result for geometry optimization as a special case.

\section{Circuits and measurements for optimal geometries}
\label{sec:methods_circuits_and_measurements}

\subsection{Possible implementation of initial-state preparation}
\label{sec:methods_circuits_and_measurements_U_ref}

\subsubsection{Outline}

As explained in the main text,
$U_{\mathrm{guess}}$ in 
Fig.~\ref{fig:circuit_for_opt_using_pite}(a)
is defined such that it acts on the initialized nuclear register to generate the superposition of $N_{q \mathrm{n}}^{3 n_{\mathrm{nucl}}}$
possible geometries having desired weights:
\begin{gather}
    U_{\mathrm{guess}}
    | 0 \rangle_{3 n_{\mathrm{nucl}} n_{q \mathrm{n}}}
    =
        \sum_{
            \boldsymbol{j}_0, \dots, 
            \boldsymbol{j}_{n_{\mathrm{nucl}} - 1}
        }
        \sqrt{w_{0 \boldsymbol{J}}}
        \underbrace{
            | \boldsymbol{j}_0 
            \rangle_{3 n_{q \mathrm{n}}}
            \otimes
            \cdots
            \otimes
            | \boldsymbol{j}_{n_{\mathrm{nucl}} - 1}
            \rangle_{3 n_{q \mathrm{n}}}
        }_{ \equiv | \boldsymbol{J} \rangle_{3 n_{\mathrm{nucl}} n_{q \mathrm{n}}} }
    ,
\end{gather}
where each term in the summation on the RHS represents the displacements of nuclei.
$\boldsymbol{J}$ represents collectively the
$3 n_{\mathrm{nucl}} n_{q \mathrm{n}}$ integers
$
\boldsymbol{j}_0, \dots, 
\boldsymbol{j}_{n_{\mathrm{nucl}} - 1}
.
$
We should adopt the weight $w_{0 \boldsymbol{J}}$ of each geometry specified by $\boldsymbol{J}$ based on an initial guess that gives rise to large weights for plausible molecular geometries.
The number $N_{\mathrm{cand}}$ of candidate geometries to which nonzero weights are assigned is arbitrary in principle.
If we have no information about which geometries are plausible,
the uniform superposition of all the possible geometries
generated by the Hadamard gate on each qubit will be suitable:
$U_{\mathrm{guess}} = H^{\otimes 3 n_{\mathrm{nucl}} n_{q \mathrm{n}}}$
for
$
w_{0 \boldsymbol{J}}
=
1/N_{q \mathrm{n}}^{3 n_{\mathrm{nucl}}}
.
$
For a physically motivated nonuniform distribution in a specific problem, on the other hand,
we should adopt a technique for amplitude encoding suitable for the desired shape of distribution.
The implementation and scaling of $U_{\mathrm{guess}}$ thus depend on various assumptions, e.g., whether the target state is sparse and how the amplitudes are calculated on the fly by referring to the nuclear positions.
For example,
Klco and Savage \cite{bib:5645} proposed a linear-depth circuit that encodes a symmetric exponential function,
which will be used in 
Appendix \ref{sec:details_of_LiH}.
In a case where $N_{\mathrm{cand}}$ is much smaller than $N_{q \mathrm{n}}^{3 n_{\mathrm{nucl}}},$
efficient techniques for a sparse state \cite{bib:5767, bib:5646} will help.

Let us move on to $U_{\mathrm{ref}}.$
The first-quantized formalism on a quantum computer itself does not ensure an antisymmetric (fermionic) wave function for multiple electrons,
in contrast to the second-quantized case.
This fact forces us to include explicit symmetrization for constructing an input state \cite{bib:4825, bib:5389}.
The term `symmetrization' of a spatial wave function means here
a transformation that makes the spatial wave function have positive or negative parity under any exchange of electrons so that the many-electron state is legitimately fermionic. 
As described in the main text,
$U_{\mathrm{ref}}$ in 
Fig.~\ref{fig:circuit_for_opt_using_pite}(b)
is defined such that it acts on the nuclear register and the initialized electronic register
to generate the desired reference electronic state
$| \psi_{\mathrm{ref}} [\boldsymbol{J}] \rangle$
for the geometry specified by $\boldsymbol{J}$:
\begin{gather}
    U_{\mathrm{ref}}
    \Big(
    \overbrace{
        | 0 \rangle_{3 n_e n_{q e}}
    }^{\mathrm{Electrons}}
        \otimes
    \overbrace{
        | \boldsymbol{J} \rangle_{3 n_{\mathrm{nucl}} n_{q \mathrm{n}}}
    }^{\mathrm{Nuclei}}
    \Big)
    =
        | \psi_{\mathrm{ref}} [\boldsymbol{J}] \rangle
        \otimes
        | \boldsymbol{J} \rangle_{3 n_{\mathrm{nucl}} n_{q \mathrm{n}}}
    .
\end{gather}

One typical and tractable choice of the initial state is the Hartree--Fock state for each geometry, i.e., the Slater determinant of orthonormalized one-electron orbitals,
as often adopted in correlated calculations on classical computers.
Here we provide an outline of possible implementation of $U_{\mathrm{ref}}$ for the Hartree--Fock state, as depicted in
Fig.~\ref{fig:initial_state_generation}(a).
That consists mainly of two parts:
$U_{\mathrm{prod}}$ for generating the product state of one-electron orbitals and the efficient symmetrization procedure proposed by Berry {\it et al.}\cite{bib:5389}.

\subsubsection{Generation of product state}

$U_{\mathrm{prod}}$ is defined to generate 
the product state $| \mathrm{Prod} [\boldsymbol{J}] \rangle$
of the same number of predetermined occupied one-electron orbitals
$\{ \phi_m [\boldsymbol{J}] \}_{m = 0}^{n_e - 1}$ as the electrons contained in the molecule by referring to the positions of nuclei:
\begin{align}
    U_{\mathrm{prod}}
    \left(
        | 0 \rangle_{3 n_e n_{q e}}
        \otimes
        | \boldsymbol{J} \rangle_{3 n_{\mathrm{nucl}} n_{q \mathrm{n}}}
    \right)
    =
        \underbrace{
        | \phi_0 \rangle
        \otimes
        \cdots
        \otimes
        | \phi_{n_e - 1} \rangle
        }_{= | \mathrm{Prod} [\boldsymbol{J}] \rangle}
        \otimes
        | \boldsymbol{J} \rangle_{3 n_{\mathrm{nucl}} n_{q \mathrm{n}}}
    ,
\end{align}
where $| \phi_m \rangle$ is the $3 n_{q e}$-qubit state that encodes $\phi_m [\boldsymbol{J}].$
The correspondence between the one-electron orbitals and the individual electrons in the register is irrelevant since
$| \mathrm{Prod} [\boldsymbol{J}] \rangle$ will be soon symmetrized.
One of the simplest forms of $U_{\mathrm{prod}}$ is for atomic-like orbitals localized at the nuclei.
Generating such a product state is possible by
applying the gate $U_{\lambda}^{(\nu)}$ that encodes a one-electron orbital $\phi_\lambda$ at the nucleus $\nu$ to each of the electrons.
$U_{\lambda}^{(\nu)}$ acts on any one of the one-electron registers by referring to the one-nucleus register $| \boldsymbol{j}_\nu \rangle_{3 n_{q n}},$
as exemplified in Fig.~\ref{fig:initial_state_generation}(b).

A one-electron orbital whose position and/or shape depend on the positions of multiple nuclei, e.g., a bonding orbital between two ions, can also be introduced by constructing a gate that refers to the registers for the multiple nuclei.
Whatever the number of nuclei referred to is,
the required techniques belong to quantum state preparation (QSP).
QSP is a collective name for techniques that generate arbitrary many-qubit states or those under some constraints.
Looking into the recent development of QSP (see, e.g., Refs.~\cite{bib:5697, bib:5767, bib:5646} and references therein),
QSP techniques specialized for first-quantized approaches have to be developed in the future for such approaches to be of practical use.

\subsubsection{Symmetrization}

As for the symmetrization procedure,
it consists of two subprocedures \cite{bib:5389}:
generation of the record of sorting and application of the reverse sort $U_{\mathrm{revsort}}$ to the product state by referring to the record.
[See Fig.~\ref{fig:initial_state_generation}(a).]
The former is nonunitary since it involves a measurement on an ancilla in the working register,
while the latter is unitary.
The application of $U_{\mathrm{revsort}}$ to the product state gives the symmetric state by assigning the appropriate signs according to the permutation of the orbitals as
\begin{align}
    U_{\mathrm{revsort}}
    \Big(
        | \mathrm{Prod} [\boldsymbol{J}] \rangle
        \otimes
        (\mathrm{working \ reg.})
    \Big)
    =
        | \psi_{\mathrm{ref}} [\boldsymbol{J}] \rangle
        \otimes
        (\mathrm{working \ reg.})
        .
\end{align}
Thanks to the superposition principle,
the reference states for all the candidate geometries are prepared at once via the circuit in Fig.~\ref{fig:initial_state_generation}(a).
The working register can be discarded after the completion of the symmetrization since it is disentangled from the electronic and nuclear registers.
The operation number and the circuit depth for the symmetrization procedure depends on $n_e$ and $n_{q e},$ independently of $n_{\mathrm{nucl}}$ and $n_{q n}.$
They also depend on the adopted sort algorithm.
If the quantum bitonic sort is adopted as well as in the original paper,
the depth for the symmetrization scales as
$\mathcal{O} ( (\log n_e)^2 \log n_{q e})$ \cite{bib:5389}.

\begin{figure}
\begin{center}
\includegraphics[width=11cm]{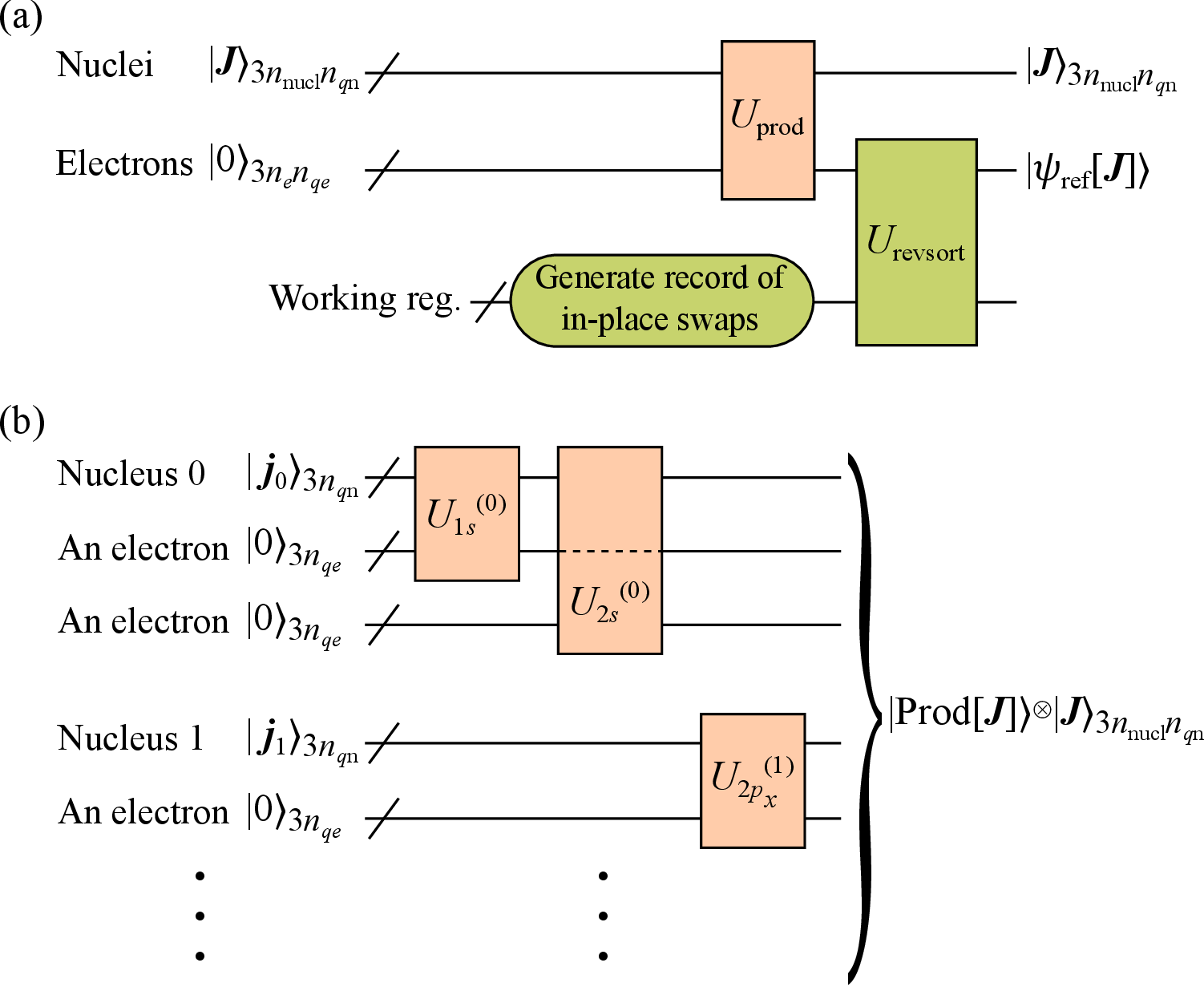}
\end{center}
\caption{
(a) Outline of possible implementation of $U_{\mathrm{ref}}$ for generating the symmetric state $| \psi_{\mathrm{ref}} [\boldsymbol{J}] \rangle$ from one-electron orbitals for a specified geometry $\boldsymbol{J}.$
$U_{\mathrm{prod}}$ gate generates the product state $| \mathrm{Prod} [\boldsymbol{J}] \rangle$ of the constituent orbitals,
while the remaining parts of this circuit represent the symmetrization procedure proposed in Ref.~\cite{bib:5389}.
(b) An example of $U_{\mathrm{prod}}$ for atomic-like orbitals.
This circuit encodes $1 s$ and $2 s$ orbitals at nucleus 0 by referring to the corresponding nuclear register.
The other orbitals are encoded similarly.
}
\label{fig:initial_state_generation}
\end{figure}

\subsection{Energy minimization for finding optimal geometry}
\label{sec:methods_circuits_and_measurements_min_energy}

We construct the circuit $\mathcal{C}_{\mathrm{opt}}$ for the entire optimization procedure within FQE, as shown in
Fig.~\ref{fig:circuit_for_opt_using_pite}(c).
This circuit first generates the input state
\begin{gather}
    | \Psi_0 \rangle
    =
        \sum_{\boldsymbol{J}}
        \sqrt{w_{0 \boldsymbol{J}}}
        | \psi_{\mathrm{ref}} [\boldsymbol{J}] \rangle
        \otimes
        | \boldsymbol{J} \rangle_{3 n_{\mathrm{nucl}} n_{q \mathrm{n}}}
    \label{struct_opt_using_FQE:initial_state}
\end{gather}
for the subsequent PITE steps implemented by the $\mathcal{C}_{\mathrm{PITE}}$ circuits.
The number $n_{\mathrm{steps}}$ of steps needs to be large enough so that all the excited states of the composite system have satisfactorily diminished.
It has also to be, however, small compared with the coherence time of quantum hardware being used.

Since we are working with the first-quantized formalism,
the coefficients of computational bases of the electronic and nuclear registers give the state of the composite system as
\begin{gather}
    | \Psi \rangle
    =
        \sum_{
            \boldsymbol{k}_0, \dots, 
            \boldsymbol{k}_{n_e - 1}
        }
        \sum_{\boldsymbol{J}}
        c_{\boldsymbol{K}, \boldsymbol{J} }
        \overbrace{
            | \boldsymbol{K} \rangle_{3 n_e n_{qe}}
        }^{\mathrm{Electrons}}
        \otimes
        \overbrace{
        | \boldsymbol{J} \rangle_{3 n_{\mathrm{nucl}} n_{q \mathrm{n}}}
        }^{\mathrm{Nuclei}}
    ,
    \label{struct_opt_using_FQE:generic_state}
\end{gather}
where
$\boldsymbol{K}$ represents collectively the
$3 n_e n_{q e}$ integers
$
\boldsymbol{k}_0, \dots, 
\boldsymbol{k}_{n_e - 1}
.
$
The normalization condition forces the coefficients to satisfy
$
\sum_{\boldsymbol{K}, \boldsymbol{J}}
|c_{\boldsymbol{K}, \boldsymbol{J} }|^2
= 1.
$
The weight of geometry $\boldsymbol{J}$ contained in $| \Psi \rangle$ is given by
$
    w_{\boldsymbol{J}}
    =
        \sum_{\boldsymbol{K}}
        |
        c_{\boldsymbol{K}, \boldsymbol{J}}
        |^2
    .
$
By defining the renormalized coefficients
$
    c_{\boldsymbol{K}} [ \boldsymbol{J} ]
    \equiv
    c_{\boldsymbol{K}, \boldsymbol{J}}/
    \sqrt{w_{\boldsymbol{J}}}
    ,
$
we write the normalized electronic state for a fixed geometry $\boldsymbol{J}$ as
\begin{gather}
    | \psi [ \boldsymbol{J} ] \rangle
    =
        \sum_{\boldsymbol{K}}
        c_{\boldsymbol{K}}
        [ \boldsymbol{J} ]
        | \boldsymbol{K}
        \rangle_{3 n_e n_{qe}}
        .
\end{gather}
We can then rewrite the state of composite system in
Eq.~(\ref{struct_opt_using_FQE:generic_state})
to the form in
Eq.~(\ref{struct_opt_using_FQE:generic_state_using_psi_J}).
By comparing this expression and 
Eq.~(\ref{struct_opt_using_FQE:initial_state}),
we understand that
$w_{\boldsymbol{J}}$ and $| \psi [ \boldsymbol{J} ] \rangle$
in the case of the input state are equal to
$w_{0 \boldsymbol{J}}$ and
$| \psi_{\mathrm{ref}} [ \boldsymbol{J} ] \rangle,$
respectively, for each $\boldsymbol{J}.$

\subsection{Optimization for classical point charges}
\label{sec:methods_circuits_and_measurements_classical}

As a special case of our scheme described above,
we can perform geometry optimization for a classical system composed of point charges.
To be specific,
the Hamiltonian for such a system is
defined by incorporating only the terms between the nuclei from the original Hamiltonian: 
$
\mathcal{H}_{\mathrm{cl}}
(\{ \hat{\boldsymbol{\mathcal{R}}}_\nu \}_\nu)
\equiv
\hat{V}_{\mathrm{nn}}
.
$
The circuit for the entire procedure in this case is depicted in Fig.~\ref{fig:pite_steps_for_classical_opt},
as a special case of 
Fig.~\ref{fig:circuit_for_opt_using_pite}(c).

We can also perform the scheme by adopting empirical anisotropic potentials of the form
$v (\boldsymbol{R}_\nu - \boldsymbol{R}_{\nu'})$
depending explicitly on the relative vector between effective nuclei.
Although such artificial potentials may lead to more complicated circuits for $e^{-i \hat{V}_{n n} \Delta t}$
than the bare-Coulomb potentials,
the scaling of circuit depths with respect to $n_{\mathrm{nucl}}$ and $n_{q \mathrm{n}}$
discussed below will not be affected.
It is because how complicated the functional forms of $v$ are has nothing to do with $n_{\mathrm{nucl}}$ and $n_{q \mathrm{n}}.$

\begin{figure}
\begin{center}
\includegraphics[width=8.5cm]{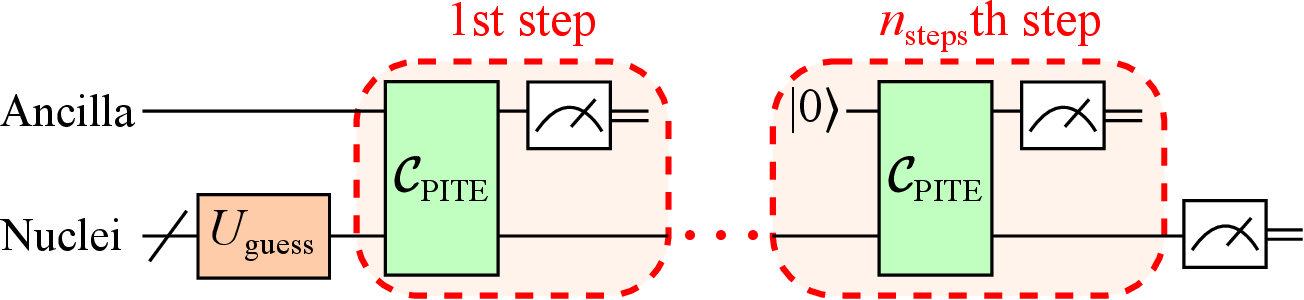}
\end{center}
\caption{
Circuit for the geometry optimization procedure
for classical point charges.
The PITE circuit $\mathcal{C}_{\mathrm{PITE}}$ in this case is for
$\mathcal{H}_{\mathrm{cl}}
(\{ \hat{\boldsymbol{\mathcal{R}}}_\nu \}_\nu).$
}
\label{fig:pite_steps_for_classical_opt}
\end{figure}

\subsection{Estimation of required number of steps}
\label{sec:estimation_of_num_of_steps}

Here we derive a rough estimation of the required number of PITE steps for the geometry optimization of an electronic system. 
To this end, we consider a case where the initial weights are nonzero only for $N_{\mathrm{cand}}$ candidate geometries and the distribution is uniform among them:
\begin{align}
    w_{0 \boldsymbol{J}}
    =
        \begin{cases}
            1/N_{\mathrm{cand}} & \mathrm{Geometry} \ \boldsymbol{J} \ \mathrm{is \ a \ candidate} \\
            0 & \mathrm{Otherwise}
        \end{cases}
        .
\end{align}
As for the electronic degrees of freedom,
we introduce an assumption that we have a good reference state
and a good estimation of the total energy for each candidate geometry $\boldsymbol{J}$, that is,
the reference state $| \psi_{\mathrm{ref}} [\boldsymbol{J}] \rangle$ contains a significant constant weight $w_e^{(\mathrm{gs})}$ of the ground state regardless of the molecule size.
This assumption is plausible if we perform sophisticated mean-field-like calculations, e.g., those based on the density functional theory, on a classical computer to find the mean-field ground states for the whole or a part of the set of candidate geometries.
We extract the one-electron orbitals from such calculations and adopt them for $U_{\mathrm{ref}}$ explained above by,
if necessary, interpolating their locations and shapes between the candidates.
The reference states constructed in this way should have large overlaps with the true ground states,
serving as good initial states for the subsequent energy minimization.
In addition, the total energies calculated by the classical computer are expected to give us a good estimation of that of the optimal geometry,
allowing us to use the energy shift technique described in Appendix \ref{sec:review_of_PITE}.

We assume that the energy difference $\Delta E_{\mathrm{cand}}$ between the optimal and second optimal candidates with the individual electronic ground states is smaller than the excitation energy in the optimal geometry.
From Eq.~(\ref{required_num_steps_for_generic_PITE}),
the required number of steps for finding the optimal geometry scales as
\begin{align}
    n_{\mathrm{steps}} (\delta)
    =
        \mathcal{O}
        \left(
            \frac{1}{\Delta E_{\mathrm{cand}} \Delta \tau}
            \log
            \frac{N_{\mathrm{cand}} }{w_e^{(\mathrm{gs})} \delta}
        \right)
\end{align}
with respect to $N_{\mathrm{cand}}$ and a tolerance $\delta.$
The expected number of measurements performed until we reach the optimal state scales as
\begin{align}
    n_{\mathrm{meas}} (\delta)
    &=
        \mathcal{O}
        \left(
            \frac{1}{\Delta E_{\mathrm{cand}} \Delta \tau}
            \frac{N_{\mathrm{cand}}}{ w_e^{(\mathrm{gs})}}
            \log
            \frac{N_{\mathrm{cand}} }{w_e^{(\mathrm{gs})} \delta}
        \right)
    ,
\end{align}
which is a special case of
Eq.~(\ref{required_num_meas_for_generic_PITE}).

\section{Implementation of phase gates in PITE circuit}
\label{sec:impl_of_phase_gates}

\subsection{Definition of phase gates}
\label{sec:impl_of_phase_gates:def_phase_gates}

\subsubsection{Electron-electron phase gates}

To implement the evolution generated by $\hat{V}_{e e}$
for the specific case of
Eq.~(\ref{def_pairwise_phase_gate}),
we define the $e$-$e$ phase gate $U_{ee}$
such that it acts on a $6 n_{qe}$-qubit state diagonally as
\begin{gather}
    U_{ee} (\Delta t)
    \overbrace{
        | \boldsymbol{k} \rangle_{3 n_{qe}}
        \otimes
        | \boldsymbol{k}' \rangle_{3 n_{qe}}
    }^{\mathrm{A \ pair \ of \ electrons}}
    \equiv
        \exp
        \left(
            -i
            v
            \left(
            \left|
                \boldsymbol{r}^{(\boldsymbol{k})}
                -
                \boldsymbol{r}^{(\boldsymbol{k}')}
            \right|
            \right)
            \Delta t
        \right)
        | \boldsymbol{k} \rangle_{3 n_{qe}}
        \otimes
        | \boldsymbol{k}' \rangle_{3 n_{qe}}
    .
    \label{def_phase_gate_for_ee_pair}
\end{gather}
The evolution can thus be implemented by applying this phase gate to each of the possible pairs of $n_e$ electrons:
\begin{gather}
    e^{-i \hat{V}_{ee} \Delta t}
    =
        \prod_{\ell > \ell'}
        \overbrace{
            U_{ee} (\Delta t)
        }^{\mathrm{On} \ \ell \mathrm{th \ and} \ \ell' \mathrm{th \ electrons}}
    ,
    \label{struct_opt_using_FQE:phase_gate_for_e-e}
\end{gather}
which requires the circuit depth on the order of $\mathcal{O} (n_e^2).$

\subsubsection{Electron-nucleus phase gates}

To implement the evolution generated by $\hat{V}_{e \mathrm{n}}$
for the specific case of
Eq.~(\ref{def_pairwise_phase_gate}),
we define the $e$-n phase gate $U_{e \mathrm{n}}^{(\nu)}$ for each nucleus $\nu$
such that it acts on a $(3 n_{qe} + 3 n_{q \mathrm{n}})$-qubit state diagonally as
\begin{gather}
    U_{e \mathrm{n}}^{(\nu)} (\Delta t)
    \overbrace{
        | \boldsymbol{k} \rangle_{3 n_{qe}}
    }^{\mathrm{An \ electron}}
    \otimes
    \overbrace{
        | \boldsymbol{j}_{\nu}
        \rangle_{3 n_{q \mathrm{n}}}
    }^{\nu \mathrm{th \ nucleus}}
    \equiv
        \exp
        \left(
            i Z_{\nu}
            v
            \left(
            \left|
            \boldsymbol{r}^{(\boldsymbol{k})}
                -
                \boldsymbol{R}_{\nu}
                (\boldsymbol{j}_\nu)
            \right|
            \right)
            \Delta t
        \right)
    | \boldsymbol{k} \rangle_{3 n_{qe}}
    \otimes
    | \boldsymbol{j}_{\nu}
    \rangle_{3 n_{q \mathrm{n}}}
    .
    \label{def_phase_gate_for_en_pair}
\end{gather}
The evolution can thus be implemented by applying this phase gate to each of the possible pairs of $n_e$ electrons and $n_{\mathrm{nucl}}$ nuclei:
\begin{align}
    e^{-i \hat{V}_{e \mathrm{n}} \Delta t}
    &=
        \prod_{\ell = 0}^{n_e - 1}
        \prod_{\nu = 0}^{n_{\mathrm{nucl}} - 1}
        \overbrace{
            U_{e \mathrm{n}}^{(\nu)}
            (\Delta t)
        }^{\mathrm{On} \ \ell \mathrm{th \ electron \ and} \ \nu \mathrm{th \ nucleus}}
    \nonumber \\
    &=
        \prod_{d = 0}^{n_e - 1}
        \prod_{\nu = 0}^{n_{\mathrm{nucl}} - 1}
        \overbrace{
            U_{e \mathrm{n}}^{(\nu)}
            (\Delta t)
        }^{\mathrm{On} \ \ell \mathrm{th \ electron \ and} \ \nu \mathrm{th \ nucleus}}
        \Bigg|_{\ell = \nu + d \ \mathrm{mod} \ n_e}
    ,
    \label{struct_opt_using_FQE:phase_gate_for_e-n}
\end{align}
where we used the generic relation
$n_e \geq n_{\mathrm{nucl}}$ for a molecule
to get the last equality.
This expression allows us to implement the evolution
with the circuit depth on the order of
$\mathcal{O} (n_e^1 n_{\mathrm{nucl}}^0)$
since the gate operators for a common $d$ can be performed simultaneously.
Such an implementation for $n_e = 4$ and $n_{\mathrm{nucl}} = 3$
is shown in Fig.~\ref{circuit:el_nucl_gate_example}
as an example.

\begin{figure}
\begin{center}
\includegraphics[width=13cm]{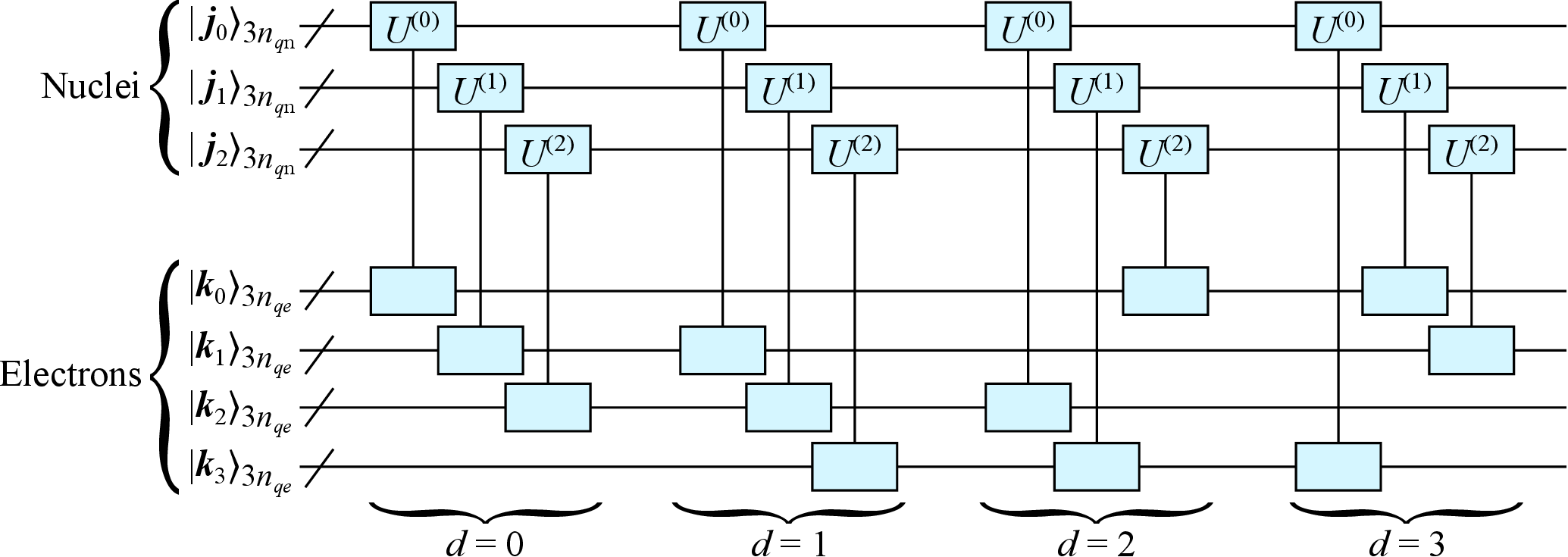}
\end{center}
\caption{
Evolution operator $e^{-i \hat{V}_{e \mathrm{n}} \Delta t}$
implemented based on
Eq.~(\ref{struct_opt_using_FQE:phase_gate_for_e-n})
for $n_e = 4$ and $n_{\mathrm{nucl}} = 3.$
$U^{(\nu)}$ in this figure is a simplified notation of
the phase gate $U_{e \mathrm{n}}^{(\nu)} (\Delta t)$
tailored for the $\nu$th nucleus.
It is depicted as connected two boxes.
}
\label{circuit:el_nucl_gate_example}
\end{figure}

\subsubsection{Nucleus-nucleus phase gates}

To implement the evolution generated by $\hat{V}_{\mathrm{n n}}$
for the specific case of
Eq.~(\ref{def_pairwise_phase_gate}),
we define the n-n phase gate $U_{\mathrm{n n}}^{(\nu, \nu')}$ for each pair of nuclei $\nu$ and $\nu'$
such that it acts on a $6 n_{q \mathrm{n}}$-qubit state diagonally as
\begin{gather}
    U_{\mathrm{nn}}^{(\nu, \nu')} (\Delta t)
    \overbrace{
        | \boldsymbol{j}_{\nu}
        \rangle_{3 n_{q \mathrm{n}}}
    }^{\nu \mathrm{th \ nucleus}}
    \otimes
    \overbrace{
        | \boldsymbol{j}_{\nu'}
        \rangle_{3 n_{q \mathrm{n}}}
    }^{\nu' \mathrm{th \ nucleus}}
    \equiv
        \exp
        \Big(
            -i Z_{\nu} Z_{\nu'}
            v
            \left(
            \left|
                \boldsymbol{R}_{\nu}
                (\boldsymbol{j}_\nu)
                -
                \boldsymbol{R}_{\nu'}
                (\boldsymbol{j}_{\nu'})
            \right|
            \right)
            \Delta t
        \Big)
    | \boldsymbol{j}_{\nu}
    \rangle_{3 n_{q \mathrm{n}}}
    \otimes
    | \boldsymbol{j}_{\nu'}
    \rangle_{3 n_{q \mathrm{n}}}
    .
    \label{def_phase_gate_for_nn_pair}
\end{gather}
The evolution can thus be implemented by applying this phase gate to each of the possible pairs of $n_{\mathrm{nucl}}$ nuclei:
\begin{align}
        e^{-i \hat{V}_{\mathrm{nn}} \Delta t}
    &=
        \prod_{\nu > \nu'}
        \overbrace{
            U_{\mathrm{nn}}^{(\nu, \nu')} (\Delta t)
        }^{\mathrm{On} \ \nu \mathrm{th \ and} \ \nu' \mathrm{th \ nuclei}}
    ,
    \label{struct_opt_using_FQE:phase_gate_for_n-n}
\end{align}
which requires the circuit depth on the order of $\mathcal{O} (n_{\mathrm{nucl}}^2).$

\subsubsection{External-field phase gates}

To implement the evolution generated by $\hat{V}_{\mathrm{ext}},$
we define the phase gate $U_{\mathrm{ext}}$ 
such that it acts on a $3 n_{q e}$-qubit state diagonally as
\begin{gather}
    U_{\mathrm{ext}} (\Delta t)
    \overbrace{
        | \boldsymbol{k} \rangle_{3 n_{qe}}
    }^{\mathrm{An \ electron}}
    \equiv
        \exp
        \left(
            -i
            v_{\mathrm{ext}}
            (
                \boldsymbol{r}^{(\boldsymbol{k})}
            )
            \Delta t
        \right)
        | \boldsymbol{k} \rangle_{3 n_{qe}}
    .
\end{gather}
The evolution can thus be implemented by applying this phase gate to each of the $n_e$ electrons:
\begin{align}
        e^{-i \hat{V}_{\mathrm{ext}} \Delta t}
    &=
        \prod_{\ell = 0}^{n_e - 1}
        \overbrace{
            U_{\mathrm{ext}} (\Delta t)
        }^{\mathrm{On} \ \ell \mathrm{th \ electron}}
        ,
\end{align}
which requires the circuit depth on the order of $\mathcal{O} (1)$
since all the gate operations can be performed simultaneously.

\subsection{Implementation using distance registers}
\label{sec:impl_of_phase_gates:impl_using_dist_regs}

\subsubsection{Electron-electron distances}

We introduce $n^{(\mathrm{d})}_{e e}$ qubits,
which we call the electron-electron distance register,
and define the unitary $U^{(\mathrm{d})}_{e e}$ 
such that it computes
the distance between two electrons as
\begin{gather}
    U^{(\mathrm{d})}_{e e}
    \left(
        | \boldsymbol{k} \rangle_{3 n_{q e}}
        \otimes
        | \boldsymbol{k}' \rangle_{3 n_{q e}}
        \otimes
        | 0 \rangle_{n^{(\mathrm{d})}_{e e}}
    \right)
    =
        | \boldsymbol{k} \rangle_{3 n_{q e}}
        \otimes
        | \boldsymbol{k}' \rangle_{3 n_{q e}}
        \otimes
        \Big|
            |
            \boldsymbol{r}^{(\boldsymbol{k})} 
            -
            \boldsymbol{r}^{(\boldsymbol{k}')}
            |
        \Big\rangle_{n^{(\mathrm{d})}_{e e}}
        ,
    \label{struct_opt_using_FQE:def_dist_gate_for_e-e}
\end{gather}
where the rightmost ket on the RHS is the state of the distance register that stores the distance within some accuracy characterized by $n^{(\mathrm{d})}_{e e}.$

\subsubsection{Electron-nucleus distances}

We define similarly the unitary
$U^{(\mathrm{d}, \nu)}_{e \mathrm{n}}$ for each $\nu$ that 
computes the distance between an electron and the $\nu$th nucleus and store it into
$n^{(\mathrm{d})}_{e \mathrm{n}}$ qubits as
\begin{gather}
    U^{(\mathrm{d}, \nu)}_{e \mathrm{n}}
    \left(
        | \boldsymbol{k} \rangle_{3 n_{q e}}
        \otimes
        | \boldsymbol{j}_\nu \rangle_{3 n_{q \mathrm{n}}}
        \otimes
        | 0 \rangle_{n^{(\mathrm{d})}_{e \mathrm{n}}}
    \right)
    =
        | \boldsymbol{k} \rangle_{3 n_{q e}}
        \otimes
        | \boldsymbol{j}_\nu \rangle_{3 n_{q \mathrm{n}}}
        \otimes
        \Big|
            |
            \boldsymbol{r}^{(\boldsymbol{k})} 
            -
            \boldsymbol{R}_{\nu} (\boldsymbol{j}_\nu)
            |
        \Big\rangle_{n^{(\mathrm{d})}_{e \mathrm{n}}}
        ,
\end{gather}
where $\boldsymbol{R}_{\nu} (\boldsymbol{j}_\nu)$
is the position of nucleus with the displacement
$\boldsymbol{j}_\nu.$

\subsubsection{Nucleus-nucleus distances}

Also, we define the unitary
$U^{(\mathrm{d}, \nu, \nu')}_{\mathrm{n n}}$ for each pair of $\nu$ and $\nu'$ that 
computes the distance between the pair of nuclei and store it into
$n^{(\mathrm{d})}_{\mathrm{n n}}$ qubits as
\begin{gather}
    U^{(\mathrm{d}, \nu, \nu')}_{\mathrm{n n}}
    \left(
        | \boldsymbol{j}_\nu \rangle_{3 n_{q \mathrm{n}}}
        \otimes
        | \boldsymbol{j}_{\nu'} \rangle_{3 n_{q \mathrm{n}}}
        \otimes
        | 0 \rangle_{n^{(\mathrm{d})}_{\mathrm{n n}}}
    \right)
    =
        | \boldsymbol{j}_\nu \rangle_{3 n_{q \mathrm{n}}}
        \otimes
        | \boldsymbol{j}_{\nu'} \rangle_{3 n_{q \mathrm{n}}}
        \otimes
        \Big|
            |
            \boldsymbol{R}_{\nu} (\boldsymbol{j}_\nu)
            -
            \boldsymbol{R}_{\nu'} (\boldsymbol{j}_{\nu'})
            |
        \Big\rangle_{n^{(\mathrm{d})}_{\mathrm{n n}}}
        .
\end{gather}
We assume that the electron-electron,
electron-nucleus, and nucleus-nucleus distance registers encode approximate distances in a common manner despite the different numbers of constituent qubits.

\subsubsection{Phase gates for interaction as a function of a distance}

We define the interaction phase gate
$U_{\mathrm{int}, \kappa} (\Delta t) \ (\kappa = ee, e \mathrm{n}, \mathrm{nn})$
for a distance register composed of
$n^{(\mathrm{d})}_\kappa$
qubits such that it acts diagonally as
\begin{align}
    U_{\mathrm{int}, \kappa} (\Delta t)
    | r \rangle_{n^{(\mathrm{d})}_\kappa}
    =
        e^{-i v (r) \Delta t}
        | r \rangle_{n^{(\mathrm{d})}_\kappa}
        ,
    \label{struct_opt_using_FQE:def_interaction_phase_gate}
\end{align}
where the ket encodes the argument of $v.$
When $v$ is given as a piecewisely defined polynomial of $r,$
the circuit for the interaction phase gate can be constructed by using the techniques in Refs.\cite{bib:5389, bib:5384},
as summarized in Appendix \ref{sec:phase_gate_for_piecewise_poly}.

\subsubsection{Circuits for pairwise phase gates}

Having introduced the unitaries for the distances between particles and the interaction $v,$
we are now able to construct the circuits for the pairwise phase gates appearing in
Eqs.~(\ref{struct_opt_using_FQE:phase_gate_for_e-e}),
(\ref{struct_opt_using_FQE:phase_gate_for_e-n}),
and
(\ref{struct_opt_using_FQE:phase_gate_for_n-n}).
Specifically,
the $e$-$e$ phase gate for an electron pair defined in
Eq.~(\ref{def_phase_gate_for_ee_pair})
can be expressed as
\begin{gather}
    U_{ee} (\Delta t)
    =
        {U^{(\mathrm{d})}_{e e}}^\dagger
        U_{\mathrm{int}, e e} (\Delta t)
        U^{(\mathrm{d})}_{e e}
        ,
\end{gather}
leading to the circuit shown in
Fig.~\ref{fig:el_el_evol_using_dist_reg}.
This circuit acts on the $6 n_{q e} + n_{e e}^{(\mathrm{d})}$ qubits via the following three steps:
computation of the distance,
generation of the phase,
and 
uncomputation for disentangling the distance register from the electronic register,
that is,
\begin{align}
        | \boldsymbol{k} \rangle_{3 n_{q e}}
        \otimes
        | \boldsymbol{k}' \rangle_{3 n_{q e}}
        \otimes
        | 0 \rangle_{n^{(\mathrm{d})}_{e e}}
    &\xmapsto{\mathrm{Distance}}
        | \boldsymbol{k} \rangle_{3 n_{q e}}
        \otimes
        | \boldsymbol{k}' \rangle_{3 n_{q e}}
        \otimes
        \Big|
            |
            \boldsymbol{r}^{(\boldsymbol{k})} 
            -
            \boldsymbol{r}^{(\boldsymbol{k}')}
            |
        \Big\rangle_{n^{(\mathrm{d})}_{e e}}
    \nonumber \\
    &\xmapsto{\mathrm{Phase}}
        | \boldsymbol{k} \rangle_{3 n_{q e}}
        \otimes
        | \boldsymbol{k}' \rangle_{3 n_{q e}}
        \otimes
        e^{i v (r) \Delta t}
        | r \rangle_{n^{(\mathrm{d})}_{e e}}
        \Big|_{
            r =
            |
            \boldsymbol{r}^{(\boldsymbol{k})} 
            -
            \boldsymbol{r}^{(\boldsymbol{k}')}
            |
        }
    \nonumber \\
    &\xmapsto{\mathrm{Distance}^\dagger}
        e^{i
        v (
            |
            \boldsymbol{r}^{(\boldsymbol{k})} 
            -
            \boldsymbol{r}^{(\boldsymbol{k}')}
            |
        )
        \Delta t}
        | \boldsymbol{k} \rangle_{3 n_{q e}}
        \otimes
        | \boldsymbol{k}' \rangle_{3 n_{q e}}
        \otimes
        | 0 \rangle_{n^{(\mathrm{d})}_{e e}}
        .
\end{align}
Since the distance register has been initialized again,
it can be recycled for other pairs of electrons.

The $e$-n phase gate for an electron-nucleus pair defined in
Eq.~(\ref{def_phase_gate_for_en_pair})
can be expressed as
\begin{gather}
    U_{e \mathrm{n}}^{(\nu)} (\Delta t)
    =
        {U^{(\mathrm{d}, \nu)}_{e \mathrm{n}}}^\dagger
        U_{\mathrm{int}, e \mathrm{n}} (- Z_\nu \Delta t)
        U^{(\mathrm{d}, \nu)}_{e \mathrm{n}}
        ,
\end{gather}
while the n-n phase gate for a nucleus pair defined in
Eq.~(\ref{def_phase_gate_for_nn_pair})
can be expressed as
\begin{gather}
    U_{\mathrm{nn}}^{(\nu, \nu')} (\Delta t)
    =
        {U^{(\mathrm{d}, \nu, \nu')}_{\mathrm{n n}}}^\dagger
        U_{\mathrm{int}, \mathrm{n n}}
        (Z_\nu Z_{\nu'} \Delta t)
        U^{(\mathrm{d}, \nu, \nu')}_{\mathrm{n n}}
        .
\end{gather}
The circuits for these pairwise phase gates can also be constructed similarly to the case of an electron pair.

\subsection{Circuit depths}
\label{sec:impl_of_phase_gates:depths}

The circuit depth for the evolution generated by $\hat{V}_{e e}$ is thus found to scale as
\begin{gather}
    \mathrm{depth}
    (e^{-i \hat{V}_{e e} \Delta t})
    =
        \mathcal{O}
        \left(
            n_e^2
            \mathrm{poly}
            \left(
                \log
                \frac{n_e^{1/3}}{\Delta x}
            \right)
        \right)
    .
    \label{circuit:depth_scaling_of_evol_e-e}
\end{gather}
[See Fig.~\ref{circuit:real_time_evol_for_opt_with_pite}]
That for $\hat{V}_{\mathrm{n n}}$ scales as
\begin{gather}
    \mathrm{depth}
    (e^{-i \hat{V}_{\mathrm{n n}} \Delta t})
    =
        \mathcal{O}
        \left(
            n_{\mathrm{nucl}}^2
            \mathrm{poly}
            \left(
                \log
                \frac{\Delta R_{\mathrm{max}}}{\Delta R}
            \right)
        \right)
    .
    \label{circuit:depth_scaling_of_evol_n-n}
\end{gather}
That for $\hat{V}_{e \mathrm{n}}$ scales as
\begin{gather}
    \mathrm{depth}
    (e^{-i \hat{V}_{e \mathrm{n}} \Delta t})
    =
        \mathcal{O}
        \left(
            n_e
            \mathrm{poly}
            \left(
                \max
                \left(
                    \log
                    \frac{n_e^{1/3}}{\Delta x}
                    ,
                    \log
                    \frac{\Delta R_{\mathrm{max}}}{\Delta R}
                \right)
            \right)
        \right)
        .
\end{gather}
Also, that for $\hat{V}_{\mathrm{ext}}$ scales as
\begin{gather}
    \mathrm{depth}
    (e^{-i \hat{V}_{\mathrm{ext}} \Delta t})
    =
        \mathcal{O}
        \left(
            \mathrm{poly}
            \left(
                \log
                \frac{n_e^{1/3}}{\Delta x}
            \right)
        \right)
    .
\end{gather}
The circuit depth of the QFT-based kinetic evolution scales as \cite{bib:5737}
\begin{gather}
    \mathrm{depth}
    (e^{-i \hat{T} \Delta t})
    =
        \mathcal{O}
        \left(
            n_e^0
            n_{q e}^2
        \right)
    =
        \mathcal{O}
        \left(
            \left(
                \log
                \frac{n_e^{1/3}}{\Delta x}
            \right)^2
        \right)
    ,
    \label{circuit:depth_scaling_of_evol_kin}
\end{gather}
for which the well known quadratic-depth implementation \cite{Nielsen_and_Chuang} of QFT is adopted.
We can also implement QFT by using the phase gradient circuits \cite{phase_gradient_circuits},
leading to a better depth $\mathcal{O} (n_e^0 n_{q e} \log n_{q e})$ instead of
Eq.~(\ref{circuit:depth_scaling_of_evol_kin}).

\subsection{Efficient implementation of electron-electron evolution using a redundant register}
\label{sec:impl_of_phase_gates:efficient_el_el_evol}

If the same number of qubits as in the electronic register are available,
we can reduce the scaling of circuit depth for the evolution
$e^{-i \hat{V}_{e e} \Delta t}$
generated by the electron-electron interactions.
Although the technique explained below is essentially the same as that in Ref.~\cite{bib:5824},
we describe it in order for this paper to be self contained.

For the electronic register consisting of $3 n_e n_{q e}$ qubits,
we construct the circuit shown in
Fig.~\ref{circuit:redundant_reg_for_el_el}(a)
by introducing an extra register consisting of $3 n_e n_{q e}$ qubits denoted as a redundant register.
The $6 n_e n_{q e}$-qubit system composed of 
the electronic register encoding a generic state in Eq.~(\ref{many_electron_state})
and the initialized redundant register undergoes the operations as
\begin{gather}
    \overbrace{
        | \psi \rangle
    }^{\mathrm{Electrons}}
    \otimes
    \overbrace{
        | 0 \rangle_{3 n_e n_{q e}}
    }^{\mathrm{Redundant}}
    \nonumber \\
    \xmapsto{\mathrm{CNOTs}}
        \Delta V^{n_e/2}
        \sum_{\boldsymbol{K}}
        \psi (
            \boldsymbol{r}^{(\boldsymbol{k}_0)},
            \dots,
            \boldsymbol{r}^{(\boldsymbol{k}_{n_e - 1})}
        )
        | \boldsymbol{K} \rangle_{3 n_e n_{q e} }
        \otimes
        | \boldsymbol{K} \rangle_{3 n_e n_{q e} }
    \nonumber \\
    \xmapsto{\mathrm{Phase}}
        \prod_{
            \substack{
                \ell > \ell' \\
                \ell \in \mathrm{Electron \ reg.} \\
                \ell' \in \mathrm{Redundant \ reg.}
            }
        }
        \overbrace{
            U_{ee} (\Delta t)
        }^{\mathrm{On} \ \ell \mathrm{th \ and} \ \ell' \mathrm{th \ electrons}}
        \Delta V^{n_e/2}
        \sum_{\boldsymbol{K}}
        \psi (
            \boldsymbol{r}^{(\boldsymbol{k}_0)},
            \dots,
            \boldsymbol{r}^{(\boldsymbol{k}_{n_e - 1})}
        )
        | \boldsymbol{K} \rangle_{3 n_e n_{q e} }
        \otimes
        | \boldsymbol{K} \rangle_{3 n_e n_{q e} }
    \nonumber \\
    =
        \prod_{
            \substack{
                \ell > \ell' \\
                \ell \in \mathrm{Electron \ reg.} \\
                \ell' \in \mathrm{Redundant \ reg.}
            }
        }
        \Delta V^{n_e/2}
        \sum_{\boldsymbol{K}}
        \psi (
            \boldsymbol{r}^{(\boldsymbol{k}_0)},
            \dots,
            \boldsymbol{r}^{(\boldsymbol{k}_{n_e - 1})}
        )
        \exp
        \left(
            -i
            v
            \left(
            \left|
                \boldsymbol{r}^{(\boldsymbol{k}_\ell)}
                -
                \boldsymbol{r}^{(\boldsymbol{k}_{\ell'})}
            \right|
            \right)
            \Delta t
        \right)
        | \boldsymbol{K} \rangle_{3 n_e n_{q e} }
        \otimes
        | \boldsymbol{K} \rangle_{3 n_e n_{q e} }
    \nonumber \\
    =
        \prod_{
            \substack{
                \ell > \ell' \\
                \ell, \ell' \in \mathrm{Electron \ reg.}
            }
        }
        \Delta V^{n_e/2}
        \sum_{\boldsymbol{K}}
        \psi (
            \boldsymbol{r}^{(\boldsymbol{k}_0)},
            \dots,
            \boldsymbol{r}^{(\boldsymbol{k}_{n_e - 1})}
        )
        \exp
        \left(
            -i
            v
            \left(
            \left|
                \boldsymbol{r}^{(\boldsymbol{k}_\ell)}
                -
                \boldsymbol{r}^{(\boldsymbol{k}_{\ell'})}
            \right|
            \right)
            \Delta t
        \right)
        | \boldsymbol{K} \rangle_{3 n_e n_{q e} }
        \otimes
        | \boldsymbol{K} \rangle_{3 n_e n_{q e} }
    \nonumber \\
    \xmapsto{\mathrm{CNOTs}}
        \prod_{
            \substack{
                \ell > \ell' \\
                \ell, \ell' \in \mathrm{Electron \ reg.}
            }
        }
        \Delta V^{n_e/2}
        \sum_{\boldsymbol{K}}
        \psi (
            \boldsymbol{r}^{(\boldsymbol{k}_0)},
            \dots,
            \boldsymbol{r}^{(\boldsymbol{k}_{n_e - 1})}
        )
        \exp
        \left(
            -i
            v
            \left(
            \left|
                \boldsymbol{r}^{(\boldsymbol{k}_\ell)}
                -
                \boldsymbol{r}^{(\boldsymbol{k}_{\ell'})}
            \right|
            \right)
            \Delta t
        \right)
        | \boldsymbol{K} \rangle_{3 n_e n_{q e} }
        \otimes
        | 0 \rangle_{3 n_e n_{q e} }
    \nonumber \\
    =
        \left(
            e^{-i \hat{V}_{e e} \Delta t}
            | \psi \rangle
        \right)
        \otimes
        | 0 \rangle_{3 n_e n_{q e} }
    ,
    \label{impl_of_el_el_evol_using_redundant}
\end{gather}
where the redundant register in the final state can be discarded safely since it is disentangled from the electronic register.
The central part of this circuit is composed of the $n_e (n_e - 1)/2$ electron-electron phase gates $U_{e e} (\Delta t).$
Their sequence can be written as
\begin{gather}
        \prod_{
            \substack{
                \ell > \ell' \\
                \ell \in \mathrm{Electron \ reg.} \\
                \ell' \in \mathrm{Redundant \ reg.}
            }
        }
        \overbrace{
            U_{ee} (\Delta t)
        }^{\mathrm{On} \ \ell \mathrm{th \ and} \ \ell' \mathrm{th \ electrons}}
    =
        \prod_{\ell' \in \mathrm{Redundant \ reg.}}
        \prod_{d = 1}^{n_e - 1 - \ell'}
        \overbrace{
            U_{ee} (\Delta t)
        }^{\mathrm{On} \ (\ell' + d) \mathrm{th \ and} \ \ell' \mathrm{th \ electrons}}
    ,
\end{gather}
which means that the $n_e - 1$ phase gates for a fixed $d$ can be performed simultaneously.
This fact enables us to implement the evolution with the depth $\mathcal{O} (n_e)$ in terms of the electron number.
Fig.~\ref{circuit:redundant_reg_for_el_el}(b)
shows the circuit for the case of $n_e = 4$ as an example.
This technique is applicable not only to electronic simulations but also to other kinds of quantum algorithms that impose all-to-all connectivity on qubits.

\begin{figure}
\begin{center}
\includegraphics[width=13cm]{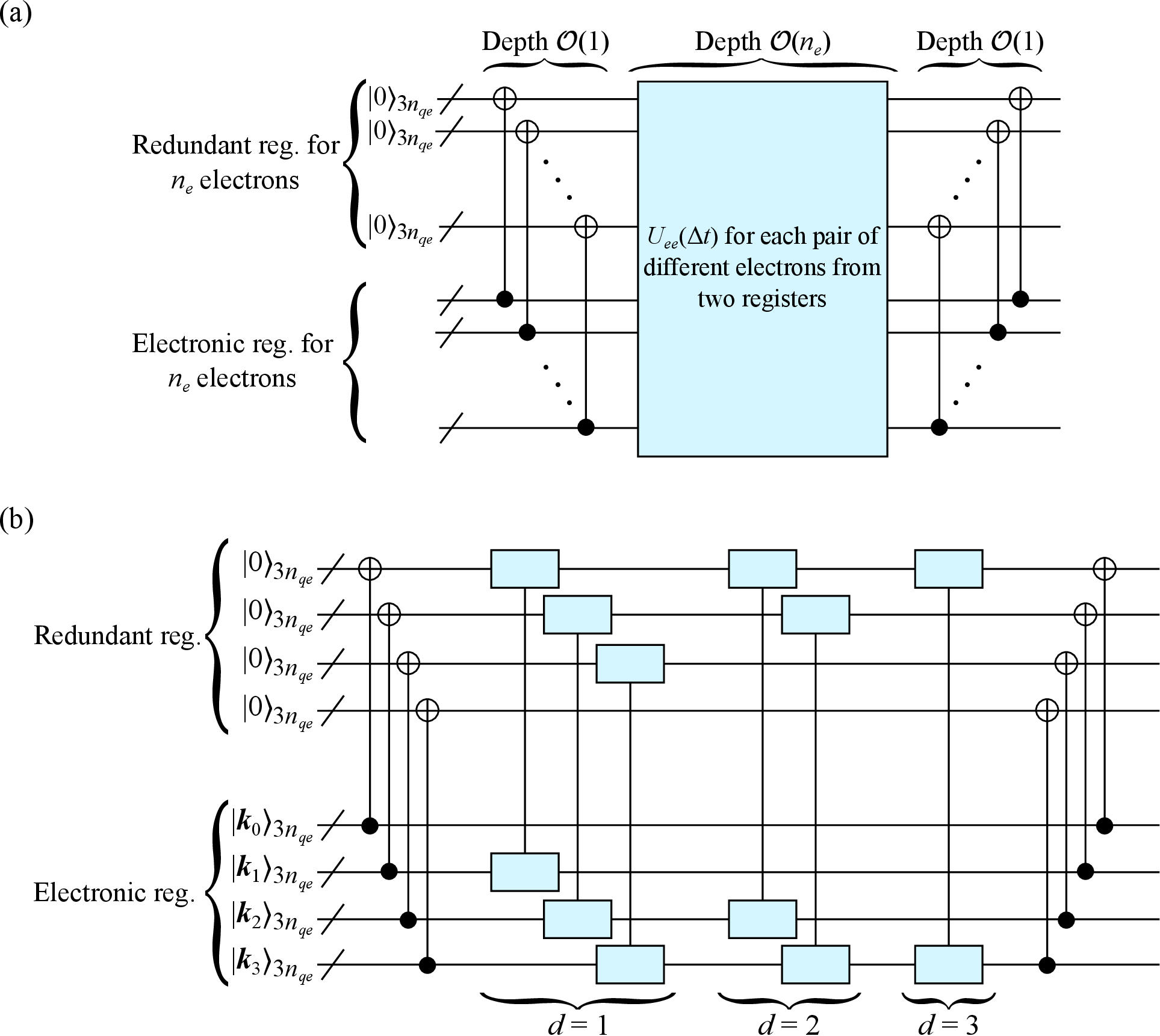}
\end{center}
\caption{
(a)
Efficient implementation of $e^{-i \hat{V}_{e e} \Delta t}$
for $n_e$ electrons
by using a redundant register consisting of the same number of qubits as in the electronic register.
Each of the CNOT symbols on the circuit means that
the $3 n_{q e}$ qubits for each electron in the electronic register are coupled via CNOT operations to the corresponding qubits in the redundant register.
The scaling of depths with respect to $n_e$ is also shown.
(b)
The implementation of evolution for $n_e = 4.$
Each pair of connected boxes on the circuit represents the electron-electron phase gate $U_{e e} (\Delta t),$ defined in
Eq.~(\ref{def_phase_gate_for_ee_pair}).
}
\label{circuit:redundant_reg_for_el_el}
\end{figure}

\section{Circuit construction of phase gate for a piecewisely defined polynomial}
\label{sec:phase_gate_for_piecewise_poly}

As stated in the main text,
one of the crucial tasks for implementing the real-time evolution for an electronic system is the construction of phase gates responsible for the interactions as functions of the distance between particles.
We assume here that the distance is encoded as a discrete value represented by the computational basis of the distance register.
In order for this paper to be self contained,
we describe the recipes for implementing the gates,
essentially the same as provided in Ref.\cite{bib:5384}

\subsection{Phase gate for a simple polynomial}

Let us consider here a variable $x$ restricted within a range $[0, L]$ and a degree-$M$ real polynomial
$f (x) = a_0 + a_1 x + \cdots + a_M x^M$
defined on the entire range.
We use $n$ qubits for representing $N \equiv 2^n$ grid points $x^{(k)} \equiv k \Delta x \ (k = 0, \dots, N - 1),$
each corresponding to the computational basis
$| k \rangle_n.$
$\Delta x \equiv L/N$ is the grid spacing.
We want to implement the polynomial phase gate $U_{\mathrm{ph}} [f]$ that acts as
\begin{gather}
| k \rangle_n
\xmapsto{U_{\mathrm{ph}} [f]}
\exp (i f (x^{(k)}) )
| k \rangle_n
\end{gather}
for a computational basis.

We define a unitary $U^{(m)} (a)$ specified by a real parameter $a$ as
$
| k \rangle_n
\xmapsto{U^{(m)} (a)}
\exp ( i a x^{(k) m} )
| k \rangle_n
$
for each $m.$
Since those for different $m$'s commute with each other,
we can write the phase gate to be implemented as
$
U_{\mathrm{ph}} [f]
=
\prod_{m = 0}^M
U^{(m)} (a_m)
.
$
For a given $k,$ let 
$k_{n - 1} \cdots k_1 k_0$ its binary representation.
The power of the coordinate is then written as
\begin{align}
    x^{(k) m}
    &=
        (\Delta x)^m
        \left(
            \sum_{\ell = 0}^{n - 1}
                2^\ell
                k_\ell
        \right)^m
    \nonumber \\
    &=
        (\Delta x)^m
        \sum_{\ell_0 = 0}^{n - 1}
        \cdots    
        \sum_{\ell_{m - 1} = 0}^{n - 1}
            2^{\ell_0 + \cdots + \ell_{m - 1}}
            k_{\ell_0}
            \dots
            k_{\ell_{m - 1}}
    \nonumber \\
    &=
        (\Delta x)^m
        \sum_{\ell_0 = 0}^{n - 1}
        \cdots    
        \sum_{\ell_{m - 1} = 0}^{n - 1}
            2^{\ell_0 + \cdots + \ell_{m - 1}}
            \delta_{k_{\ell_0}, 1}
            \dots
            \delta_{k_{\ell_{m - 1}}, 1}
            .
    \label{phase_gate_for_polynomial:x_to_power_of_m}
\end{align}
For each term on the RHS in the equation above,
we define a multiply controlled phase gate 
$\mathrm{C} Z_{\ell_0, \dots, \ell_{m - 1}} (\theta)$
that acts on an arbitrary single qubit
as 
$Z (\theta) = | 0 \rangle \langle 0 | + e^{i \theta} | 1 \rangle \langle 1 |$
among at most $m$ qubits
$| q_{\ell_0} \rangle, \dots, | q_{\ell_{m - 1}} \rangle$
(repeated indices allowed)
with the other at most $m - 1$ as the control bits.
From these phase gates and
Eq.~(\ref{phase_gate_for_polynomial:x_to_power_of_m}),
we can write the action of the degree-$m$ contribution as
\begin{align}
    U^{(m)} (a_m)
    | k \rangle_n
    &=
        \exp
        \left(
            i
            a_m
            (\Delta x)^m
            \sum_{\ell_0 = 0}^{n - 1}
            \cdots    
            \sum_{\ell_{m - 1} = 0}^{n - 1}
                2^{\ell_0 + \cdots + \ell_{m - 1}}
                \delta_{k_{\ell_0}, 1}
                \dots
                \delta_{k_{\ell_{m - 1}}, 1}
        \right)
        | k \rangle_n
    \nonumber \\
    &=
        \prod_{\ell_0 = 0}^{n - 1}
        \cdots    
        \prod_{\ell_{m - 1} = 0}^{n - 1}
        \exp
        \left(
            i
            a_m
            (\Delta x)^m
            2^{\ell_0 + \cdots + \ell_{m - 1}}
            \delta_{k_{\ell_0}, 1}
            \dots
            \delta_{k_{\ell_{m - 1}}, 1}
        \right)
        | k \rangle_n
    \nonumber \\
    &=
        \prod_{\ell_0 = 0}^{n - 1}
        \cdots    
        \prod_{\ell_{m - 1} = 0}^{n - 1}
        \mathrm{C} 
        Z_{ \ell_0, \dots, \ell_{m - 1} }
        \left(
            a_m
            (\Delta x)^m
            2^{\ell_0 + \cdots + \ell_{m - 1}}
        \right)
        | k \rangle_n
        .
        \label{phase_gate_for_polynomial:U_m_impl_by_ctrl_phase_gates}
\end{align}
The equation above indicates that $U^{(m)} (a_m)$ can be implemented from the $n^m$ multiply controlled phase gates.
Since $U_{\mathrm{ph}} [f]$ can be implemented by concatenating $U^{(m)} (a_m)$ for all the $m$'s,
the number of involved controlled phase gates is
$\sum_{m = 0}^M n^m = \mathcal{O} (n^M).$
If we adopt the techniques for a generic multiply controlled single-qubit gate proposed by Silva and Park \cite{bib:5662},
each $\mathrm{C} Z_{\ell_0, \dots, \ell_{m - 1}} (\theta)$
is implemented with a depth linear in $m$
with respect to single-qubit and CNOT gates.
The depth for the polynomial phase gate in such a case is thus
\begin{gather}
    \mathrm{depth} 
    (U_{\mathrm{ph}} [f])
    =
        \mathcal{O} (M n^M)
    \label{phase_gate_for_polynomial:depth_of_simple_poly}
\end{gather}
since $m$ reaches $M.$

Finding all the controlled phase gates in
Eq.~(\ref{phase_gate_for_polynomial:U_m_impl_by_ctrl_phase_gates})
commute with each other,
we can merge those involving the same combination of qubits.
Although this fact does not reduce the scaling of depth in
Eq.~(\ref{phase_gate_for_polynomial:depth_of_simple_poly}),
it allows one to implement $U_{\mathrm{ph}} [f]$ as a shallower circuit.
We show a possible implementation for $n = 4$ and $f(x) = a_3 x^3$
in Fig.~\ref{fig:phase_gate_poly_example} as an example.

\begin{figure}
\begin{center}
\includegraphics[width=13cm]{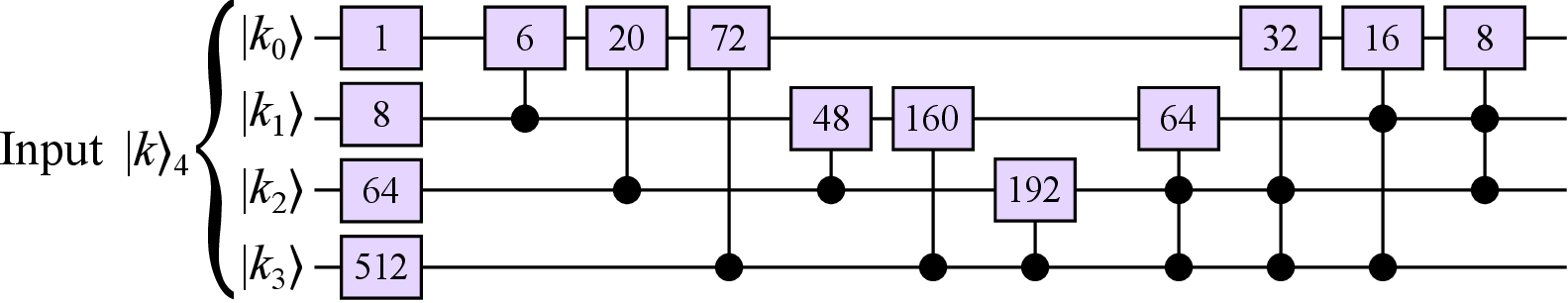}
\end{center}
\caption{
Implementation of $U_{\mathrm{ph}} [f]$ for a polynomial $f (x) = a_3 x^3$
for $n = 4$ qubits.
Each box containing an integer $s$ in this circuit represents a single-qubit phase gate
$Z (a_3 (\Delta x)^3 s).$
}
\label{fig:phase_gate_poly_example}
\end{figure}

\subsection{Phase gate for a piecewisely defined polynomial}

We next consider a real continuous function $g$
which is given as a polynomial on each of 
$n_{\mathrm{pieces}}$ non-overlapping ranges.
We assume the $p$th range
$(p = 0, \dots, n_{\mathrm{pieces}} - 1)$
to be
$[ x^{(k_{\mathrm{min}, p})}, x^{(k_{\mathrm{max}, p})} ],$
for which the polynomial $g_p$ is known,
so that the small ranges cover the entire range, $[0, L].$

For implementing the phase gate for $g,$
we define a comparator $V_p$ for the $p$th range $(p \geq 1)$
as a unitary acting on the computational basis as follows:
\begin{gather}
    | k \rangle_n
    \otimes
    \overbrace{
        | 0 \rangle
    }^{\mathrm{Ancilla}}
    \xmapsto{V_p}
        | k \rangle_n
        \otimes
        | k \geq k_{\mathrm{min}, p} \rangle
        ,
    \label{phase_gate_for_polynomial:def_comparator_Vp}
\end{gather}
where the output ancillary state is $| 1 \rangle$
if $k \geq k_{\mathrm{min}, p},$ otherwise $| 0 \rangle.$
By using the comparator for two two-bit integers as a building block
(explicitly provided by Berry {\it et al.} \cite{bib:5389} using only the Fredkin and CNOT gates),
the comparator in
Eq.~(\ref{phase_gate_for_polynomial:def_comparator_Vp})
for an input $n$-bit integer can be constructed by fixing the other $n$-bit integer at $k_{\mathrm{min}, p}.$
We then define a unitary $W_p$ from $V_p$ and
$U_{\mathrm{ph}} [g_p - g_{p - 1}],$
as shown in Fig.~\ref{fig:phase_gates_for_poly}(a).
We have to recall here that,
even though the expression $g_p - g_{p - 1}$ is mathematically invalid due to the different domains of the two polynomials, the gate operation itself is possible since their expressions are known.
When $| k \rangle_n$ is input to $W_p,$
it acquires a phase factor 
$\exp (i g_p (x^{(k)}) - i g_{p - 1} (x^{(k)}) )$
if $x^{(k)} \geq x^{(k_{\mathrm{min}, p})},$
otherwise no phase factor is acquired.

The phase gate $U_{\mathrm{ph}}^{(\mathrm{pw})} [g]$
for the piecewisely defined polynomial $g$ is now implementable by putting the phase gate for $g_0$ and
concatenating $V_p$ with increasing $p,$ as shown in
Fig.~\ref{fig:phase_gates_for_poly}(b).
One can easily confirm that the circuit works as expected:
\begin{gather}
U_{\mathrm{ph}}^{(\mathrm{pw})} [g] 
(| k \rangle_n \otimes | 0 \rangle)
=
\exp(i g (x^{(k)}))
(| k \rangle_n \otimes | 0 \rangle)
.  
\end{gather}
From the circuit depth for the comparator,
$\mathrm{depth} (V_p) = \mathcal{O} (\log n)$ \cite{bib:5389},
and that for a simple polynomial in
Eq.~(\ref{phase_gate_for_polynomial:depth_of_simple_poly}),
we find that the depth for the piecewisely defined polynomial scales as
\begin{align}
    \mathrm{depth}
    \left(
        U_{\mathrm{ph}}^{(\mathrm{pw})} [g]
    \right)
    &=
        \mathcal{O} 
        \left(
            n_{\mathrm{pieces}}
            \max
            \left(
                V_p,  U_{\mathrm{ph}} [g_p - g_{p - 1}]
            \right)
        \right)
    \nonumber \\
    &=
        \mathcal{O}
        \left(
            n_{\mathrm{pieces}}
            M
            n^M
        \right)
        .
    \label{phase_gate_for_polynomial:depth_of_polynomials}
\end{align}
The exponential scaling of 
Eq.~(\ref{phase_gate_for_polynomial:depth_of_polynomials})
in terms of $M$ suggests introducing more narrow ranges rather than increasing $M.$ 
If the ranges are narrow enough that the target function $g$ is approximated accurately by a low-order polynomial ($M = 1$ or 2 for example) over each range,
we can suppress the problematic scaling with maintaining the accuracy.

\begin{figure}
\begin{center}
\includegraphics[width=15cm]{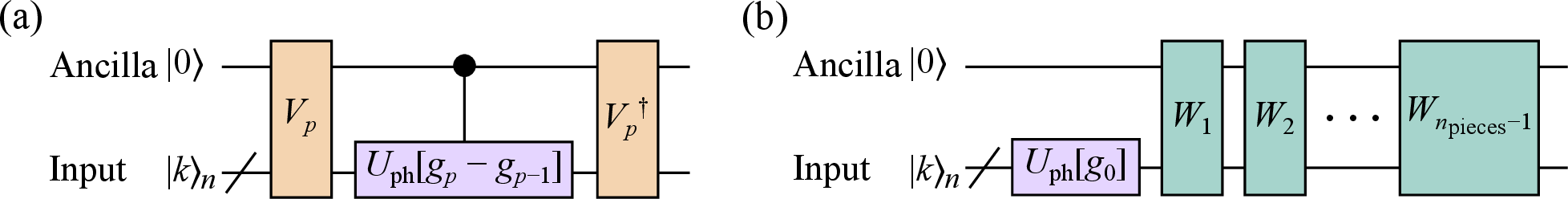}
\end{center}
\caption{
(a)
Definition of $W_p$ for $p \geq 1$
from the comparator $V_p$ \cite{bib:5389} and the phase gate $U_{\mathrm{ph}}$ for a simple polynomial.
The resultant ancillary state is $| 0 \rangle$ regardless of $| k \rangle_n$ thanks to the uncomputation.
(b)
Definition of $U_{\mathrm{ph}}^{(\mathrm{pw})} [g]$
for the piecewisely defined polynomial.
}
\label{fig:phase_gates_for_poly}
\end{figure}

\section{Review of VITE approach}
\label{sec:review_of_VITE}

For solving the imaginary-time Schr\"odinger equation in the VQE manner,
we approximate the solution as
$
|\Psi(\tau)\rangle
\approx
|\Phi(\boldsymbol{\theta}(\tau))\rangle
$
expressed with $n_{\mathrm{p}}$ real parameters $\boldsymbol{\theta}(\tau)$
specified by a single real parameter $\tau,$
referred to as an imaginary time.
In the language of quantum computation,
$\boldsymbol{\theta}$ are the circuit parameters characterizing an adopted ansatz.
Here, we use the McLachlan's variational principle \cite{bib:4803} to derive the equation of motion for $\boldsymbol{\theta}.$
By using the expected energy
$
E_\tau
\equiv
\langle \Phi (\boldsymbol{\theta} (\tau)) |
\mathcal{H}
| \Phi (\boldsymbol{\theta} (\tau)) \rangle
$
for a fixed state $|\Phi(\boldsymbol{\theta}(\tau))\rangle$
at a given $\tau$ for the Hamiltonian $\mathcal{H},$
the requirement that the distance between derivative 
$
\partial|\Phi(\boldsymbol{\theta}(\tau))\rangle / \partial \tau
$
and 
$
-(\mathcal{H} - E_\tau ) |\Phi(\boldsymbol{\theta}(\tau))\rangle 
$
should be a stationary value is expressed as \cite{bib:4807}
\begin{gather}
	\delta \left \|
	    \left(
	        \frac{\partial}{\partial \tau}
	        +
	        \mathcal{H}
	        -
	        E_\tau
	    \right)
        |\Phi(\boldsymbol{\theta}(\tau))\rangle 
    \right\|^2=0
    .
\end{gather}
This equation is solved under the constraint
$
\| |\Phi(\boldsymbol{\theta}(\tau))\rangle \|^2 = 1
$
for normalization.
Accordingly, by executing the variation, we obtain the equation governing the evolution of the parameters:
\begin{gather}
	\mathcal{M} (\tau)
	\frac{d\boldsymbol{\theta}(\tau)}{d \tau} 
	= 
	\boldsymbol{\mathcal{V}} (\tau)
\label{eom_vite}
\end{gather}
where the matrix
\begin{gather}
    \mathcal{M}_{j j'}(\tau)
    =
    \operatorname{Re}
    \left\langle
    \frac{\partial \Phi(\boldsymbol{\theta})}{\partial \theta_j}
    \middle|
    \frac{\partial \Phi(\boldsymbol{\theta})}{\partial \theta_{j'}}
    \right\rangle
    \Bigg|_{\boldsymbol{\theta} = \boldsymbol{\theta} (\tau)}
\end{gather} 
and the vector
\begin{gather}
    \mathcal{V}_j (\tau)
    =
    -
    \operatorname{Re}
    \left\langle
    \frac{\partial \Phi(\boldsymbol{\theta})}{\partial \theta_j}
    \middle|
    \mathcal{H}
    \middle|
    \Phi(\boldsymbol{\theta})
    \right\rangle
    \Bigg|_{\boldsymbol{\theta} = \boldsymbol{\theta} (\tau)}
    \label{sec:review_of_VITE_def_V}
\end{gather}
for $j, j' = 0, \dots, n_{\mathrm{p}} - 1$ were introduced.
From Eq (\ref{eom_vite}),
the parameters are updated from a finite difference $\Delta \tau$ of the imaginary time as
$
\boldsymbol{\theta}(\tau+\Delta\tau) 
=
\boldsymbol{\theta}(\tau)
+
\Delta\tau \mathcal{M}^{-1}(\tau) \boldsymbol{\mathcal{V}} (\tau)
$
within the first-order accuracy of $\Delta \tau.$

Next, we have to evaluate $\mathcal{M}(\tau)$ and $\mathcal{V}(\tau)$ for Eq.~(\ref{eom_vite}) by using the quantum computer.
We consider here a case where the variational parameters for a trial $n_q$-qubit wave function enter only via single-qubit rotation gates
$R_\mu (\theta) = e^{-i \theta \sigma_\mu/2}$ with $\mu = x, y, z.$
The analytic gradients of a parametrized unitary are presented in Ref \cite{bib:5838}.
Specifically, the derivative of an $n$-parameter unitary of the form
$
\mathcal{U} (\boldsymbol{\theta}) 
=
\mathcal{U}' (\theta_1, \ldots, \theta_{k-1}) 
R_\mu (\theta_k) 
\mathcal{U}'' (\theta_{k+1}, \ldots, \theta_{n_{\mathrm{p}}})
$
with respect to the $k$th parameter is given obviously as
\begin{gather}
    \frac{\partial \mathcal{U} (\boldsymbol{\theta})}{\partial \theta_k}
    =
    -
    \frac{i}{2}
    \mathcal{U}' (\theta_1, \ldots, \theta_{k-1}) 
    \sigma_\mu
    R_{\mu}(\theta_k) 
    \mathcal{U}'' (\theta_{k+1}, \ldots, \theta_{n_{\mathrm{p}}})  .  
\end{gather}
The inner product of different circuits for obtaining
$\mathcal{M}_{j j'} (\tau)$ can be evaluated using, for example, the Hadamard test \cite{bib:5839}.

The remaining quantities we have to compute are $E_\tau$ and
$\boldsymbol{\mathcal{V}} (\tau).$
We briefly explain the methodology presented by
Ollitrault et al.~\cite{Ollitrault2022arXiv}, but of course,
the method presented in Ref.~\cite{bib:5737} can be adopted.
The Hamiltonian $\mathcal{H}$ in our case is 
decomposed into the kinetic and potential parts as in
Eq.~(\ref{struct_opt_using_FQE:def_Hamiltonian}).
Since the potential term is diagonal for the computational basis, the expectation value for the potential part can be computed from repeated measurements on the trial state using computational bases.
On the contrary, the kinetic part is diagonal for the momentum basis.
Thus, we can evaluate the expectation value for the kinetic part by applying the QFT to the trial state immediately before performing a measurement.
In the following, we discuss the details for evaluating the matrix element for a potential operator $\hat{V}.$
First, we write the derivative of the parameterized circuit as
$
|\partial_j \mathcal{U} (\boldsymbol{\theta}) \rangle 
\equiv
-(i/2) \mathcal{W}_j (\boldsymbol{\theta}) |0\rangle_{n_q}
.
$ 
What we have to evaluate is then,
from Eq.~(\ref{sec:review_of_VITE_def_V}),
$
\operatorname{Im} \langle  0|_{n_q} 
    \mathcal{W}_j (\boldsymbol{\theta} (\tau))^{\dagger} 
    \hat{V} 
    \mathcal{U} (\boldsymbol{\theta} (\tau))
    |0\rangle_{n_q}
.
$
It can be obtained from the circuit shown in
Fig.~\ref{fig:circuit_for_matel_in_vite}
that uses one ancillary qubit.
For simplicity,
we denote $\mathcal{U}(\boldsymbol{\theta} (\tau))$ and
$\mathcal{W}_j(\boldsymbol{\theta} (\tau))$
by $\mathcal{U}$ and $\mathcal{W}_j$, respectively. 
The $Z_{\phi}$ is the phase gate.
The input state $|0\rangle \otimes |0\rangle_{n_q}$ changes throughout the circuit as
\begin{gather}
    \frac{1}{2} \sum_{s = 0, 1} |s\rangle 
    \otimes
    \left(
    \mathcal{W}_j + (-1)^{s}e^{i\phi} \mathcal{U}
    \right)
    |0\rangle_{n_q}
    .
\end{gather}
The probability for finding
$
| s \rangle \otimes | k \rangle_{n_q}
\
(k = 0, \dots, 2^{n_q} - 1)
$
when we measure all the qubits is thus calculated as 
\begin{gather}
    \mathbb{P}_{s, k}^{(j)}
    =
    \frac{1}{4} 
    \left|
        \langle k |_{n_q} \mathcal{W}_j | 0\rangle_{n_q}
    \right|^{2}
    +
    \frac{1}{4} 
    \left|
        \langle k |_{n_q} \mathcal{U} | 0\rangle_{n_q}
    \right|^{2}
    +
    \frac{(-1)^{s}}{2}
    \left(
        \cos \phi
        \operatorname{Re}
        v_k^{(j)}
        -
        \sin \phi
        \operatorname{Im}
        v_k^{(j)}
    \right)
    ,
\end{gather}
where
$
v_k^{(j)}
\equiv
\langle 0 |_{n_q} \mathcal{W}_j^{\dagger}  | k \rangle_{n_q}
\langle k |_{n_q} \mathcal{U} | 0\rangle_{n_q}
$
Finally, by setting $\phi=\pi/2$, we get the target value from the probabilities $\mathbb{P}_{s, k}^{(j)}$
for all the combinations of $s$ and $k$ as follows:
\begin{gather}
    \sum_{s,k} (-1)^{s}
    \langle k |_{n_q} \hat{V} | k \rangle_{n_q}
    \mathbb{P}_{s,k}^{(j)}
    =
    -
    \operatorname{Im}
    \langle  0|_{n_q} 
        \mathcal{W}_j^{\dagger} 
        \hat{V}
        \mathcal{U} 
    |0 \rangle_{n_q} .
\end{gather}

\begin{figure}
\begin{center}
\includegraphics[width=7cm]{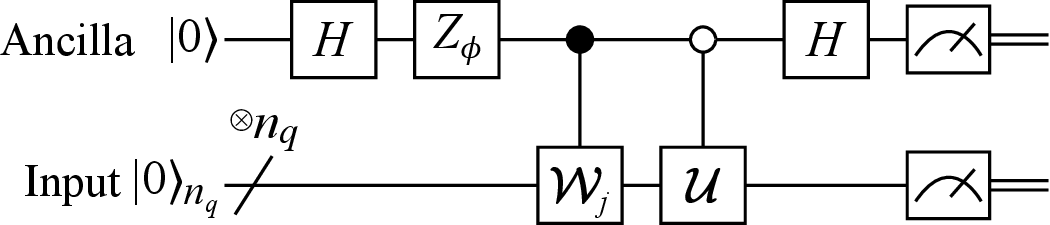}
\end{center}
\caption{
Circuit for calculation of 
$
\operatorname{Im} \langle  0|_{n_q} 
    \mathcal{W}_j^{\dagger} 
    \hat{V} 
    \mathcal{U}
    |0\rangle_{n_q}
.
$
$H$ is the Hadamard gate.
$Z_\phi = \mathrm{diag} (1, e^{i \phi})$ is a phase gate.
}
\label{fig:circuit_for_matel_in_vite}
\end{figure}

\section{Details of numerical simulations}

\subsection{PITE for a model LiH molecule}
\label{sec:details_of_LiH}

\subsubsection{Hamiltonian matrix for a fixed geometry}

As explained in the main text,
we adopted $n_{q e} = 6$ for $N_{q e} = 64$ grid points for each of the $n_e = 2$ electrons in the one-dimensional space.
For a fixed molecular geometry, 
the Hamiltonian matrix in the position representation of the two electrons has dimension of $64^{n_e} = 4096.$
Since the kinetic-energy operator for the two electrons is written as
$\hat{T} = \hat{T}_{1 e} \otimes \hat{I}_{1 e} + \hat{I}_{1 e} \otimes \hat{T}_{1 e}$
with the kinetic-energy $\hat{T}_{1 e}$ and the identity $\hat{I}_{1 e}$ operators for a single electron,
the matrix element is
\begin{align}
    \langle k_0, k_1 |_{2 n_{q e}}
    \hat{T}
    | k_0', k_1' \rangle_{2 n_{q e}}
    =
        \langle k_0 |_{n_{q e}}
        \hat{T}_{1 e} 
        | k_0' \rangle_{n_{q e}}
        \delta_{k_1 k_1'}
        +
        \delta_{k_0 k_0'}
        \langle k_1 |_{n_{q e}}
        \hat{T}_{1 e} 
        | k_1' \rangle_{n_{q e}}
        ,
    \label{details_of_LiH:mat_elem_kin}
\end{align}
where the matrix element for a single electron is calculated explicitly as
[see Appendix C in Ref.~\cite{bib:5737}]
\begin{align}
    \langle k |_{n_e} \hat{T}_{1 e} | k' \rangle_{n_e}
    =
        \frac{e^{-i \pi (k - k')}}{N_{q e}}
        \sum_{s = 0}^{N_{q e} - 1}
            E_s
            \exp \frac{2 \pi i (k - k') s}{N_{q e}}
        .
\end{align}
$E_s \equiv (s - N_{q e}/2)^2 (\Delta p)^2/2$
is the discretized kinetic energy with the momentum step
$\Delta p \equiv 2 \pi/L.$
The matrix element for the electron-nucleus interactions is
\begin{align}
    \langle k_0, k_1 |_{2 n_{q e}}
    \hat{V}_{e \mathrm{n}}
    | k_0', k_1' \rangle_{2 n_{q e}}
    =
        \left(
            v_{e \mathrm{n}} (x^{(k_0)})
            +
            v_{e \mathrm{n}} (x^{(k_1)})
        \right)
        \delta_{k_0 k_0'}
        \delta_{k_1 k_1'}
    \label{details_of_LiH:mat_elem_V_en}
\end{align}
for the potential $v_{e \mathrm{n}}$ felt by each electron.
That for the electron-electron interaction is
\begin{align}
    \langle k_0, k_1 |_{2 n_{q e}}
    \hat{V}_{ee}
    | k_0', k_1' \rangle_{2 n_{q e}}
    =
        v_{e e} (| x^{(k_0)} - x^{(k_1)} |)
        \delta_{k_0 k_0'}
        \delta_{k_1 k_1'}
        .
    \label{details_of_LiH:mat_elem_V_ee}
\end{align}
By gathering
Eqs.~(\ref{details_of_LiH:mat_elem_kin}),
(\ref{details_of_LiH:mat_elem_V_en}), and
(\ref{details_of_LiH:mat_elem_V_ee}) and
the repulsion energy between the Li and H ions,
we can construct the Hamiltonian matrix for the fixed geometry.
All the energy eigenstates are obtained via numerical diagonalization of it.

\subsubsection{Geometry optimization based on PITE}

As explained in the main text,
we adopted $n_{q \mathrm{n}} = 3$ for 8 candidate geometries.
Although we have to treat the multiple geometries,
the Hamiltonian matrix in the position representation for the electrons and nuclei takes the block diagonal form as
\begin{align}
    \mathcal{H}
    =
    \begin{pmatrix}
        \mathcal{H}_0 & \\
        {}            & \ddots \\ 
        {}            & {}     & \mathcal{H}_7
    \end{pmatrix}
    ,
\end{align}
where the submatrix $\mathcal{H}_J$ of dimension 4096 is for 
the $J$th candidate geometry.
This form allowed us to treat the nuclear subspaces one by one. 
We obtained the exact energy eigenstates for the candidate geometries via numerical diagonalization of the submatrices for the comparison with the optimized states via the simulated PITE steps.

Our simulation was performed by tracking the trial states undergoing the PITE steps in
Fig.~\ref{fig:circuit_for_opt_using_pite}(c).
We prepared the reference states in
Eqs.~(\ref{init_wave_func_symm}) and (\ref{init_wave_func_antisymm})
simply by setting the state vectors to them
because our optimization scheme does not assume specific implementation of QSP.
Since the total Hamiltonian is block diagonal,
the PITE steps can be simulated via matrix-vector multiplication for
the eight 4096-component vectors,
each of which represents the trial electronic state for one of the candidates.
The kinetic-evolution operator for the two electrons is written as
$
e^{-i \hat{T} \Delta t}
=
( e^{-i \hat{T}_{1 e} \Delta t} \otimes \hat{I}_{1 e} )
( \hat{I}_{1 e} \otimes e^{-i \hat{T}_{1 e} \Delta t})
.
$
It is thus simulated by multiplying the state vector by the two matrices,
each of which is a Kronecker product of two 64-dimensional matrices.
The matrix element of $e^{-i \hat{T}_{1 e} \Delta t}$
is calculated explicitly as [see Appendix C in Ref.~\cite{bib:5737}]
\begin{gather}
    \langle k |
    e^{-i \hat{T}_{1 e} \Delta t}
    | k'  \rangle_n
    =
        \frac{e^{-i \pi (k - k')}}{N_{q e}}        
        \sum_{s = 0}^{N_{q e} - 1}
            \exp
            \left(
                -i
                E_s
                \Delta t
                +
                \frac{2 \pi i (k - k') s}{N_{q e}}
            \right)
        .
\end{gather}
As for the position-dependent evolution operators
$
e^{ -i \hat{V}_{e \mathrm{n}} \Delta t},
e^{ -i \hat{V}_{e e} \Delta t},
$
and
$e^{ -i \hat{V}_{\mathrm{n n}} \Delta t},$
they are simulated straightforwardly since they act diagonally for the position representation.

\subsubsection{Optimization starting from a nonuniform weight distribution}

In addition to the geometry optimization using the uniform initial weight distribution provided in the main text,
we performed that using a nonuniform distribution.
The linear-depth circuit proposed by Klco and Savage \cite{bib:5645} encodes an exponential function.
We adopted it as $U_{\mathrm{guess}}$ for the three-qubit nuclear register responsible for the bond length between the two nuclei,
as shown in Fig.~\ref{fig:LiH_opt_gs_unif_and_exp_weights}(a),
where $\theta_\ell \equiv \arctan \exp (2^\ell \alpha) \ (\ell = 0, 1)$ are the angle parameters for an exponent $\alpha.$
It is easily confirmed that this circuit transforms the initialized register to
$\sum_{J = 0}^7 \sqrt{w_{0 J}} | J \rangle_3,$ where 
\begin{align}
    w_{0 J}
    =
        \frac{1}{2 (1 + \lambda + \lambda^2 + \lambda^3)}
        \cdot
        \begin{cases}
            \lambda^{J} & J < 4 \\
            \lambda^{7 - J} & J \geq 4
        \end{cases}
\end{align}
with $\lambda \equiv e^{2 \alpha}.$
This circuit allows us to start the optimization from the nonuniform distribution symmetric around $J = 3.5$ for the weights.
We tried the candidate geometries specified by the bond lengths 
$d_J = 0.05 + 0.5 J \ (J = 0, \dots, 7).$
We used the same imaginary-time steps $\Delta \tau$ as in the main text.
Figure \ref{fig:LiH_opt_gs_unif_and_exp_weights}(b) and (c)
show the weight of each geometry during the steps starting from the uniform and nonuniform $(\lambda = 2)$ distributions of weights, respectively.
The two initial peaks at $J = 3$ and $4$ in the exponential distribution had the same value,
whose relative heights changed undergoing the steps.
When the nine steps were done,
the weight of the optimal geometry $(J = 3)$ was found to be larger than that for $J = 4,$ as seen in the right panel of
Fig.~\ref{fig:LiH_opt_gs_unif_and_exp_weights}(c),
by an amount larger than that for the uniform initial distribution in the right panel of
Fig.~\ref{fig:LiH_opt_gs_unif_and_exp_weights}(b).
The overall shapes of weight distributions have become asymmetric after the steps in both cases,
reflecting the asymmetric energy curve.

\begin{figure}
\begin{center}
\includegraphics[width=13cm]{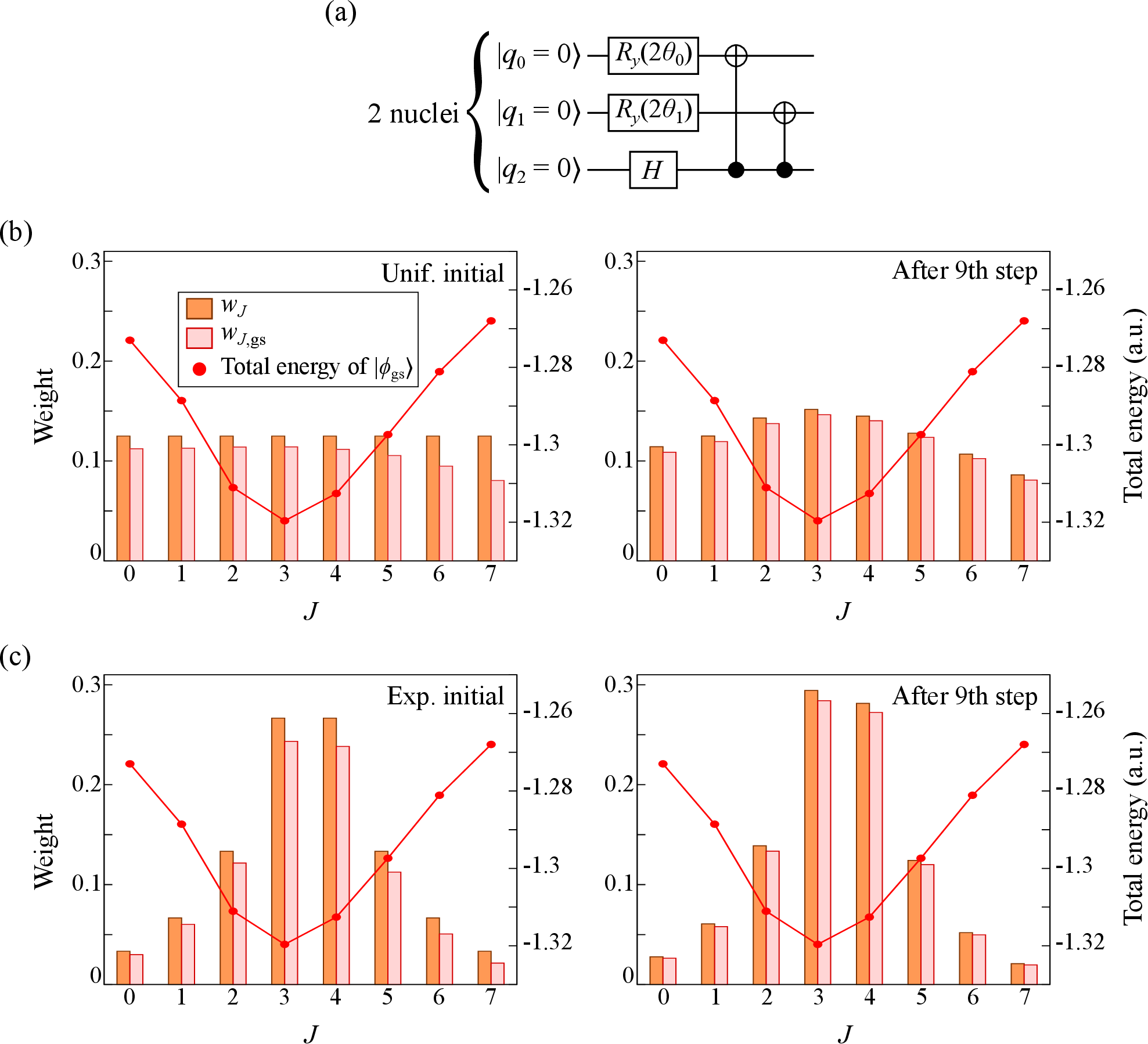}
\end{center}
\caption{
(a)
$U_{\mathrm{guess}}$ for generating a nonuniform distribution of the initial weights of
$| J \rangle_3 = | 2^2 q_2 + 2^1 q_1 + 2^0 q_0 \rangle_3.$
$R_y (\theta) \equiv e^{-i \theta \sigma_y/2}$ is the single-qubit $y$ rotation.
The angle parameters $\theta_\ell$ are calculated from $\alpha$ for the exponential function.
(b) and (c) are the results of geometry optimization for the LiH molecule starting from the uniform and nonuniform distributions of weights, respectively. 
}
\label{fig:LiH_opt_gs_unif_and_exp_weights}
\end{figure}

\subsection{VITE and PITE for a model H$_2^+$ molecule}
\label{sec:details_of_H2p}

In the VITE-based optimization scheme,
the initial weights of candidates have random values to be optimized as a part of the ansatz, as seen in
Fig.~\ref{fig:circuit_for_opt_of_H2_using_vite}.
To see typical histograms of the geometry weights for the H$_2^+$ molecule,
we show those during the steps in
Fig.~\ref{fig:H2_opt_weights_vite_and_pite}(a).
The initial distribution has only a tiny weight for the ground state in the optimal geometry ($J = 2$).
The distribution after the 200th steps already has a significant weight of the optimal state.
It is, however, smaller than that of $J = 3.$
The distribution after the 800th steps has prominent peaks at $J = 2$ and $3,$
while the other candidates now look hopeless.

We also performed PITE-based optimization for the same system.
We provide its results here.
We adopted the same numbers of qubits for the electronic and nuclear registers as in the VITE case.
To obtain the ground states for each candidate geometry,
we used the initial spatial function
$\Psi (x) \propto \exp(-(x - X_{\mathrm{m}})^2/w^2)$
centered at the midpoint
$X_{\mathrm{m}} \equiv (X_{\mathrm{H} \alpha} + X_{\mathrm{H} \beta})/2$
of the H ions.
$w \equiv 3$ is its width.
We adopted the same scheduling of steps $\Delta \tau$ as for the LiH molecule (see the main text).
Figure \ref{fig:H2_opt_weights_vite_and_pite}(b)
shows the weight of each geometry during the steps starting from a uniform distribution of weights.
The peak indicating the optimal geometry is already detectable after the fourth step.
The direct comparison of the numbers of steps for finding the optimal geometry between the PITE- and VITE-based schemes is,
however, not fair since the former used $U_{\mathrm{ref}}$ for the plausible ground states, while the latter started from the completely random state.

\begin{figure}
\begin{center}
\includegraphics[width=13cm]{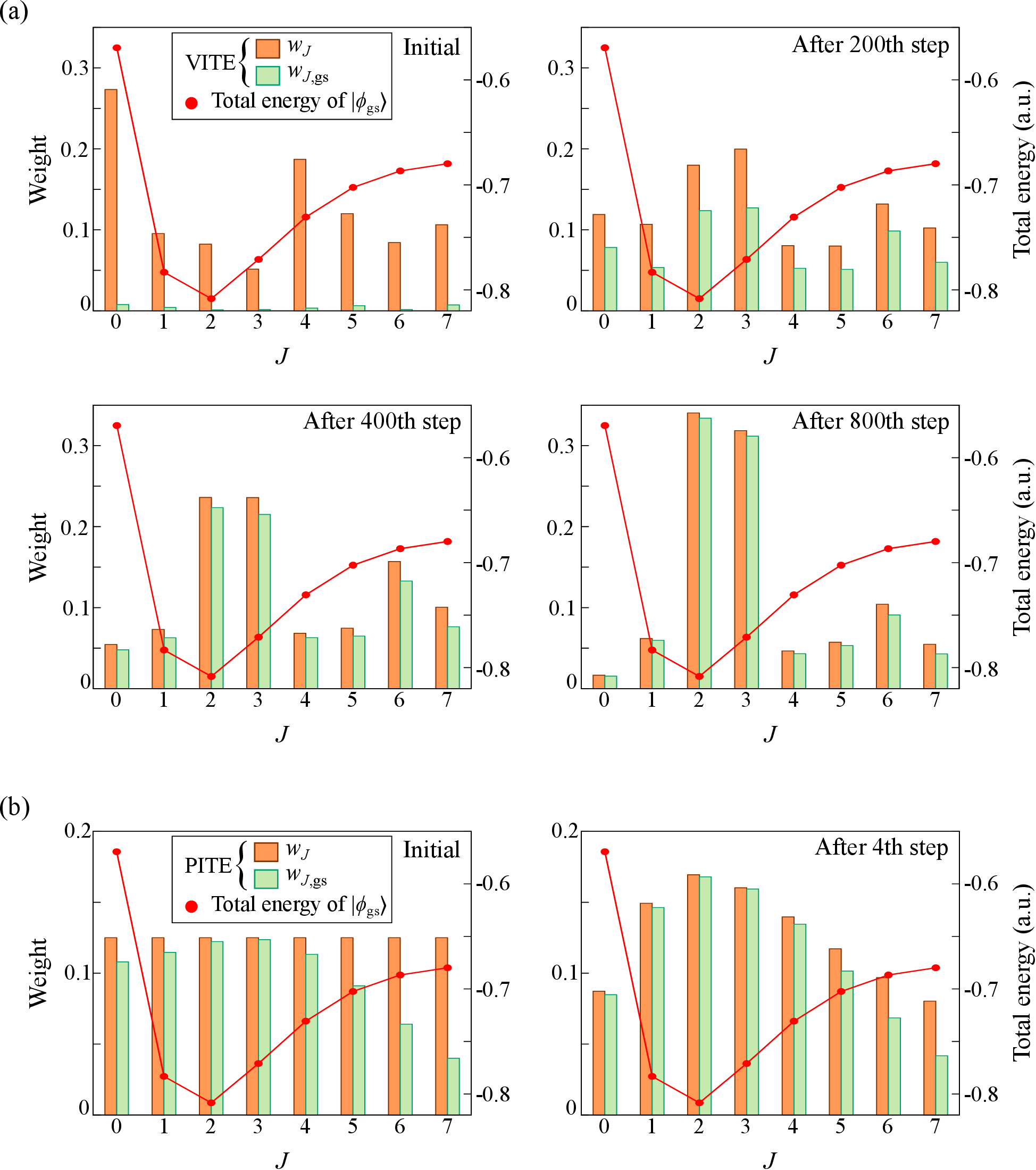}
\end{center}
\caption{
(a)
Results of VITE-based geometry optimization for the H$_2^+$ molecule starting from a random distribution of the weights of the candidates $J.$
(b)
Those of PITE-based geometry optimization starting from a uniform weight distribution.
}
\label{fig:H2_opt_weights_vite_and_pite}
\end{figure}

\subsection{PITE for a classical C$_6$H$_6$-Ar system}
\label{sec:details_of_C6H6_Ar}

\subsubsection{Potential parameters}

The interaction energy between a hydrocarbon molecule and a rare-gas atom $a$ based on the ILJ potentials \cite{bib:5834} is calculated from the additive pairwise contributions:
$E_{\mathrm{int}} = \sum_{b} V^{(\mathrm{ILJ})}_{a b},$
where $b$ runs over the bonds forming the molecule.
The expression for each atom-bond pair is given by
\begin{gather}
    V^{(\mathrm{ILJ})}_{a b}
    =
        \frac{D}{n (s) - m}
        \left(
            \frac{m}{s^{n (s)}}
            -
            \frac{n (s)}{s^m}
        \right)
    .
\end{gather}
$
D
\equiv
D_{a b \perp} \sin^2 \theta_{a b}
+
D_{a b \parallel} \cos^2 \theta_{a b}
$
and
$
\lambda_{a b}
\equiv
\lambda_{a b \perp} \sin^2 \theta_{a b}
+
\lambda_{a b \parallel} \cos^2 \theta_{a b}
$
have been introduced for taking into account
the anisotropicity of relative position of $a$ and $b$ with the Jacobi angle $\theta_{a b}$ between the bond vector and the vector connecting the bond center $\boldsymbol{R}_b$ and the atom $a$ at $\boldsymbol{R}_a.$  
$s \equiv |\boldsymbol{R}_a - \boldsymbol{R}_b|/\lambda_{a b}$
is the dimensionless argument for the radial potential.
$D_{a b \parallel}$ and $\lambda_{a b \parallel}$
($D_{a b \perp}$ and $\lambda_{a b \perp}$)
are the depth of potential well and its location, respectively,
when $\theta = 0$ ($\theta = \pi/2$).
We adopted $\beta = 10$ for $n (s) \equiv \beta + 4 s^2$
and $m = 6$ for the C$_6$H$_6$-Ar system as in the original paper.
The parameters for bond-atom pairs for this system are as follows \cite{bib:5834}:
$\lambda_{\mathrm{Ar, CC} \perp} = 3.879$ \AA,
$\lambda_{\mathrm{Ar, CC} \parallel} = 4.189$ \AA,
$D_{\mathrm{Ar, CC} \perp} = 3.895$ meV,
$D_{\mathrm{Ar, CC} \parallel} = 4.910$ meV,
$\lambda_{\mathrm{Ar, CH} \perp} = 3.641$ \AA,
$\lambda_{\mathrm{Ar, CH} \parallel} = 3.851$ \AA,
$D_{\mathrm{Ar, CH} \perp} = 4.814$ meV, and
$D_{\mathrm{Ar, CH} \parallel} = 3.981$ meV.

\subsubsection{Geometry optimization based on PITE}

As explained in the main text,
we adopted $n_{q \mathrm{n}} = 3$ in the $x$ and $z$ directions for 64 candidate geometries.
The electronic degrees of freedom are absent in this case and our simulation was thus performed by tracking the trial states undergoing the PITE steps in
Fig.~\ref{fig:pite_steps_for_classical_opt}.
Specifically,
the action of the evolution operator
$
e^{-i \hat{V}_{\mathrm{nn}} \Delta t}
=
e^{-i E_{\mathrm{int}} \Delta t}
$
at each step was simulated by letting the 64 candidate geometries acquire the phase factors,
that are constant throughout the iterations.

\end{widetext}

\bibliography{ref}

\begin{thebibliography}{64}%
\makeatletter
\providecommand \@ifxundefined [1]{%
 \@ifx{#1\undefined}
}%
\providecommand \@ifnum [1]{%
 \ifnum #1\expandafter \@firstoftwo
 \else \expandafter \@secondoftwo
 \fi
}%
\providecommand \@ifx [1]{%
 \ifx #1\expandafter \@firstoftwo
 \else \expandafter \@secondoftwo
 \fi
}%
\providecommand \natexlab [1]{#1}%
\providecommand \enquote  [1]{``#1''}%
\providecommand \bibnamefont  [1]{#1}%
\providecommand \bibfnamefont [1]{#1}%
\providecommand \citenamefont [1]{#1}%
\providecommand \href@noop [0]{\@secondoftwo}%
\providecommand \href [0]{\begingroup \@sanitize@url \@href}%
\providecommand \@href[1]{\@@startlink{#1}\@@href}%
\providecommand \@@href[1]{\endgroup#1\@@endlink}%
\providecommand \@sanitize@url [0]{\catcode `\\12\catcode `\$12\catcode
  `\&12\catcode `\#12\catcode `\^12\catcode `\_12\catcode `\%12\relax}%
\providecommand \@@startlink[1]{}%
\providecommand \@@endlink[0]{}%
\providecommand \url  [0]{\begingroup\@sanitize@url \@url }%
\providecommand \@url [1]{\endgroup\@href {#1}{\urlprefix }}%
\providecommand \urlprefix  [0]{URL }%
\providecommand \Eprint [0]{\href }%
\providecommand \doibase [0]{https://doi.org/}%
\providecommand \selectlanguage [0]{\@gobble}%
\providecommand \bibinfo  [0]{\@secondoftwo}%
\providecommand \bibfield  [0]{\@secondoftwo}%
\providecommand \translation [1]{[#1]}%
\providecommand \BibitemOpen [0]{}%
\providecommand \bibitemStop [0]{}%
\providecommand \bibitemNoStop [0]{.\EOS\space}%
\providecommand \EOS [0]{\spacefactor3000\relax}%
\providecommand \BibitemShut  [1]{\csname bibitem#1\endcsname}%
\let\auto@bib@innerbib\@empty
%</preamble>
\bibitem [{\citenamefont {Axelrod}\ \emph {et~al.}(2022)\citenamefont
  {Axelrod}, \citenamefont {Schwalbe-Koda}, \citenamefont {Mohapatra},
  \citenamefont {Damewood}, \citenamefont {Greenman},\ and\ \citenamefont
  {G{\'o}mez-Bombarelli}}]{bib:5744}%
  \BibitemOpen
  \bibfield  {author} {\bibinfo {author} {\bibfnamefont {S.}~\bibnamefont
  {Axelrod}}, \bibinfo {author} {\bibfnamefont {D.}~\bibnamefont
  {Schwalbe-Koda}}, \bibinfo {author} {\bibfnamefont {S.}~\bibnamefont
  {Mohapatra}}, \bibinfo {author} {\bibfnamefont {J.}~\bibnamefont {Damewood}},
  \bibinfo {author} {\bibfnamefont {K.~P.}\ \bibnamefont {Greenman}},\ and\
  \bibinfo {author} {\bibfnamefont {R.}~\bibnamefont {G{\'o}mez-Bombarelli}},\
  }\bibfield  {title} {\bibinfo {title} {Learning matter: Materials design with
  machine learning and atomistic simulations},\ }\href
  {https://doi.org/10.1021/accountsmr.1c00238} {\bibfield  {journal} {\bibinfo
  {journal} {Accounts of Materials Research}\ }\textbf {\bibinfo {volume}
  {3}},\ \bibinfo {pages} {343} (\bibinfo {year} {2022})}\BibitemShut {NoStop}%
\bibitem [{\citenamefont {Pereira}\ \emph {et~al.}(2021)\citenamefont
  {Pereira}, \citenamefont {Vieira},\ and\ \citenamefont {Santos}}]{bib:5745}%
  \BibitemOpen
  \bibfield  {author} {\bibinfo {author} {\bibfnamefont {J.~M.}\ \bibnamefont
  {Pereira}}, \bibinfo {author} {\bibfnamefont {M.}~\bibnamefont {Vieira}},\
  and\ \bibinfo {author} {\bibfnamefont {S.~M.}\ \bibnamefont {Santos}},\
  }\bibfield  {title} {\bibinfo {title} {Step-by-step design of proteins for
  small molecule interaction: A review on recent milestones},\ }\href
  {https://doi.org/https://doi.org/10.1002/pro.4098} {\bibfield  {journal}
  {\bibinfo  {journal} {Protein Science}\ }\textbf {\bibinfo {volume} {30}},\
  \bibinfo {pages} {1502} (\bibinfo {year} {2021})},\ \Eprint
  {https://arxiv.org/abs/https://onlinelibrary.wiley.com/doi/pdf/10.1002/pro.4098}
  {https://onlinelibrary.wiley.com/doi/pdf/10.1002/pro.4098} \BibitemShut
  {NoStop}%
\bibitem [{\citenamefont {Pandey}\ \emph {et~al.}(2022)\citenamefont {Pandey},
  \citenamefont {Fernandez}, \citenamefont {Gentile}, \citenamefont {Isayev},
  \citenamefont {Tropsha}, \citenamefont {Stern},\ and\ \citenamefont
  {Cherkasov}}]{bib:5747}%
  \BibitemOpen
  \bibfield  {author} {\bibinfo {author} {\bibfnamefont {M.}~\bibnamefont
  {Pandey}}, \bibinfo {author} {\bibfnamefont {M.}~\bibnamefont {Fernandez}},
  \bibinfo {author} {\bibfnamefont {F.}~\bibnamefont {Gentile}}, \bibinfo
  {author} {\bibfnamefont {O.}~\bibnamefont {Isayev}}, \bibinfo {author}
  {\bibfnamefont {A.}~\bibnamefont {Tropsha}}, \bibinfo {author} {\bibfnamefont
  {A.~C.}\ \bibnamefont {Stern}},\ and\ \bibinfo {author} {\bibfnamefont
  {A.}~\bibnamefont {Cherkasov}},\ }\bibfield  {title} {\bibinfo {title} {The
  transformational role of gpu computing and deep learning in drug discovery},\
  }\href {https://doi.org/10.1038/s42256-022-00463-x} {\bibfield  {journal}
  {\bibinfo  {journal} {Nature Machine Intelligence}\ }\textbf {\bibinfo
  {volume} {4}},\ \bibinfo {pages} {211} (\bibinfo {year} {2022})}\BibitemShut
  {NoStop}%
\bibitem [{\citenamefont {Hohenberg}\ and\ \citenamefont
  {Kohn}(1964)}]{bib:76}%
  \BibitemOpen
  \bibfield  {author} {\bibinfo {author} {\bibfnamefont {P.}~\bibnamefont
  {Hohenberg}}\ and\ \bibinfo {author} {\bibfnamefont {W.}~\bibnamefont
  {Kohn}},\ }\bibfield  {title} {\bibinfo {title} {Inhomogeneous electron
  gas},\ }\href {https://doi.org/10.1103/PhysRev.136.B864} {\bibfield
  {journal} {\bibinfo  {journal} {Phys. Rev.}\ }\textbf {\bibinfo {volume}
  {136}},\ \bibinfo {pages} {B864} (\bibinfo {year} {1964})}\BibitemShut
  {NoStop}%
\bibitem [{\citenamefont {Kohn}\ and\ \citenamefont {Sham}(1965)}]{bib:77}%
  \BibitemOpen
  \bibfield  {author} {\bibinfo {author} {\bibfnamefont {W.}~\bibnamefont
  {Kohn}}\ and\ \bibinfo {author} {\bibfnamefont {L.~J.}\ \bibnamefont
  {Sham}},\ }\bibfield  {title} {\bibinfo {title} {Self-consistent equations
  including exchange and correlation effects},\ }\href
  {https://doi.org/10.1103/PhysRev.140.A1133} {\bibfield  {journal} {\bibinfo
  {journal} {Phys. Rev.}\ }\textbf {\bibinfo {volume} {140}},\ \bibinfo {pages}
  {A1133} (\bibinfo {year} {1965})}\BibitemShut {NoStop}%
\bibitem [{\citenamefont {Helgaker}\ \emph {et~al.}(2000)\citenamefont
  {Helgaker}, \citenamefont {J{\o}rgensen},\ and\ \citenamefont
  {Olsen}}]{Helgaker}%
  \BibitemOpen
  \bibfield  {author} {\bibinfo {author} {\bibfnamefont {T.}~\bibnamefont
  {Helgaker}}, \bibinfo {author} {\bibfnamefont {P.}~\bibnamefont
  {J{\o}rgensen}},\ and\ \bibinfo {author} {\bibfnamefont {J.}~\bibnamefont
  {Olsen}},\ }\href@noop {} {\emph {\bibinfo {title} {Molecular
  Electronic-Structure Theory}}}\ (\bibinfo  {publisher} {Wiley},\ \bibinfo
  {year} {2000})\BibitemShut {NoStop}%
\bibitem [{\citenamefont {Feynman}(1939)}]{bib:32}%
  \BibitemOpen
  \bibfield  {author} {\bibinfo {author} {\bibfnamefont {R.~P.}\ \bibnamefont
  {Feynman}},\ }\bibfield  {title} {\bibinfo {title} {Forces in molecules},\
  }\href {https://doi.org/10.1103/PhysRev.56.340} {\bibfield  {journal}
  {\bibinfo  {journal} {Phys. Rev.}\ }\textbf {\bibinfo {volume} {56}},\
  \bibinfo {pages} {340} (\bibinfo {year} {1939})}\BibitemShut {NoStop}%
\bibitem [{\citenamefont {Pulay}(1969)}]{bib:5743}%
  \BibitemOpen
  \bibfield  {author} {\bibinfo {author} {\bibfnamefont {P.}~\bibnamefont
  {Pulay}},\ }\bibfield  {title} {\bibinfo {title} {Ab initio calculation of
  force constants and equilibrium geometries in polyatomic molecules},\ }\href
  {https://doi.org/10.1080/00268976900100941} {\bibfield  {journal} {\bibinfo
  {journal} {Molecular Physics}\ }\textbf {\bibinfo {volume} {17}},\ \bibinfo
  {pages} {197} (\bibinfo {year} {1969})},\ \Eprint
  {https://arxiv.org/abs/https://doi.org/10.1080/00268976900100941}
  {https://doi.org/10.1080/00268976900100941} \BibitemShut {NoStop}%
\bibitem [{\citenamefont {Schlegel}(2011)}]{bib:5742}%
  \BibitemOpen
  \bibfield  {author} {\bibinfo {author} {\bibfnamefont {H.~B.}\ \bibnamefont
  {Schlegel}},\ }\bibfield  {title} {\bibinfo {title} {Geometry optimization},\
  }\href {https://doi.org/https://doi.org/10.1002/wcms.34} {\bibfield
  {journal} {\bibinfo  {journal} {WIREs Computational Molecular Science}\
  }\textbf {\bibinfo {volume} {1}},\ \bibinfo {pages} {790} (\bibinfo {year}
  {2011})},\ \Eprint
  {https://arxiv.org/abs/https://wires.onlinelibrary.wiley.com/doi/pdf/10.1002/wcms.34}
  {https://wires.onlinelibrary.wiley.com/doi/pdf/10.1002/wcms.34} \BibitemShut
  {NoStop}%
\bibitem [{\citenamefont {Feynman}(1982)}]{bib:4828}%
  \BibitemOpen
  \bibfield  {author} {\bibinfo {author} {\bibfnamefont {R.~P.}\ \bibnamefont
  {Feynman}},\ }\bibfield  {title} {\bibinfo {title} {Simulating physics with
  computers},\ }\href {https://doi.org/10.1007/BF02650179} {\bibfield
  {journal} {\bibinfo  {journal} {International Journal of Theoretical
  Physics}\ }\textbf {\bibinfo {volume} {21}},\ \bibinfo {pages} {467}
  (\bibinfo {year} {1982})}\BibitemShut {NoStop}%
\bibitem [{\citenamefont {Hirai}\ \emph {et~al.}(2022)\citenamefont {Hirai},
  \citenamefont {Horiba}, \citenamefont {Shirai}, \citenamefont {Kanno},
  \citenamefont {Omiya}, \citenamefont {Nakagawa},\ and\ \citenamefont
  {Koh}}]{bib:5752}%
  \BibitemOpen
  \bibfield  {author} {\bibinfo {author} {\bibfnamefont {H.}~\bibnamefont
  {Hirai}}, \bibinfo {author} {\bibfnamefont {T.}~\bibnamefont {Horiba}},
  \bibinfo {author} {\bibfnamefont {S.}~\bibnamefont {Shirai}}, \bibinfo
  {author} {\bibfnamefont {K.}~\bibnamefont {Kanno}}, \bibinfo {author}
  {\bibfnamefont {K.}~\bibnamefont {Omiya}}, \bibinfo {author} {\bibfnamefont
  {Y.~O.}\ \bibnamefont {Nakagawa}},\ and\ \bibinfo {author} {\bibfnamefont
  {S.}~\bibnamefont {Koh}},\ }\bibfield  {title} {\bibinfo {title} {Molecular
  structure optimization based on electrons–nuclei quantum dynamics
  computation},\ }\href {https://doi.org/10.1021/acsomega.2c01546} {\bibfield
  {journal} {\bibinfo  {journal} {ACS Omega}\ }\textbf {\bibinfo {volume}
  {7}},\ \bibinfo {pages} {19784} (\bibinfo {year} {2022})},\ \Eprint
  {https://arxiv.org/abs/https://doi.org/10.1021/acsomega.2c01546}
  {https://doi.org/10.1021/acsomega.2c01546} \BibitemShut {NoStop}%
\bibitem [{\citenamefont {{Wiesner}}(1996)}]{bib:5373}%
  \BibitemOpen
  \bibfield  {author} {\bibinfo {author} {\bibfnamefont {S.}~\bibnamefont
  {{Wiesner}}},\ }\bibfield  {title} {\bibinfo {title} {{Simulations of
  Many-Body Quantum Systems by a Quantum Computer}},\ }\href@noop {} {\bibfield
   {journal} {\bibinfo  {journal} {arXiv e-prints}\ ,\ \bibinfo {eid}
  {quant-ph/9603028}} (\bibinfo {year} {1996})},\ \Eprint
  {https://arxiv.org/abs/quant-ph/9603028} {arXiv:quant-ph/9603028 [quant-ph]}
  \BibitemShut {NoStop}%
\bibitem [{\citenamefont {Zalka}(1998)}]{bib:5372}%
  \BibitemOpen
  \bibfield  {author} {\bibinfo {author} {\bibfnamefont {C.}~\bibnamefont
  {Zalka}},\ }\bibfield  {title} {\bibinfo {title} {Simulating quantum systems
  on a quantum computer},\ }\href {https://doi.org/10.1098/rspa.1998.0162}
  {\bibfield  {journal} {\bibinfo  {journal} {Proceedings of the Royal Society
  of London. Series A: Mathematical, Physical and Engineering Sciences}\
  }\textbf {\bibinfo {volume} {454}},\ \bibinfo {pages} {313} (\bibinfo {year}
  {1998})},\ \Eprint
  {https://arxiv.org/abs/https://royalsocietypublishing.org/doi/pdf/10.1098/rspa.1998.0162}
  {https://royalsocietypublishing.org/doi/pdf/10.1098/rspa.1998.0162}
  \BibitemShut {NoStop}%
\bibitem [{\citenamefont {Kassal}\ \emph {et~al.}(2008)\citenamefont {Kassal},
  \citenamefont {Jordan}, \citenamefont {Love}, \citenamefont {Mohseni},\ and\
  \citenamefont {Aspuru-Guzik}}]{bib:5328}%
  \BibitemOpen
  \bibfield  {author} {\bibinfo {author} {\bibfnamefont {I.}~\bibnamefont
  {Kassal}}, \bibinfo {author} {\bibfnamefont {S.~P.}\ \bibnamefont {Jordan}},
  \bibinfo {author} {\bibfnamefont {P.~J.}\ \bibnamefont {Love}}, \bibinfo
  {author} {\bibfnamefont {M.}~\bibnamefont {Mohseni}},\ and\ \bibinfo {author}
  {\bibfnamefont {A.}~\bibnamefont {Aspuru-Guzik}},\ }\bibfield  {title}
  {\bibinfo {title} {Polynomial-time quantum algorithm for the simulation of
  chemical dynamics},\ }\href {https://doi.org/10.1073/pnas.0808245105}
  {\bibfield  {journal} {\bibinfo  {journal} {Proceedings of the National
  Academy of Sciences}\ }\textbf {\bibinfo {volume} {105}},\ \bibinfo {pages}
  {18681} (\bibinfo {year} {2008})},\ \Eprint
  {https://arxiv.org/abs/https://www.pnas.org/content/105/48/18681.full.pdf}
  {https://www.pnas.org/content/105/48/18681.full.pdf} \BibitemShut {NoStop}%
\bibitem [{\citenamefont {Jones}\ \emph {et~al.}(2019)\citenamefont {Jones},
  \citenamefont {Endo}, \citenamefont {McArdle}, \citenamefont {Yuan},\ and\
  \citenamefont {Benjamin}}]{bib:4797}%
  \BibitemOpen
  \bibfield  {author} {\bibinfo {author} {\bibfnamefont {T.}~\bibnamefont
  {Jones}}, \bibinfo {author} {\bibfnamefont {S.}~\bibnamefont {Endo}},
  \bibinfo {author} {\bibfnamefont {S.}~\bibnamefont {McArdle}}, \bibinfo
  {author} {\bibfnamefont {X.}~\bibnamefont {Yuan}},\ and\ \bibinfo {author}
  {\bibfnamefont {S.~C.}\ \bibnamefont {Benjamin}},\ }\bibfield  {title}
  {\bibinfo {title} {Variational quantum algorithms for discovering hamiltonian
  spectra},\ }\href {https://doi.org/10.1103/PhysRevA.99.062304} {\bibfield
  {journal} {\bibinfo  {journal} {Phys. Rev. A}\ }\textbf {\bibinfo {volume}
  {99}},\ \bibinfo {pages} {062304} (\bibinfo {year} {2019})}\BibitemShut
  {NoStop}%
\bibitem [{\citenamefont {McArdle}\ \emph {et~al.}(2019)\citenamefont
  {McArdle}, \citenamefont {Jones}, \citenamefont {Endo}, \citenamefont {Li},
  \citenamefont {Benjamin},\ and\ \citenamefont {Yuan}}]{bib:4802}%
  \BibitemOpen
  \bibfield  {author} {\bibinfo {author} {\bibfnamefont {S.}~\bibnamefont
  {McArdle}}, \bibinfo {author} {\bibfnamefont {T.}~\bibnamefont {Jones}},
  \bibinfo {author} {\bibfnamefont {S.}~\bibnamefont {Endo}}, \bibinfo {author}
  {\bibfnamefont {Y.}~\bibnamefont {Li}}, \bibinfo {author} {\bibfnamefont
  {S.~C.}\ \bibnamefont {Benjamin}},\ and\ \bibinfo {author} {\bibfnamefont
  {X.}~\bibnamefont {Yuan}},\ }\bibfield  {title} {\bibinfo {title}
  {Variational ansatz-based quantum simulation of imaginary time evolution},\
  }\href {https://doi.org/10.1038/s41534-019-0187-2} {\bibfield  {journal}
  {\bibinfo  {journal} {npj Quantum Information}\ }\textbf {\bibinfo {volume}
  {5}},\ \bibinfo {pages} {75} (\bibinfo {year} {2019})}\BibitemShut {NoStop}%
\bibitem [{\citenamefont {Yuan}\ \emph {et~al.}(2019)\citenamefont {Yuan},
  \citenamefont {Endo}, \citenamefont {Zhao}, \citenamefont {Li},\ and\
  \citenamefont {Benjamin}}]{bib:4807}%
  \BibitemOpen
  \bibfield  {author} {\bibinfo {author} {\bibfnamefont {X.}~\bibnamefont
  {Yuan}}, \bibinfo {author} {\bibfnamefont {S.}~\bibnamefont {Endo}}, \bibinfo
  {author} {\bibfnamefont {Q.}~\bibnamefont {Zhao}}, \bibinfo {author}
  {\bibfnamefont {Y.}~\bibnamefont {Li}},\ and\ \bibinfo {author}
  {\bibfnamefont {S.~C.}\ \bibnamefont {Benjamin}},\ }\bibfield  {title}
  {\bibinfo {title} {Theory of variational quantum simulation},\ }\href
  {https://doi.org/10.22331/q-2019-10-07-191} {\bibfield  {journal} {\bibinfo
  {journal} {{Quantum}}\ }\textbf {\bibinfo {volume} {3}},\ \bibinfo {pages}
  {191} (\bibinfo {year} {2019})}\BibitemShut {NoStop}%
\bibitem [{\citenamefont {Peruzzo}\ \emph {et~al.}(2014)\citenamefont
  {Peruzzo}, \citenamefont {McClean}, \citenamefont {Shadbolt}, \citenamefont
  {Yung}, \citenamefont {Zhou}, \citenamefont {Love}, \citenamefont
  {Aspuru-Guzik},\ and\ \citenamefont {O'Brien}}]{bib:4470}%
  \BibitemOpen
  \bibfield  {author} {\bibinfo {author} {\bibfnamefont {A.}~\bibnamefont
  {Peruzzo}}, \bibinfo {author} {\bibfnamefont {J.}~\bibnamefont {McClean}},
  \bibinfo {author} {\bibfnamefont {P.}~\bibnamefont {Shadbolt}}, \bibinfo
  {author} {\bibfnamefont {M.-H.}\ \bibnamefont {Yung}}, \bibinfo {author}
  {\bibfnamefont {X.-Q.}\ \bibnamefont {Zhou}}, \bibinfo {author}
  {\bibfnamefont {P.~J.}\ \bibnamefont {Love}}, \bibinfo {author}
  {\bibfnamefont {A.}~\bibnamefont {Aspuru-Guzik}},\ and\ \bibinfo {author}
  {\bibfnamefont {J.~L.}\ \bibnamefont {O'Brien}},\ }\bibfield  {title}
  {\bibinfo {title} {A variational eigenvalue solver on a photonic quantum
  processor},\ }\href {https://doi.org/10.1038/ncomms5213} {\bibfield
  {journal} {\bibinfo  {journal} {Nature Communications}\ }\textbf {\bibinfo
  {volume} {5}},\ \bibinfo {pages} {4213 EP } (\bibinfo {year} {2014})},\
  \bibinfo {note} {article}\BibitemShut {NoStop}%
\bibitem [{\citenamefont {McClean}\ \emph {et~al.}(2016)\citenamefont
  {McClean}, \citenamefont {Romero}, \citenamefont {Babbush},\ and\
  \citenamefont {Aspuru-Guzik}}]{bib:4517}%
  \BibitemOpen
  \bibfield  {author} {\bibinfo {author} {\bibfnamefont {J.~R.}\ \bibnamefont
  {McClean}}, \bibinfo {author} {\bibfnamefont {J.}~\bibnamefont {Romero}},
  \bibinfo {author} {\bibfnamefont {R.}~\bibnamefont {Babbush}},\ and\ \bibinfo
  {author} {\bibfnamefont {A.}~\bibnamefont {Aspuru-Guzik}},\ }\bibfield
  {title} {\bibinfo {title} {The theory of variational hybrid quantum-classical
  algorithms},\ }\href {https://doi.org/10.1088/1367-2630/18/2/023023}
  {\bibfield  {journal} {\bibinfo  {journal} {New Journal of Physics}\ }\textbf
  {\bibinfo {volume} {18}},\ \bibinfo {pages} {023023} (\bibinfo {year}
  {2016})}\BibitemShut {NoStop}%
\bibitem [{\citenamefont {Kassal}\ and\ \citenamefont
  {Aspuru-Guzik}(2009)}]{bib:6046}%
  \BibitemOpen
  \bibfield  {author} {\bibinfo {author} {\bibfnamefont {I.}~\bibnamefont
  {Kassal}}\ and\ \bibinfo {author} {\bibfnamefont {A.}~\bibnamefont
  {Aspuru-Guzik}},\ }\bibfield  {title} {\bibinfo {title} {Quantum algorithm
  for molecular properties and geometry optimization},\ }\href
  {https://doi.org/10.1063/1.3266959} {\bibfield  {journal} {\bibinfo
  {journal} {The Journal of Chemical Physics}\ }\textbf {\bibinfo {volume}
  {131}},\ \bibinfo {pages} {224102} (\bibinfo {year} {2009})},\ \Eprint
  {https://arxiv.org/abs/https://doi.org/10.1063/1.3266959}
  {https://doi.org/10.1063/1.3266959} \BibitemShut {NoStop}%
\bibitem [{\citenamefont {Kosugi}\ \emph {et~al.}(2022)\citenamefont {Kosugi},
  \citenamefont {Nishiya}, \citenamefont {Nishi},\ and\ \citenamefont
  {Matsushita}}]{bib:5737}%
  \BibitemOpen
  \bibfield  {author} {\bibinfo {author} {\bibfnamefont {T.}~\bibnamefont
  {Kosugi}}, \bibinfo {author} {\bibfnamefont {Y.}~\bibnamefont {Nishiya}},
  \bibinfo {author} {\bibfnamefont {H.}~\bibnamefont {Nishi}},\ and\ \bibinfo
  {author} {\bibfnamefont {Y.-i.}\ \bibnamefont {Matsushita}},\ }\bibfield
  {title} {\bibinfo {title} {Imaginary-time evolution using forward and
  backward real-time evolution with a single ancilla: First-quantized
  eigensolver algorithm for quantum chemistry},\ }\href
  {https://doi.org/10.1103/PhysRevResearch.4.033121} {\bibfield  {journal}
  {\bibinfo  {journal} {Phys. Rev. Research}\ }\textbf {\bibinfo {volume}
  {4}},\ \bibinfo {pages} {033121} (\bibinfo {year} {2022})}\BibitemShut
  {NoStop}%
\bibitem [{\citenamefont {Kosugi}\ and\ \citenamefont
  {Matsushita}(2020{\natexlab{a}})}]{bib:5005}%
  \BibitemOpen
  \bibfield  {author} {\bibinfo {author} {\bibfnamefont {T.}~\bibnamefont
  {Kosugi}}\ and\ \bibinfo {author} {\bibfnamefont {Y.-i.}\ \bibnamefont
  {Matsushita}},\ }\bibfield  {title} {\bibinfo {title} {Construction of
  green's functions on a quantum computer: Quasiparticle spectra of
  molecules},\ }\href {https://doi.org/10.1103/PhysRevA.101.012330} {\bibfield
  {journal} {\bibinfo  {journal} {Phys. Rev. A}\ }\textbf {\bibinfo {volume}
  {101}},\ \bibinfo {pages} {012330} (\bibinfo {year}
  {2020}{\natexlab{a}})}\BibitemShut {NoStop}%
\bibitem [{\citenamefont {Kosugi}\ and\ \citenamefont
  {Matsushita}(2020{\natexlab{b}})}]{bib:5163}%
  \BibitemOpen
  \bibfield  {author} {\bibinfo {author} {\bibfnamefont {T.}~\bibnamefont
  {Kosugi}}\ and\ \bibinfo {author} {\bibfnamefont {Y.-i.}\ \bibnamefont
  {Matsushita}},\ }\bibfield  {title} {\bibinfo {title} {Linear-response
  functions of molecules on a quantum computer: Charge and spin responses and
  optical absorption},\ }\href
  {https://doi.org/10.1103/PhysRevResearch.2.033043} {\bibfield  {journal}
  {\bibinfo  {journal} {Phys. Rev. Research}\ }\textbf {\bibinfo {volume}
  {2}},\ \bibinfo {pages} {033043} (\bibinfo {year}
  {2020}{\natexlab{b}})}\BibitemShut {NoStop}%
\bibitem [{\citenamefont {Jones}\ \emph {et~al.}(2012)\citenamefont {Jones},
  \citenamefont {Whitfield}, \citenamefont {McMahon}, \citenamefont {Yung},
  \citenamefont {Meter}, \citenamefont {Aspuru-Guzik},\ and\ \citenamefont
  {Yamamoto}}]{bib:5824}%
  \BibitemOpen
  \bibfield  {author} {\bibinfo {author} {\bibfnamefont {N.~C.}\ \bibnamefont
  {Jones}}, \bibinfo {author} {\bibfnamefont {J.~D.}\ \bibnamefont
  {Whitfield}}, \bibinfo {author} {\bibfnamefont {P.~L.}\ \bibnamefont
  {McMahon}}, \bibinfo {author} {\bibfnamefont {M.-H.}\ \bibnamefont {Yung}},
  \bibinfo {author} {\bibfnamefont {R.~V.}\ \bibnamefont {Meter}}, \bibinfo
  {author} {\bibfnamefont {A.}~\bibnamefont {Aspuru-Guzik}},\ and\ \bibinfo
  {author} {\bibfnamefont {Y.}~\bibnamefont {Yamamoto}},\ }\bibfield  {title}
  {\bibinfo {title} {Faster quantum chemistry simulation on fault-tolerant
  quantum computers},\ }\href {https://doi.org/10.1088/1367-2630/14/11/115023}
  {\bibfield  {journal} {\bibinfo  {journal} {New Journal of Physics}\ }\textbf
  {\bibinfo {volume} {14}},\ \bibinfo {pages} {115023} (\bibinfo {year}
  {2012})}\BibitemShut {NoStop}%
\bibitem [{\citenamefont {Chan}\ \emph {et~al.}(2023)\citenamefont {Chan},
  \citenamefont {Meister}, \citenamefont {Jones}, \citenamefont {Tew},\ and\
  \citenamefont {Benjamin}}]{bib:5658}%
  \BibitemOpen
  \bibfield  {author} {\bibinfo {author} {\bibfnamefont {H.~H.~S.}\
  \bibnamefont {Chan}}, \bibinfo {author} {\bibfnamefont {R.}~\bibnamefont
  {Meister}}, \bibinfo {author} {\bibfnamefont {T.}~\bibnamefont {Jones}},
  \bibinfo {author} {\bibfnamefont {D.~P.}\ \bibnamefont {Tew}},\ and\ \bibinfo
  {author} {\bibfnamefont {S.~C.}\ \bibnamefont {Benjamin}},\ }\bibfield
  {title} {\bibinfo {title} {Grid-based methods for chemistry simulations on a
  quantum computer},\ }\href {https://doi.org/10.1126/sciadv.abo7484}
  {\bibfield  {journal} {\bibinfo  {journal} {Science Advances}\ }\textbf
  {\bibinfo {volume} {9}},\ \bibinfo {pages} {eabo7484} (\bibinfo {year}
  {2023})},\ \Eprint
  {https://arxiv.org/abs/https://www.science.org/doi/pdf/10.1126/sciadv.abo7484}
  {https://www.science.org/doi/pdf/10.1126/sciadv.abo7484} \BibitemShut
  {NoStop}%
\bibitem [{\citenamefont {Abrams}\ and\ \citenamefont
  {Lloyd}(1997)}]{bib:4825}%
  \BibitemOpen
  \bibfield  {author} {\bibinfo {author} {\bibfnamefont {D.~S.}\ \bibnamefont
  {Abrams}}\ and\ \bibinfo {author} {\bibfnamefont {S.}~\bibnamefont {Lloyd}},\
  }\bibfield  {title} {\bibinfo {title} {Simulation of many-body fermi systems
  on a universal quantum computer},\ }\href
  {https://doi.org/10.1103/PhysRevLett.79.2586} {\bibfield  {journal} {\bibinfo
   {journal} {Phys. Rev. Lett.}\ }\textbf {\bibinfo {volume} {79}},\ \bibinfo
  {pages} {2586} (\bibinfo {year} {1997})}\BibitemShut {NoStop}%
\bibitem [{\citenamefont {Berry}\ \emph {et~al.}(2018)\citenamefont {Berry},
  \citenamefont {Kieferov{\'a}}, \citenamefont {Scherer}, \citenamefont
  {Sanders}, \citenamefont {Low}, \citenamefont {Wiebe}, \citenamefont
  {Gidney},\ and\ \citenamefont {Babbush}}]{bib:5389}%
  \BibitemOpen
  \bibfield  {author} {\bibinfo {author} {\bibfnamefont {D.~W.}\ \bibnamefont
  {Berry}}, \bibinfo {author} {\bibfnamefont {M.}~\bibnamefont
  {Kieferov{\'a}}}, \bibinfo {author} {\bibfnamefont {A.}~\bibnamefont
  {Scherer}}, \bibinfo {author} {\bibfnamefont {Y.~R.}\ \bibnamefont
  {Sanders}}, \bibinfo {author} {\bibfnamefont {G.~H.}\ \bibnamefont {Low}},
  \bibinfo {author} {\bibfnamefont {N.}~\bibnamefont {Wiebe}}, \bibinfo
  {author} {\bibfnamefont {C.}~\bibnamefont {Gidney}},\ and\ \bibinfo {author}
  {\bibfnamefont {R.}~\bibnamefont {Babbush}},\ }\bibfield  {title} {\bibinfo
  {title} {Improved techniques for preparing eigenstates of fermionic
  hamiltonians},\ }\href {https://doi.org/10.1038/s41534-018-0071-5} {\bibfield
   {journal} {\bibinfo  {journal} {npj Quantum Information}\ }\textbf {\bibinfo
  {volume} {4}},\ \bibinfo {pages} {22} (\bibinfo {year} {2018})}\BibitemShut
  {NoStop}%
\bibitem [{\citenamefont {Nishi}\ \emph {et~al.}(2023)\citenamefont {Nishi},
  \citenamefont {Hamada}, \citenamefont {Nishiya}, \citenamefont {Kosugi},\
  and\ \citenamefont {Matsushita}}]{nishi2023analyzing}%
  \BibitemOpen
  \bibfield  {author} {\bibinfo {author} {\bibfnamefont {H.}~\bibnamefont
  {Nishi}}, \bibinfo {author} {\bibfnamefont {K.}~\bibnamefont {Hamada}},
  \bibinfo {author} {\bibfnamefont {Y.}~\bibnamefont {Nishiya}}, \bibinfo
  {author} {\bibfnamefont {T.}~\bibnamefont {Kosugi}},\ and\ \bibinfo {author}
  {\bibfnamefont {Y.-i.}\ \bibnamefont {Matsushita}},\ }\href
  {https://doi.org/10.1103/PhysRevResearch.5.043048} {\bibinfo {title} {Optimal
  scheduling in probabilistic imaginary-time evolution on a quantum computer}}
  (\bibinfo {year} {2023})\BibitemShut {NoStop}%
\bibitem [{\citenamefont {{Somma}}(2015)}]{bib:5391}%
  \BibitemOpen
  \bibfield  {author} {\bibinfo {author} {\bibfnamefont {R.~D.}\ \bibnamefont
  {{Somma}}},\ }\bibfield  {title} {\bibinfo {title} {{Quantum simulations of
  one dimensional quantum systems}},\ }\href@noop {} {\bibfield  {journal}
  {\bibinfo  {journal} {arXiv e-prints}\ ,\ \bibinfo {eid} {arXiv:1503.06319}}
  (\bibinfo {year} {2015})},\ \Eprint {https://arxiv.org/abs/1503.06319}
  {arXiv:1503.06319 [quant-ph]} \BibitemShut {NoStop}%
\bibitem [{\citenamefont {Ollitrault}\ \emph {et~al.}(2020)\citenamefont
  {Ollitrault}, \citenamefont {Mazzola},\ and\ \citenamefont
  {Tavernelli}}]{bib:5384}%
  \BibitemOpen
  \bibfield  {author} {\bibinfo {author} {\bibfnamefont {P.~J.}\ \bibnamefont
  {Ollitrault}}, \bibinfo {author} {\bibfnamefont {G.}~\bibnamefont
  {Mazzola}},\ and\ \bibinfo {author} {\bibfnamefont {I.}~\bibnamefont
  {Tavernelli}},\ }\bibfield  {title} {\bibinfo {title} {Nonadiabatic molecular
  quantum dynamics with quantum computers},\ }\href
  {https://doi.org/10.1103/PhysRevLett.125.260511} {\bibfield  {journal}
  {\bibinfo  {journal} {Phys. Rev. Lett.}\ }\textbf {\bibinfo {volume} {125}},\
  \bibinfo {pages} {260511} (\bibinfo {year} {2020})}\BibitemShut {NoStop}%
\bibitem [{\citenamefont {Childs}\ \emph {et~al.}(2021)\citenamefont {Childs},
  \citenamefont {Su}, \citenamefont {Tran}, \citenamefont {Wiebe},\ and\
  \citenamefont {Zhu}}]{bib:6196}%
  \BibitemOpen
  \bibfield  {author} {\bibinfo {author} {\bibfnamefont {A.~M.}\ \bibnamefont
  {Childs}}, \bibinfo {author} {\bibfnamefont {Y.}~\bibnamefont {Su}}, \bibinfo
  {author} {\bibfnamefont {M.~C.}\ \bibnamefont {Tran}}, \bibinfo {author}
  {\bibfnamefont {N.}~\bibnamefont {Wiebe}},\ and\ \bibinfo {author}
  {\bibfnamefont {S.}~\bibnamefont {Zhu}},\ }\bibfield  {title} {\bibinfo
  {title} {Theory of trotter error with commutator scaling},\ }\href
  {https://doi.org/10.1103/PhysRevX.11.011020} {\bibfield  {journal} {\bibinfo
  {journal} {Phys. Rev. X}\ }\textbf {\bibinfo {volume} {11}},\ \bibinfo
  {pages} {011020} (\bibinfo {year} {2021})}\BibitemShut {NoStop}%
\bibitem [{\citenamefont {{Draper}}(2000)}]{bib:5379}%
  \BibitemOpen
  \bibfield  {author} {\bibinfo {author} {\bibfnamefont {T.~G.}\ \bibnamefont
  {{Draper}}},\ }\bibfield  {title} {\bibinfo {title} {{Addition on a Quantum
  Computer}},\ }\href@noop {} {\bibfield  {journal} {\bibinfo  {journal} {arXiv
  e-prints}\ ,\ \bibinfo {eid} {quant-ph/0008033}} (\bibinfo {year} {2000})},\
  \Eprint {https://arxiv.org/abs/quant-ph/0008033} {arXiv:quant-ph/0008033
  [quant-ph]} \BibitemShut {NoStop}%
\bibitem [{\citenamefont {{Draper}}\ \emph {et~al.}(2004)\citenamefont
  {{Draper}}, \citenamefont {{Kutin}}, \citenamefont {{Rains}},\ and\
  \citenamefont {{Svore}}}]{bib:5396}%
  \BibitemOpen
  \bibfield  {author} {\bibinfo {author} {\bibfnamefont {T.~G.}\ \bibnamefont
  {{Draper}}}, \bibinfo {author} {\bibfnamefont {S.~A.}\ \bibnamefont
  {{Kutin}}}, \bibinfo {author} {\bibfnamefont {E.~M.}\ \bibnamefont
  {{Rains}}},\ and\ \bibinfo {author} {\bibfnamefont {K.~M.}\ \bibnamefont
  {{Svore}}},\ }\bibfield  {title} {\bibinfo {title} {{A logarithmic-depth
  quantum carry-lookahead adder}},\ }\href@noop {} {\bibfield  {journal}
  {\bibinfo  {journal} {arXiv e-prints}\ ,\ \bibinfo {eid} {quant-ph/0406142}}
  (\bibinfo {year} {2004})},\ \Eprint {https://arxiv.org/abs/quant-ph/0406142}
  {arXiv:quant-ph/0406142 [quant-ph]} \BibitemShut {NoStop}%
\bibitem [{\citenamefont {{Cuccaro}}\ \emph {et~al.}(2004)\citenamefont
  {{Cuccaro}}, \citenamefont {{Draper}}, \citenamefont {{Kutin}},\ and\
  \citenamefont {{Petrie Moulton}}}]{bib:5397}%
  \BibitemOpen
  \bibfield  {author} {\bibinfo {author} {\bibfnamefont {S.~A.}\ \bibnamefont
  {{Cuccaro}}}, \bibinfo {author} {\bibfnamefont {T.~G.}\ \bibnamefont
  {{Draper}}}, \bibinfo {author} {\bibfnamefont {S.~A.}\ \bibnamefont
  {{Kutin}}},\ and\ \bibinfo {author} {\bibfnamefont {D.}~\bibnamefont {{Petrie
  Moulton}}},\ }\bibfield  {title} {\bibinfo {title} {{A new quantum
  ripple-carry addition circuit}},\ }\href
  {https://doi.org/10.48550/arXiv.quant-ph/0410184} {\bibfield  {journal}
  {\bibinfo  {journal} {arXiv e-prints}\ ,\ \bibinfo {eid} {quant-ph/0410184}}
  (\bibinfo {year} {2004})},\ \Eprint {https://arxiv.org/abs/quant-ph/0410184}
  {arXiv:quant-ph/0410184 [quant-ph]} \BibitemShut {NoStop}%
\bibitem [{\citenamefont {Kowada}\ \emph {et~al.}(2006)\citenamefont {Kowada},
  \citenamefont {Portugal},\ and\ \citenamefont {de~Figueiredo}}]{bib:5405}%
  \BibitemOpen
  \bibfield  {author} {\bibinfo {author} {\bibfnamefont {L.~A.~B.}\
  \bibnamefont {Kowada}}, \bibinfo {author} {\bibfnamefont {R.}~\bibnamefont
  {Portugal}},\ and\ \bibinfo {author} {\bibfnamefont {C.~M.~H.}\ \bibnamefont
  {de~Figueiredo}},\ }\bibfield  {title} {\bibinfo {title} {Reversible
  karatsuba's algorithm},\ }\href@noop {} {\bibfield  {journal} {\bibinfo
  {journal} {Journal of Universal Computer Science}\ }\textbf {\bibinfo
  {volume} {12}},\ \bibinfo {pages} {499} (\bibinfo {year} {2006})}\BibitemShut
  {NoStop}%
\bibitem [{\citenamefont {{Parent}}\ \emph {et~al.}(2017)\citenamefont
  {{Parent}}, \citenamefont {{Roetteler}},\ and\ \citenamefont
  {{Mosca}}}]{bib:5398}%
  \BibitemOpen
  \bibfield  {author} {\bibinfo {author} {\bibfnamefont {A.}~\bibnamefont
  {{Parent}}}, \bibinfo {author} {\bibfnamefont {M.}~\bibnamefont
  {{Roetteler}}},\ and\ \bibinfo {author} {\bibfnamefont {M.}~\bibnamefont
  {{Mosca}}},\ }\bibfield  {title} {\bibinfo {title} {{Improved reversible and
  quantum circuits for Karatsuba-based integer multiplication}},\ }\href@noop
  {} {\bibfield  {journal} {\bibinfo  {journal} {arXiv e-prints}\ ,\ \bibinfo
  {eid} {arXiv:1706.03419}} (\bibinfo {year} {2017})},\ \Eprint
  {https://arxiv.org/abs/1706.03419} {arXiv:1706.03419 [quant-ph]} \BibitemShut
  {NoStop}%
\bibitem [{\citenamefont {Dutta}\ \emph {et~al.}(2018)\citenamefont {Dutta},
  \citenamefont {Bhattacharjee},\ and\ \citenamefont
  {Chattopadhyay}}]{bib:5399}%
  \BibitemOpen
  \bibfield  {author} {\bibinfo {author} {\bibfnamefont {S.}~\bibnamefont
  {Dutta}}, \bibinfo {author} {\bibfnamefont {D.}~\bibnamefont
  {Bhattacharjee}},\ and\ \bibinfo {author} {\bibfnamefont {A.}~\bibnamefont
  {Chattopadhyay}},\ }\bibfield  {title} {\bibinfo {title} {Quantum circuits
  for toom-cook multiplication},\ }\href
  {https://doi.org/10.1103/PhysRevA.98.012311} {\bibfield  {journal} {\bibinfo
  {journal} {Phys. Rev. A}\ }\textbf {\bibinfo {volume} {98}},\ \bibinfo
  {pages} {012311} (\bibinfo {year} {2018})}\BibitemShut {NoStop}%
\bibitem [{\citenamefont {{Hadfield}}(2018)}]{Hadfield}%
  \BibitemOpen
  \bibfield  {author} {\bibinfo {author} {\bibfnamefont {S.}~\bibnamefont
  {{Hadfield}}},\ }\bibfield  {title} {\bibinfo {title} {{Quantum Algorithms
  for Scientific Computing and Approximate Optimization}},\ }\href@noop {}
  {\bibfield  {journal} {\bibinfo  {journal} {arXiv e-prints}\ ,\ \bibinfo
  {eid} {arXiv:1805.03265}} (\bibinfo {year} {2018})},\ \Eprint
  {https://arxiv.org/abs/1805.03265} {arXiv:1805.03265 [quant-ph]} \BibitemShut
  {NoStop}%
\bibitem [{\citenamefont {Benenti}\ and\ \citenamefont
  {Strini}(2008)}]{bib:5390}%
  \BibitemOpen
  \bibfield  {author} {\bibinfo {author} {\bibfnamefont {G.}~\bibnamefont
  {Benenti}}\ and\ \bibinfo {author} {\bibfnamefont {G.}~\bibnamefont
  {Strini}},\ }\bibfield  {title} {\bibinfo {title} {Quantum simulation of the
  single-particle schrödinger equation},\ }\href
  {https://doi.org/10.1119/1.2894532} {\bibfield  {journal} {\bibinfo
  {journal} {American Journal of Physics}\ }\textbf {\bibinfo {volume} {76}},\
  \bibinfo {pages} {657} (\bibinfo {year} {2008})},\ \Eprint
  {https://arxiv.org/abs/https://doi.org/10.1119/1.2894532}
  {https://doi.org/10.1119/1.2894532} \BibitemShut {NoStop}%
\bibitem [{\citenamefont {{Brassard}}\ and\ \citenamefont
  {{Hoyer}}(1997)}]{bib:4884}%
  \BibitemOpen
  \bibfield  {author} {\bibinfo {author} {\bibfnamefont {G.}~\bibnamefont
  {{Brassard}}}\ and\ \bibinfo {author} {\bibfnamefont {P.}~\bibnamefont
  {{Hoyer}}},\ }\bibfield  {title} {\bibinfo {title} {An exact quantum
  polynomial-time algorithm for simon's problem},\ }in\ \href
  {https://doi.org/10.1109/ISTCS.1997.595153} {\emph {\bibinfo {booktitle}
  {Proceedings of the Fifth Israeli Symposium on Theory of Computing and
  Systems}}}\ (\bibinfo {year} {1997})\ pp.\ \bibinfo {pages}
  {12--23}\BibitemShut {NoStop}%
\bibitem [{\citenamefont {{Brassard}}\ \emph {et~al.}(2000)\citenamefont
  {{Brassard}}, \citenamefont {{Hoyer}}, \citenamefont {{Mosca}},\ and\
  \citenamefont {{Tapp}}}]{bib:4878}%
  \BibitemOpen
  \bibfield  {author} {\bibinfo {author} {\bibfnamefont {G.}~\bibnamefont
  {{Brassard}}}, \bibinfo {author} {\bibfnamefont {P.}~\bibnamefont {{Hoyer}}},
  \bibinfo {author} {\bibfnamefont {M.}~\bibnamefont {{Mosca}}},\ and\ \bibinfo
  {author} {\bibfnamefont {A.}~\bibnamefont {{Tapp}}},\ }\bibfield  {title}
  {\bibinfo {title} {{Quantum Amplitude Amplification and Estimation}},\
  }\href@noop {} {\bibfield  {journal} {\bibinfo  {journal} {arXiv e-prints}\
  ,\ \bibinfo {eid} {quant-ph/0005055}} (\bibinfo {year} {2000})},\ \Eprint
  {https://arxiv.org/abs/quant-ph/0005055} {arXiv:quant-ph/0005055 [quant-ph]}
  \BibitemShut {NoStop}%
\bibitem [{\citenamefont {{Nishi}}\ \emph {et~al.}(2022)\citenamefont
  {{Nishi}}, \citenamefont {{Kosugi}}, \citenamefont {{Nishiya}},\ and\
  \citenamefont {{Matsushita}}}]{Nishi_QAA}%
  \BibitemOpen
  \bibfield  {author} {\bibinfo {author} {\bibfnamefont {H.}~\bibnamefont
  {{Nishi}}}, \bibinfo {author} {\bibfnamefont {T.}~\bibnamefont {{Kosugi}}},
  \bibinfo {author} {\bibfnamefont {Y.}~\bibnamefont {{Nishiya}}},\ and\
  \bibinfo {author} {\bibfnamefont {Y.-i.}\ \bibnamefont {{Matsushita}}},\
  }\bibfield  {title} {\bibinfo {title} {{Acceleration of probabilistic
  imaginary-time evolution method combined with quantum amplitude
  amplification}},\ }\href {https://doi.org/10.48550/arXiv.2212.13816}
  {\bibfield  {journal} {\bibinfo  {journal} {arXiv e-prints}\ ,\ \bibinfo
  {eid} {arXiv:2212.13816}} (\bibinfo {year} {2022})},\ \Eprint
  {https://arxiv.org/abs/2212.13816} {arXiv:2212.13816 [quant-ph]} \BibitemShut
  {NoStop}%
\bibitem [{\citenamefont {{Nishi}}\ \emph {et~al.}(2023)\citenamefont
  {{Nishi}}, \citenamefont {{Kosugi}}, \citenamefont {{Nishiya}},\ and\
  \citenamefont {{Matsushita}}}]{2023arXiv230803605N}%
  \BibitemOpen
  \bibfield  {author} {\bibinfo {author} {\bibfnamefont {H.}~\bibnamefont
  {{Nishi}}}, \bibinfo {author} {\bibfnamefont {T.}~\bibnamefont {{Kosugi}}},
  \bibinfo {author} {\bibfnamefont {Y.}~\bibnamefont {{Nishiya}}},\ and\
  \bibinfo {author} {\bibfnamefont {Y.-i.}\ \bibnamefont {{Matsushita}}},\
  }\bibfield  {title} {\bibinfo {title} {{Quadratic speedups of multi-step
  probabilistic algorithms in state preparation}},\ }\href@noop {} {\bibfield
  {journal} {\bibinfo  {journal} {arXiv e-prints}\ ,\ \bibinfo {eid}
  {arXiv:2308.03605}} (\bibinfo {year} {2023})},\ \Eprint
  {https://arxiv.org/abs/2308.03605} {arXiv:2308.03605 [quant-ph]} \BibitemShut
  {NoStop}%
\bibitem [{\citenamefont {Tempel}\ \emph {et~al.}(2009)\citenamefont {Tempel},
  \citenamefont {Mart{\'i}nez},\ and\ \citenamefont {Maitra}}]{bib:3919}%
  \BibitemOpen
  \bibfield  {author} {\bibinfo {author} {\bibfnamefont {D.~G.}\ \bibnamefont
  {Tempel}}, \bibinfo {author} {\bibfnamefont {T.~J.}\ \bibnamefont
  {Mart{\'i}nez}},\ and\ \bibinfo {author} {\bibfnamefont {N.~T.}\ \bibnamefont
  {Maitra}},\ }\bibfield  {title} {\bibinfo {title} {Revisiting molecular
  dissociation in density functional theory: A simple model},\ }\href
  {https://doi.org/10.1021/ct800535c} {\bibfield  {journal} {\bibinfo
  {journal} {Journal of Chemical Theory and Computation}\ }\textbf {\bibinfo
  {volume} {5}},\ \bibinfo {pages} {770} (\bibinfo {year} {2009})}\BibitemShut
  {NoStop}%
\bibitem [{\citenamefont {Li}(2021)}]{bib:5679}%
  \BibitemOpen
  \bibfield  {author} {\bibinfo {author} {\bibfnamefont {C.}~\bibnamefont
  {Li}},\ }\bibfield  {title} {\bibinfo {title} {Exact analytical solution of
  the ground-state hydrogenic problem with soft coulomb potential},\ }\href
  {https://doi.org/10.1021/acs.jpca.1c00698} {\bibfield  {journal} {\bibinfo
  {journal} {The Journal of Physical Chemistry A}\ }\textbf {\bibinfo {volume}
  {125}},\ \bibinfo {pages} {5146} (\bibinfo {year} {2021})},\ \bibinfo {note}
  {pMID: 34096283},\ \Eprint
  {https://arxiv.org/abs/https://doi.org/10.1021/acs.jpca.1c00698}
  {https://doi.org/10.1021/acs.jpca.1c00698} \BibitemShut {NoStop}%
\bibitem [{\citenamefont {Wagner}\ \emph {et~al.}(2012)\citenamefont {Wagner},
  \citenamefont {Stoudenmire}, \citenamefont {Burke},\ and\ \citenamefont
  {White}}]{bib:5799}%
  \BibitemOpen
  \bibfield  {author} {\bibinfo {author} {\bibfnamefont {L.~O.}\ \bibnamefont
  {Wagner}}, \bibinfo {author} {\bibfnamefont {E.~M.}\ \bibnamefont
  {Stoudenmire}}, \bibinfo {author} {\bibfnamefont {K.}~\bibnamefont {Burke}},\
  and\ \bibinfo {author} {\bibfnamefont {S.~R.}\ \bibnamefont {White}},\
  }\bibfield  {title} {\bibinfo {title} {Reference electronic structure
  calculations in one dimension},\ }\href {https://doi.org/10.1039/C2CP24118H}
  {\bibfield  {journal} {\bibinfo  {journal} {Phys. Chem. Chem. Phys.}\
  }\textbf {\bibinfo {volume} {14}},\ \bibinfo {pages} {8581} (\bibinfo {year}
  {2012})}\BibitemShut {NoStop}%
\bibitem [{\citenamefont {Kandala}\ \emph {et~al.}(2017)\citenamefont
  {Kandala}, \citenamefont {Mezzacapo}, \citenamefont {Temme}, \citenamefont
  {Takita}, \citenamefont {Brink}, \citenamefont {Chow},\ and\ \citenamefont
  {Gambetta}}]{bib:4292}%
  \BibitemOpen
  \bibfield  {author} {\bibinfo {author} {\bibfnamefont {A.}~\bibnamefont
  {Kandala}}, \bibinfo {author} {\bibfnamefont {A.}~\bibnamefont {Mezzacapo}},
  \bibinfo {author} {\bibfnamefont {K.}~\bibnamefont {Temme}}, \bibinfo
  {author} {\bibfnamefont {M.}~\bibnamefont {Takita}}, \bibinfo {author}
  {\bibfnamefont {M.}~\bibnamefont {Brink}}, \bibinfo {author} {\bibfnamefont
  {J.~M.}\ \bibnamefont {Chow}},\ and\ \bibinfo {author} {\bibfnamefont
  {J.~M.}\ \bibnamefont {Gambetta}},\ }\bibfield  {title} {\bibinfo {title}
  {Hardware-efficient variational quantum eigensolver for small molecules and
  quantum magnets},\ }\href
  {https://doi.org/https://doi.org/10.1038/nature23879} {\bibfield  {journal}
  {\bibinfo  {journal} {Nature}\ }\textbf {\bibinfo {volume} {549}},\ \bibinfo
  {pages} {242} (\bibinfo {year} {2017})}\BibitemShut {NoStop}%
\bibitem [{\citenamefont {Suzuki}\ \emph {et~al.}(2021)\citenamefont {Suzuki},
  \citenamefont {Kawase}, \citenamefont {Masumura}, \citenamefont {Hiraga},
  \citenamefont {Nakadai}, \citenamefont {Chen}, \citenamefont {Nakanishi},
  \citenamefont {Mitarai}, \citenamefont {Imai}, \citenamefont {Tamiya},
  \citenamefont {Yamamoto}, \citenamefont {Yan}, \citenamefont {Kawakubo},
  \citenamefont {Nakagawa}, \citenamefont {Ibe}, \citenamefont {Zhang},
  \citenamefont {Yamashita}, \citenamefont {Yoshimura}, \citenamefont
  {Hayashi},\ and\ \citenamefont {Fujii}}]{Suzuki2021qulacsfast}%
  \BibitemOpen
  \bibfield  {author} {\bibinfo {author} {\bibfnamefont {Y.}~\bibnamefont
  {Suzuki}}, \bibinfo {author} {\bibfnamefont {Y.}~\bibnamefont {Kawase}},
  \bibinfo {author} {\bibfnamefont {Y.}~\bibnamefont {Masumura}}, \bibinfo
  {author} {\bibfnamefont {Y.}~\bibnamefont {Hiraga}}, \bibinfo {author}
  {\bibfnamefont {M.}~\bibnamefont {Nakadai}}, \bibinfo {author} {\bibfnamefont
  {J.}~\bibnamefont {Chen}}, \bibinfo {author} {\bibfnamefont {K.~M.}\
  \bibnamefont {Nakanishi}}, \bibinfo {author} {\bibfnamefont {K.}~\bibnamefont
  {Mitarai}}, \bibinfo {author} {\bibfnamefont {R.}~\bibnamefont {Imai}},
  \bibinfo {author} {\bibfnamefont {S.}~\bibnamefont {Tamiya}}, \bibinfo
  {author} {\bibfnamefont {T.}~\bibnamefont {Yamamoto}}, \bibinfo {author}
  {\bibfnamefont {T.}~\bibnamefont {Yan}}, \bibinfo {author} {\bibfnamefont
  {T.}~\bibnamefont {Kawakubo}}, \bibinfo {author} {\bibfnamefont {Y.~O.}\
  \bibnamefont {Nakagawa}}, \bibinfo {author} {\bibfnamefont {Y.}~\bibnamefont
  {Ibe}}, \bibinfo {author} {\bibfnamefont {Y.}~\bibnamefont {Zhang}}, \bibinfo
  {author} {\bibfnamefont {H.}~\bibnamefont {Yamashita}}, \bibinfo {author}
  {\bibfnamefont {H.}~\bibnamefont {Yoshimura}}, \bibinfo {author}
  {\bibfnamefont {A.}~\bibnamefont {Hayashi}},\ and\ \bibinfo {author}
  {\bibfnamefont {K.}~\bibnamefont {Fujii}},\ }\bibfield  {title} {\bibinfo
  {title} {Qulacs: a fast and versatile quantum circuit simulator for research
  purpose},\ }\href {https://doi.org/10.22331/q-2021-10-06-559} {\bibfield
  {journal} {\bibinfo  {journal} {{Quantum}}\ }\textbf {\bibinfo {volume}
  {5}},\ \bibinfo {pages} {559} (\bibinfo {year} {2021})}\BibitemShut {NoStop}%
\bibitem [{\citenamefont {Pirani}\ \emph {et~al.}(2004)\citenamefont {Pirani},
  \citenamefont {Alberti}, \citenamefont {Castro}, \citenamefont {Teixidor},\
  and\ \citenamefont {Cappelletti}}]{bib:5834}%
  \BibitemOpen
  \bibfield  {author} {\bibinfo {author} {\bibfnamefont {F.}~\bibnamefont
  {Pirani}}, \bibinfo {author} {\bibfnamefont {M.}~\bibnamefont {Alberti}},
  \bibinfo {author} {\bibfnamefont {A.}~\bibnamefont {Castro}}, \bibinfo
  {author} {\bibfnamefont {M.}~\bibnamefont {Teixidor}},\ and\ \bibinfo
  {author} {\bibfnamefont {D.}~\bibnamefont {Cappelletti}},\ }\bibfield
  {title} {\bibinfo {title} {Atom–bond pairwise additive representation for
  intermolecular potential energy surfaces},\ }\href
  {https://doi.org/https://doi.org/10.1016/j.cplett.2004.06.100} {\bibfield
  {journal} {\bibinfo  {journal} {Chemical Physics Letters}\ }\textbf {\bibinfo
  {volume} {394}},\ \bibinfo {pages} {37} (\bibinfo {year} {2004})}\BibitemShut
  {NoStop}%
\bibitem [{\citenamefont {Pirani}\ \emph {et~al.}(2008)\citenamefont {Pirani},
  \citenamefont {Brizi}, \citenamefont {Roncaratti}, \citenamefont
  {Casavecchia}, \citenamefont {Cappellettigege},\ and\ \citenamefont
  {Vecchiocattivi}}]{bib:5835}%
  \BibitemOpen
  \bibfield  {author} {\bibinfo {author} {\bibfnamefont {F.}~\bibnamefont
  {Pirani}}, \bibinfo {author} {\bibfnamefont {S.}~\bibnamefont {Brizi}},
  \bibinfo {author} {\bibfnamefont {L.~F.}\ \bibnamefont {Roncaratti}},
  \bibinfo {author} {\bibfnamefont {P.}~\bibnamefont {Casavecchia}}, \bibinfo
  {author} {\bibfnamefont {D.}~\bibnamefont {Cappellettigege}},\ and\ \bibinfo
  {author} {\bibfnamefont {F.}~\bibnamefont {Vecchiocattivi}},\ }\bibfield
  {title} {\bibinfo {title} {Beyond the lennard-jones model: a simple and
  accurate potential function probed by high resolution scattering data useful
  for molecular dynamics simulations},\ }\href
  {https://doi.org/10.1039/B808524B} {\bibfield  {journal} {\bibinfo  {journal}
  {Physical Chemistry Chemical Physics}\ }\textbf {\bibinfo {volume} {10}},\
  \bibinfo {pages} {5489} (\bibinfo {year} {2008})}\BibitemShut {NoStop}%
\bibitem [{\citenamefont {Pirani}\ \emph {et~al.}(2003)\citenamefont {Pirani},
  \citenamefont {Porrini}, \citenamefont {Cavalli}, \citenamefont
  {Bartolomei},\ and\ \citenamefont {Cappellettigaga}}]{bib:5837}%
  \BibitemOpen
  \bibfield  {author} {\bibinfo {author} {\bibfnamefont {F.}~\bibnamefont
  {Pirani}}, \bibinfo {author} {\bibfnamefont {M.}~\bibnamefont {Porrini}},
  \bibinfo {author} {\bibfnamefont {S.}~\bibnamefont {Cavalli}}, \bibinfo
  {author} {\bibfnamefont {M.}~\bibnamefont {Bartolomei}},\ and\ \bibinfo
  {author} {\bibfnamefont {D.}~\bibnamefont {Cappellettigaga}},\ }\bibfield
  {title} {\bibinfo {title} {Potential energy surfaces for the benzene–rare
  gas systems},\ }\href
  {https://doi.org/https://doi.org/10.1016/S0009-2614(02)01540-3} {\bibfield
  {journal} {\bibinfo  {journal} {Chemical Physics Letters}\ }\textbf {\bibinfo
  {volume} {367}},\ \bibinfo {pages} {405} (\bibinfo {year}
  {2003})}\BibitemShut {NoStop}%
\bibitem [{\citenamefont {Martyn}\ \emph {et~al.}(2021)\citenamefont {Martyn},
  \citenamefont {Rossi}, \citenamefont {Tan},\ and\ \citenamefont
  {Chuang}}]{bib:5728}%
  \BibitemOpen
  \bibfield  {author} {\bibinfo {author} {\bibfnamefont {J.~M.}\ \bibnamefont
  {Martyn}}, \bibinfo {author} {\bibfnamefont {Z.~M.}\ \bibnamefont {Rossi}},
  \bibinfo {author} {\bibfnamefont {A.~K.}\ \bibnamefont {Tan}},\ and\ \bibinfo
  {author} {\bibfnamefont {I.~L.}\ \bibnamefont {Chuang}},\ }\bibfield  {title}
  {\bibinfo {title} {Grand unification of quantum algorithms},\ }\href
  {https://doi.org/10.1103/PRXQuantum.2.040203} {\bibfield  {journal} {\bibinfo
   {journal} {PRX Quantum}\ }\textbf {\bibinfo {volume} {2}},\ \bibinfo {pages}
  {040203} (\bibinfo {year} {2021})}\BibitemShut {NoStop}%
\bibitem [{\citenamefont {Kosugi}\ \emph {et~al.}(2023)\citenamefont {Kosugi},
  \citenamefont {Nishi},\ and\ \citenamefont {ichiro Matsushita}}]{bib:6103}%
  \BibitemOpen
  \bibfield  {author} {\bibinfo {author} {\bibfnamefont {T.}~\bibnamefont
  {Kosugi}}, \bibinfo {author} {\bibfnamefont {H.}~\bibnamefont {Nishi}},\ and\
  \bibinfo {author} {\bibfnamefont {Y.}~\bibnamefont {ichiro Matsushita}},\
  }\bibfield  {title} {\bibinfo {title} {First-quantized eigensolver for ground
  and excited states of electrons under a uniform magnetic field},\ }\href
  {https://doi.org/10.35848/1347-4065/acddc0} {\bibfield  {journal} {\bibinfo
  {journal} {Japanese Journal of Applied Physics}\ }\textbf {\bibinfo {volume}
  {62}},\ \bibinfo {pages} {062004} (\bibinfo {year} {2023})}\BibitemShut
  {NoStop}%
\bibitem [{\citenamefont {Klco}\ and\ \citenamefont {Savage}(2020)}]{bib:5645}%
  \BibitemOpen
  \bibfield  {author} {\bibinfo {author} {\bibfnamefont {N.}~\bibnamefont
  {Klco}}\ and\ \bibinfo {author} {\bibfnamefont {M.~J.}\ \bibnamefont
  {Savage}},\ }\bibfield  {title} {\bibinfo {title} {Minimally entangled state
  preparation of localized wave functions on quantum computers},\ }\href
  {https://doi.org/10.1103/PhysRevA.102.012612} {\bibfield  {journal} {\bibinfo
   {journal} {Phys. Rev. A}\ }\textbf {\bibinfo {volume} {102}},\ \bibinfo
  {pages} {012612} (\bibinfo {year} {2020})}\BibitemShut {NoStop}%
\bibitem [{\citenamefont {Mozafari}\ \emph {et~al.}(2022)\citenamefont
  {Mozafari}, \citenamefont {De~Micheli},\ and\ \citenamefont
  {Yang}}]{bib:5767}%
  \BibitemOpen
  \bibfield  {author} {\bibinfo {author} {\bibfnamefont {F.}~\bibnamefont
  {Mozafari}}, \bibinfo {author} {\bibfnamefont {G.}~\bibnamefont
  {De~Micheli}},\ and\ \bibinfo {author} {\bibfnamefont {Y.}~\bibnamefont
  {Yang}},\ }\bibfield  {title} {\bibinfo {title} {Efficient deterministic
  preparation of quantum states using decision diagrams},\ }\href
  {https://doi.org/10.1103/PhysRevA.106.022617} {\bibfield  {journal} {\bibinfo
   {journal} {Phys. Rev. A}\ }\textbf {\bibinfo {volume} {106}},\ \bibinfo
  {pages} {022617} (\bibinfo {year} {2022})}\BibitemShut {NoStop}%
\bibitem [{\citenamefont {Zhang}\ \emph {et~al.}(2022)\citenamefont {Zhang},
  \citenamefont {Li},\ and\ \citenamefont {Yuan}}]{bib:5646}%
  \BibitemOpen
  \bibfield  {author} {\bibinfo {author} {\bibfnamefont {X.-M.}\ \bibnamefont
  {Zhang}}, \bibinfo {author} {\bibfnamefont {T.}~\bibnamefont {Li}},\ and\
  \bibinfo {author} {\bibfnamefont {X.}~\bibnamefont {Yuan}},\ }\bibfield
  {title} {\bibinfo {title} {Quantum state preparation with optimal circuit
  depth: Implementations and applications},\ }\href
  {https://doi.org/10.1103/PhysRevLett.129.230504} {\bibfield  {journal}
  {\bibinfo  {journal} {Phys. Rev. Lett.}\ }\textbf {\bibinfo {volume} {129}},\
  \bibinfo {pages} {230504} (\bibinfo {year} {2022})}\BibitemShut {NoStop}%
\bibitem [{\citenamefont {{Sun}}\ \emph {et~al.}(2021)\citenamefont {{Sun}},
  \citenamefont {{Tian}}, \citenamefont {{Yang}}, \citenamefont {{Yuan}},\ and\
  \citenamefont {{Zhang}}}]{bib:5697}%
  \BibitemOpen
  \bibfield  {author} {\bibinfo {author} {\bibfnamefont {X.}~\bibnamefont
  {{Sun}}}, \bibinfo {author} {\bibfnamefont {G.}~\bibnamefont {{Tian}}},
  \bibinfo {author} {\bibfnamefont {S.}~\bibnamefont {{Yang}}}, \bibinfo
  {author} {\bibfnamefont {P.}~\bibnamefont {{Yuan}}},\ and\ \bibinfo {author}
  {\bibfnamefont {S.}~\bibnamefont {{Zhang}}},\ }\bibfield  {title} {\bibinfo
  {title} {{Asymptotically Optimal Circuit Depth for Quantum State Preparation
  and General Unitary Synthesis}},\ }\href
  {https://doi.org/10.48550/arXiv.2108.06150} {\bibfield  {journal} {\bibinfo
  {journal} {arXiv e-prints}\ ,\ \bibinfo {eid} {arXiv:2108.06150}} (\bibinfo
  {year} {2021})},\ \Eprint {https://arxiv.org/abs/2108.06150}
  {arXiv:2108.06150 [quant-ph]} \BibitemShut {NoStop}%
\bibitem [{\citenamefont {Nielsen}\ and\ \citenamefont
  {Chuang}(2011)}]{Nielsen_and_Chuang}%
  \BibitemOpen
  \bibfield  {author} {\bibinfo {author} {\bibfnamefont {M.~A.}\ \bibnamefont
  {Nielsen}}\ and\ \bibinfo {author} {\bibfnamefont {I.~L.}\ \bibnamefont
  {Chuang}},\ }\href@noop {} {\emph {\bibinfo {title} {Quantum Computation and
  Quantum Information: 10th Anniversary Edition}}},\ \bibinfo {edition} {10th}\
  ed.\ (\bibinfo  {publisher} {Cambridge University Press},\ \bibinfo {address}
  {New York, NY, USA},\ \bibinfo {year} {2011})\BibitemShut {NoStop}%
\bibitem [{\citenamefont {Gidney}(2017)}]{phase_gradient_circuits}%
  \BibitemOpen
  \bibfield  {author} {\bibinfo {author} {\bibfnamefont {C.}~\bibnamefont
  {Gidney}},\ }\href@noop {} {\bibinfo {title} {Efficient controlled phase
  gradients}},\ \bibinfo {howpublished} {\url{https://algassert.com/}}
  (\bibinfo {year} {2017})\BibitemShut {NoStop}%
\bibitem [{\citenamefont {da~Silva}\ and\ \citenamefont
  {Park}(2022)}]{bib:5662}%
  \BibitemOpen
  \bibfield  {author} {\bibinfo {author} {\bibfnamefont {A.~J.}\ \bibnamefont
  {da~Silva}}\ and\ \bibinfo {author} {\bibfnamefont {D.~K.}\ \bibnamefont
  {Park}},\ }\bibfield  {title} {\bibinfo {title} {Linear-depth quantum
  circuits for multiqubit controlled gates},\ }\href
  {https://doi.org/10.1103/PhysRevA.106.042602} {\bibfield  {journal} {\bibinfo
   {journal} {Phys. Rev. A}\ }\textbf {\bibinfo {volume} {106}},\ \bibinfo
  {pages} {042602} (\bibinfo {year} {2022})}\BibitemShut {NoStop}%
\bibitem [{\citenamefont {McLachlan}(1964)}]{bib:4803}%
  \BibitemOpen
  \bibfield  {author} {\bibinfo {author} {\bibfnamefont {A.}~\bibnamefont
  {McLachlan}},\ }\bibfield  {title} {\bibinfo {title} {A variational solution
  of the time-dependent schrodinger equation},\ }\href
  {https://doi.org/https://doi.org/10.1080/00268976400100041} {\bibfield
  {journal} {\bibinfo  {journal} {Molecular Physics}\ }\textbf {\bibinfo
  {volume} {8}},\ \bibinfo {pages} {39} (\bibinfo {year} {1964})}\BibitemShut
  {NoStop}%
\bibitem [{\citenamefont {Schuld}\ \emph {et~al.}(2019)\citenamefont {Schuld},
  \citenamefont {Bergholm}, \citenamefont {Gogolin}, \citenamefont {Izaac},\
  and\ \citenamefont {Killoran}}]{bib:5838}%
  \BibitemOpen
  \bibfield  {author} {\bibinfo {author} {\bibfnamefont {M.}~\bibnamefont
  {Schuld}}, \bibinfo {author} {\bibfnamefont {V.}~\bibnamefont {Bergholm}},
  \bibinfo {author} {\bibfnamefont {C.}~\bibnamefont {Gogolin}}, \bibinfo
  {author} {\bibfnamefont {J.}~\bibnamefont {Izaac}},\ and\ \bibinfo {author}
  {\bibfnamefont {N.}~\bibnamefont {Killoran}},\ }\bibfield  {title} {\bibinfo
  {title} {Evaluating analytic gradients on quantum hardware},\ }\href
  {https://doi.org/10.1103/PhysRevA.99.032331} {\bibfield  {journal} {\bibinfo
  {journal} {Phys. Rev. A}\ }\textbf {\bibinfo {volume} {99}},\ \bibinfo
  {pages} {032331} (\bibinfo {year} {2019})}\BibitemShut {NoStop}%
\bibitem [{\citenamefont {Mitarai}\ and\ \citenamefont
  {Fujii}(2019)}]{bib:5839}%
  \BibitemOpen
  \bibfield  {author} {\bibinfo {author} {\bibfnamefont {K.}~\bibnamefont
  {Mitarai}}\ and\ \bibinfo {author} {\bibfnamefont {K.}~\bibnamefont
  {Fujii}},\ }\bibfield  {title} {\bibinfo {title} {Methodology for replacing
  indirect measurements with direct measurements},\ }\href
  {https://doi.org/10.1103/PhysRevResearch.1.013006} {\bibfield  {journal}
  {\bibinfo  {journal} {Phys. Rev. Research}\ }\textbf {\bibinfo {volume}
  {1}},\ \bibinfo {pages} {013006} (\bibinfo {year} {2019})}\BibitemShut
  {NoStop}%
\bibitem [{\citenamefont {Ollitrault}\ \emph {et~al.}(2022)\citenamefont
  {Ollitrault}, \citenamefont {Jandura}, \citenamefont {Miessen}, \citenamefont
  {Burghardt}, \citenamefont {Martinazzo}, \citenamefont {Tacchino},\ and\
  \citenamefont {Tavernelli}}]{Ollitrault2022arXiv}%
  \BibitemOpen
  \bibfield  {author} {\bibinfo {author} {\bibfnamefont {P.~J.}\ \bibnamefont
  {Ollitrault}}, \bibinfo {author} {\bibfnamefont {S.}~\bibnamefont {Jandura}},
  \bibinfo {author} {\bibfnamefont {A.}~\bibnamefont {Miessen}}, \bibinfo
  {author} {\bibfnamefont {I.}~\bibnamefont {Burghardt}}, \bibinfo {author}
  {\bibfnamefont {R.}~\bibnamefont {Martinazzo}}, \bibinfo {author}
  {\bibfnamefont {F.}~\bibnamefont {Tacchino}},\ and\ \bibinfo {author}
  {\bibfnamefont {I.}~\bibnamefont {Tavernelli}},\ }\bibfield  {title}
  {\bibinfo {title} {Quantum algorithms for grid-based variational time
  evolution},\ }\bibfield  {journal} {\bibinfo  {journal} {arXiv preprint
  arXiv:2203.02521}\ }\href {https://doi.org/10.48550/ARXIV.2203.02521}
  {10.48550/ARXIV.2203.02521} (\bibinfo {year} {2022})\BibitemShut {NoStop}%
\end{thebibliography}%

\end{document}